\documentclass{jfm}

\usepackage{bm}
\usepackage{graphicx}
\usepackage{subfigure}
\usepackage{amsmath}
\usepackage{color}

\usepackage{amssymb}
\usepackage{natbib}

\ifCUPmtlplainloaded \else
  \iffontfound
    \IfFileExists{upmath.sty}
      {\typeout{^^JFound AMS Euler Roman fonts on the system,
                   using the 'upmath' package.^^J}%
       \usepackage{upmath}}
      {\typeout{^^JFound AMS Euler Roman fonts on the system, but you
                   dont seem to have the}%
       \typeout{'upmath' package installed. JFM.cls can take advantage
                 of these fonts,^^Jif you use 'upmath' package.^^J}%
      }
  \else
  \fi
\fi

\ifCUPmtlplainloaded \else
  \iffontfound
    \IfFileExists{amssymb.sty}
      {\typeout{^^JFound AMS Symbol fonts on the system, using the
                'amssymb' package.^^J}%
       \usepackage{amssymb}%
         \let\leq=\leqslant
         \let\geq=\geqslant
      }{}
  \fi
\fi

\ifCUPmtlplainloaded \else
  \IfFileExists{amsbsy.sty}
    {\typeout{^^JFound the 'amsbsy' package on the system, using it.^^J}%
     \usepackage{amsbsy}}
    {\providecommand\boldsymbol[1]{\mbox{\boldmath $##1$}}}
\fi

\newcommand\Real{\mbox{Re}} 
\newcommand\Imag{\mbox{Im}} 
\newcommand\Rey{\mbox{\textit{Re}}}  
\newcommand\Pm{\mbox{\textit{Pm}}} 
\newcommand\Reym{\mbox{\textit{Rm}}}  

 \newcommand\Aone{\protect $\mathcal{A}_1$}
 \newcommand\SNone{\protect $SN_1$}
 \newcommand\SNtwo{\protect $SN_2$}
\newcommand\Wu{\protect W^u}
\newcommand\Ws{\protect W^s}
\newcommand{\f}[2]{\frac{#1}{#2}}
\newcommand{\dpart}[2]{\frac{\partial #1}{\partial #2}}

\renewcommand{\vec}[1]{\boldsymbol{#1}}
\newcommand{\grad}[1]{\vec{\nabla}{#1}}
\newcommand{\curl}[1]{\vec{\nabla}\times{#1}}
\renewcommand{\div}[1]{\vec{\nabla}\cdot{#1}}

\newsavebox{\astrutbox}
\sbox{\astrutbox}{\rule[-5pt]{0pt}{20pt}}

\newcommand\Ba{\overline{\vec{B}}}
\newcommand\Bax{\overline{B}_x}

\newcommand\Bazero{\overline{\vec{B}}_0}
\newcommand\Bazerox{\overline{B}_{0x}}
\newcommand\Bazeroy{\overline{B}_{0y}}

\title[Global bifurcations to MRI dynamo action in Keplerian shear
flow]{Global bifurcations to subcritical magnetorotational
  dynamo action\\ in Keplerian shear flow}

\author[\ \ A. Riols, F. Rincon, C. Cossu, G. Lesur,  P.-Y. Longaretti,
G. I. Ogilvie, J. Herault]%
{A.\ns R\ls I\ls O\ls L\ls S$^{1,2}$,
 F.\ns R\ls I\ls N\ls C\ls O\ls N$^{1,2}$ \thanks{Email
   address for correspondence: francois.rincon@irap.omp.eu}, 
  C.\ns C\ls O\ls S\ls S\ls U$^{3}$,\\
  G.\ns  L\ls E\ls S\ls U\ls R\ls$^{4}$,
  P.-Y.\ns \ls L\ls O\ls N\ls G\ls A\ls R\ls E\ls T\ls T\ls I\ls
  $^{4}$,  
  G.\ns I.\ns O\ls G\ls I\ls L\ls V\ls I\ls E$^{5}$\ns \and\ns 
  J.\ns H\ls E\ls R\ls A\ls U\ls L\ls T$^{6}$}
  
\affiliation{$^1$Universit\'e de Toulouse; UPS-OMP; IRAP; Toulouse, France\\[\affilskip]
$^2$CNRS; IRAP; 14 avenue Edouard Belin, F-31400 Toulouse, France\\[\affilskip] 
$^3$CNRS-Institut de M\'ecanique des Fluides de Toulouse (IMFT),\\ All\'ee du Professeur Camille Soula, 31400 Toulouse, France\\[\affilskip]
$^4$UJF-Grenoble 1 / CNRS-INSU, Institut de Plan\'etologie et
d'Astrophysique de Grenoble (IPAG) UMR 5274, Grenoble, F-38041,
France\\[\affilskip]
$^5$Department of Applied Mathematics and Theoretical Physics,
University of Cambridge,\\ Centre for Mathematical Sciences, Wilberforce
Road, Cambridge CB3 0WA, United Kingdom\\[\affilskip]
$^6$Laboratoire de Physique Statistique de l'Ecole Normale Sup\'erieure, CNRS UMR 8550,\\
24 Rue Lhomond, 75231 Paris Cedex 05, France
}

\date{?; revised ?; accepted ?. - To be entered by editorial office}
\begin{document}

\maketitle

\begin{abstract}
Magnetorotational dynamo action in Keplerian shear flow is a
three-dimensio\-nal, nonlinear magnetohydrodynamic process
whose study is relevant to the understanding of accretion processes
and magnetic field generation in astrophysics. Transition to this form
of dynamo action is subcritical and shares many characteristics of
transition to turbulence in non-rotating hydrodynamic shear
flows. This suggests that these different fluid systems become active
through similar generic bifurcation mechanisms, which in both cases
have eluded detailed understanding so far. In this paper, we
build on recent work on the two problems to investigate numerically
the bifurcation mechanisms at work in the incompressible Keplerian
magnetorotational dynamo problem in the shearing box framework. Using
numerical techniques imported from dynamical
systems research, we show that the onset of chaotic dynamo action at
magnetic Prandtl numbers larger than unity is primarily associated with global
homoclinic and heteroclinic bifurcations of nonlinear
magnetorotational dynamo cycles. These
global bifurcations are found to be supplemented by local bifurcations
of cycles marking the beginning of period-doubling cascades.
The results suggest that nonlinear magnetorotational
dynamo cycles provide the pathway to turbulent injection of both
kinetic and magnetic energy in incompressible magnetohydrodynamic
Keplerian shear flow in the absence of an externally imposed magnetic
field. Studying the nonlinear physics and bifurcations of these
cycles in different regimes and configurations may subsequently help
to better understand the physical conditions of excitation of
magnetohydrodynamic turbulence and instability-driven dynamos in a
variety of astrophysical systems and laboratory experiments. The
detailed characterization of global bifurcations provided for this
three-dimensional subcritical fluid dynamics problem may also prove
useful for the problem of transition to turbulence in
hydrodynamic shear flows.
\end{abstract}

\section{Introduction\label{intro}}

\subsection{Scientific context and motivations\label{intro-context}}
Magnetorotational dynamo action is a three-dimensional, nonlinear
magnetohydrodynamic (MHD) process whose study is most directly
relevant to the understanding of accretion processes and magnetic
field generation in astrophysics. Accretion takes place in young
stellar objects, interacting binary systems and galactic nuclei and is known
in many cases to be mediated by a fluid disk
\citep{pringle81,papaloizou95,lin96,frank02,hartmann09,armitage10}.
Typical luminosity variation timescales inferred from observations of
these systems indicate that they accrete at a fairly high rate, which
requires that vertical angular momentum be
efficiently transported outwards through the disk
\citep{lyndenbell74}. Identifying and characterizing efficient transport
mechanisms in disks is one of the most important
questions of astrophysical fluid dynamics
\citep{balbus98,balbus03}.

 Turbulent transport is a most plausible candidate in this respect and
 is thought to be related to the nonlinear development of the
 magnetorotational instability (MRI), 
a linear MHD instability occurring in differentially rotating flows
whose rotation rate $\Omega$ decreases with the distance to the rotation
axis \citep{velikhov59,chandra60,balbus91}. Keplerian flow,
$\Omega(r)\propto 1/r^{3/2}$, is the most studied
MRI-unstable flow. The MRI draws energy from the background shear but
requires the presence of a magnetic field $\vec{B}$, as the
instability is mediated by magnetic tension. For instance,
plane waves with a wavenumber $\vec{k}$ can only be MRI-unstable if $\vec{k}\cdot\vec{B}\neq 0$. 
In the presence of a uniform background magnetic field threading the
flow, the MRI acts as a genuine linear instability which proceeds
exponentially in time and breaks down into MHD turbulence in both two and
three dimensions \citep{hawley91,hawley95}. How efficient this
turbulence is at transporting angular momentum remains a
matter of debate \citep{fromang07a,lesur07,longaretti10}
and may significantly depend on the physical characteristics
of the disk plasma, such as its magnetic Prandtl number $\Pm$
\citep{balbus08}.  Another important issue is that this simple
configuration is  probably not relevant to all kinds of accreting
systems. In the absence of an externally imposed field (henceforth
referred to as a zero net-flux configuration), one may wonder whether
MRI-driven MHD turbulence can be sustained at all in disks. For
this to be possible, vigorous  magnetic induction must be present to
sustain a zero net-flux, MRI-mediating field against ohmic diffusion,
but the Keplerian flow itself is not a dynamo. One possibility is that 
the field is sustained by the MRI-driven turbulence itself, leading  
to a dynamo in which magnetic field generation and turbulence
activation processes are intrinsically coupled. This is the MRI
\textit{dynamo} problem. 

Early numerical simulations indicated
that this form of dynamo can indeed be excited in Keplerian
flow, but only in three dimensions (as expected from Cowling's
anti-dynamo theorem), only for initial perturbations of finite
amplitude, and only if the dynamical effects of the magnetic field 
on the flow (magnetic tension) are taken into account
 \citep{branden95,hawley96}. The latter
requirements clearly suggest that some form of MRI is active
in the process and that the dynamo is not kinematic. 
This built-in nonlinearity is one of the main reasons why it has so
far proven very difficult to understand the dynamo transition even
qualitatively. In particular, the very existence of an MRI dynamo
mechanism at low $\Pm$ (the most common dissipative regime in disks)
remains a widely debated issue
\citep{fromang07b,davis10,bodo11,kapyla11,simon11,oishi11}.

The MRI dynamo actually provides a clear illustration  of a whole class of
subcritical dynamos in (possibly rotating) shear flows prone to MHD
instabilities, sometimes referred to as instability-driven dynamos
\citep[e. g.][]{spruit02,cline03,rincon08}. These dynamos are
very interesting from the astrophysical point of view because they
seem to naturally generate time-dependent, system-scale coherent magnetic
fields but appear to be different in nature from all kinds of
linear kinematic dynamos commonly invoked in the astrophysical context
\citep{branden95}, such as the  mean-field \citep{steenbeck66,moffatt77}
and fluctuation (turbulent, small-scale) dynamos \citep{zeldo84,childress95}.
The MRI dynamo is also a good candidate for detection in laboratory
experiments, which makes it a particularly worthwhile object of study
in the broader context of dynamo theory.

Uncovering the exact bifurcation mechanisms underlying the MRI dynamo
transition has not yet proven possible but is clearly an essential
prerequisite to making progress on a variety of astrophysical and MHD
problems. The objective of this paper is to address
this transition problem from a fundamental fluid dynamics perspective,
mostly leaving aside astrophysical applications. As will be explained in 
detail shortly, the problem is actually similar to that of hydrodynamic
transition to turbulence of non-rotating linearly stable shear flows,
about which much has been learned in recent years. The approach taken on
in this paper is based on the assumption that similar progress on the
MRI dynamo problem may be possible along the same lines. To motivate
this approach, we find it useful to offer a short summary of the status
of research on hydrodynamic shear flows in the next paragraph. Readers
familiar with the problem may jump directly to
\S\ref{principles-SSPMHD}, where the connections with the MRI dynamo
are explained in detail.

\subsection{Hydrodynamic transition in shear flows\label{principles-hydro}}
\begin{figure}
   \centering
\includegraphics[width=\textwidth]{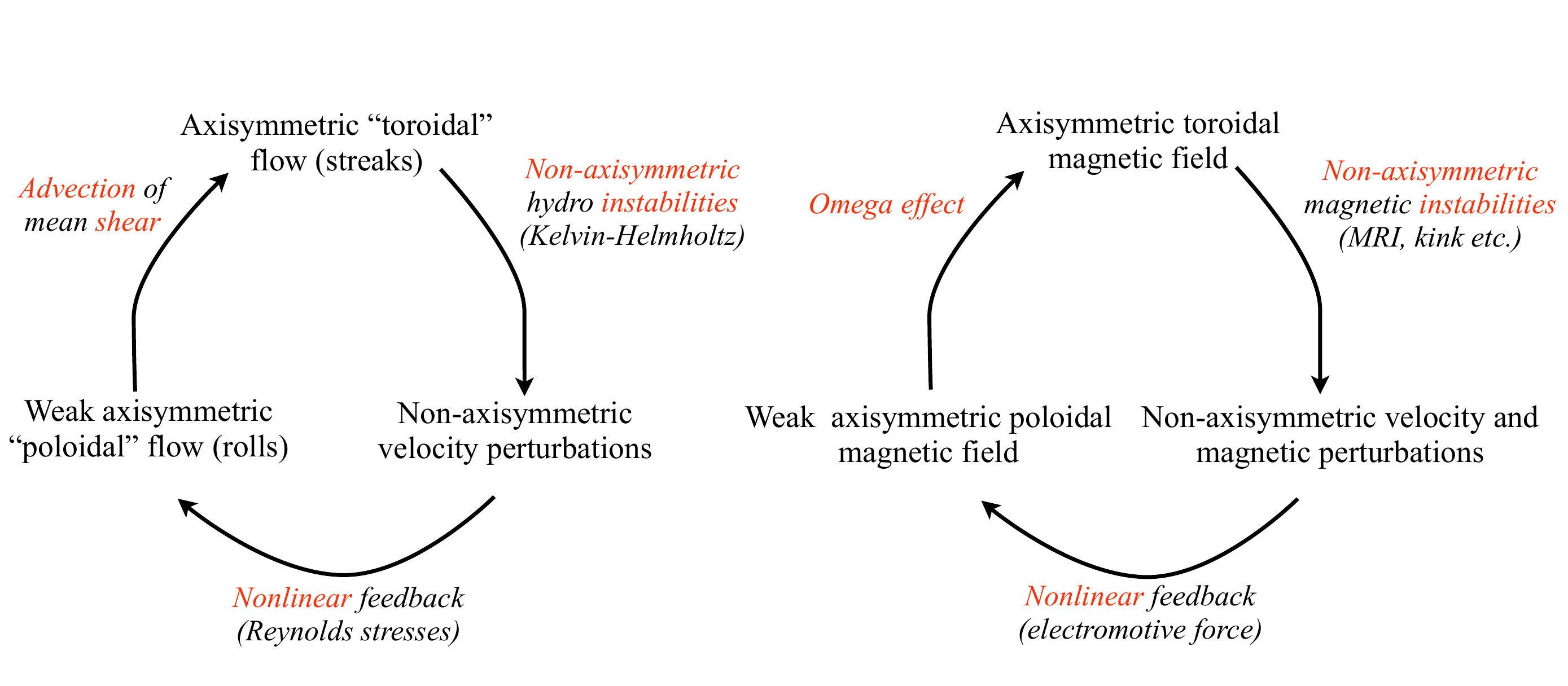}
\caption{Hydrodynamic self-sustaining process
  \citep{waleffe97} and self-sustaining dynamo processes in shear flows
prone to MHD instabilities \citep{rincon07b,rincon08}.\label{SSP}}
\end{figure}

The origins of hydrodynamic turbulence in linearly stable non-rotating shear
flows is a long-standing problem \citep{reynolds83} which has been extensively
reviewed \citep{grossmann2000,kerswell05,eckhardt07,eckhardt09,kawahara12}.
Transition in such flows is known both experimentally and numerically
to be subcritical \citep{darbyshire95,dauchot95a,hof03}, and
numerical and theoretical work conducted in the 1990s indicated
that the transitional dynamics relies on a nonlinear hydrodynamic
self-sustaining process (SSP)
\citep{hamilton95,waleffe95a,waleffe95b,waleffe97}.
Let us denote the direction aligned with the base flow as the
streamwise direction. The self-sustaining process, illustrated in
figure~\ref{SSP} (left), consists in an interplay between non-normal
transient amplification of streamwise-independent velocity structures (the
so-called streaks) by the lift-up mechanism  acting on weak
streamwise-independent streamwise vortices
\citep{landahl80,schmid00,dauchot95b}, 
and streamwise-dependent unstable perturbations of the streaks.
These perturbations feed back nonlinearly on the streamwise vortices, 
making it possible for the flow as a whole to be self-sustaining
\citep{waleffe95b}. This mechanism is believed to be active even in
fully developed turbulent flows \citep{hwang2010,hwang2011}.

Another characteristic of hydrodynamic shear flow turbulence in many
laboratory and numerical experiments is its transient nature,
which suggests that the transitional dynamics is structured around a chaotic
saddle, at least in constrained geometries\footnote{The situation may be
  different in spatially extended systems in which spatiotemporal
  chaos can develop, see discussion in \S\ref{hydrodiscussion}.}
\citep{faisst04,kerswell05,hof06,peixinho06,willis07,hof08,schneider08a}. 
The border in phase space separating initial
conditions leading to rapid relaminarization of the flow from initial
conditions leading to longer lived (but still ultimately decaying)
chaotic states is a complicated surface commonly referred to as the
``edge of chaos'', on which flow trajectories seem to converge to
a relative fractal
attractor. \citep{Schmiegel1997,Moehlis2004a,Moehlis2004b,skufca06,schneider06,schneider07a,
 eckhardt07,willis07,schneider08b,kim08,duguet08b,eckhardt09,mellibovsky09,Vollmer2009,lozar12}.

The causes of the transition have mostly been investigated using direct
numerical simulations of reduced dynamical models
 \citep{moehlis02,Moehlis2005,Vollmer2009,lebovitz2009,lebovitz2012},
of pipe flow \citep{schneider07b,kerswell07,duguet08a,duguet08b,willis2009,
mellibovsky11,mellibovsky12,willis13} and of plane Couette flow 
\citep{wang07,gibson08,halcrow09,vanveen11,kreilos12}.
Simple, three-dimensional nonlinear invariant solutions (fixed points,
travelling waves, periodic and relative periodic orbits) of the
Navier-Stokes equations whose  dynamics relies on the self-sustaining process  
seem to play a critical role in the transition
\citep{nagata90,waleffe98,waleffe01,itano2001,kawahara01,waleffe03,faisst03,hof04,wedin04,pringle07,
viswanath07,duguet08a,gibson08,gibson09,pringle09}. However, a detailed understanding of
the underlying bifurcation mechanisms and of how these solutions
relate to the edge of chaos is still lacking. One of the most favoured
hypotheses put forward by
  \cite{Schmiegel1997} is the occurrence of chaos-generating global
  bifurcations involving the stable and unstable manifolds of such
  invariant "edge states". Indirect evidence for such bifurcations 
  in the form of homoclinic and heteroclinic orbits has only recently
  been reported in plane Couette flow
  \citep{gibson08,halcrow09,vanveen11}. The most direct numerical
  evidence so far  \citep{kreilos12} suggests that both local
  (period-doubling cascades) and global bifurcations \citep[boundary crises,
  see][]{grebogi87, ott02} may be important, but the field remains
  a very active area of research.

\subsection{Self-sustaining dynamo processes, transition and invariant solutions\label{principles-SSPMHD}}
The idea that the MRI dynamo could be based on an analogous
self-sustaining dynamo process was put forward by
\cite{rincon07b}. Let us introduce some
terminology to explain the analogy in detail. In dynamo
theory, the shearwise and spanwise  projections of vector fields are
usually referred to as their poloidal components and their streamwise
projection as their toroidal component. ``Streamwise-independent''
fields are often referred to as ``axisymmetric'' fields in dynamo
theory (even in cartesian coordinate systems), while
``streamwise-dependent'' translates into ``non-axisymmetric''.

The self-sustaining MRI dynamo process is illustrated in
figure~\ref{SSP} (right). The first part is the
stretching by the differential rotation of a weak, zero net-flux
axisymmetric poloidal magnetic field into a stronger
axisymmetric toroidal field. This so-called $\Omega$ effect
is the analogue of the lift-up mechanism. The second part of the dynamo
SSP is the amplification of  non-axisymmetric perturbations by the MRI
\citep{balbus92,ogilvie96,terquem96,branden06,lesur08b}. Recall that
magnetic tension is essential to the MRI. A non-axisymmetric
version of the instability is naturally expected in the presence of a
dominant toroidal axisymmetric magnetic field, which is just what
the $\Omega$ effect produces. The non-axisymmetric MRI of a toroidal
field is the analogue in the dynamo SSP of the streamwise-dependent
unstable perturbations of the streaks in the
hydrodynamic SSP. The dynamo loop is closed thanks to the nonlinear
interactions of MRI-amplified perturbations feeding back on the
original axisymmetric field components through 
a nonlinear induction term \citep{lesur08b,lesur08}. This
is the dynamo equivalent of the nonlinear self-advection of 
streamwise-dependent unstable perturbations feeding back on
streamwise vortices in the hydrodynamic problem. This scenario 
may also be relevant to dynamos involving other MHD
instabilities \citep{rincon08}, such as the kink instability
\citep{spruit02}, magnetic buoyancy coupled to
Kelvin-Helmholtz instability \citep{cline03,davies10,tobias11},
and magnetoshear instabilities \citep{miesch07}.

Recurrent forms of MRI dynamo action analogous to hydrodynamic
regeneration cycles in non-rotating shear flows have been spotted in
many simulations \citep{branden95,stone96,lesur08,gressel10,davis10,simon12},
and incompressible MRI dynamo turbulence has also been found
numerically to be supertransient \citep{rempel10}. A steady
nonlinear three-dimensional MRI dynamo solution was computed 
in Keplerian plane Couette flow  by \cite{rincon07b} along the lines
of homotopy-based computations of nonlinear solutions in
hydrodynamic shear flows \citep{nagata90,waleffe03,wedin04}, and
an unstable nonlinear MRI dynamo periodic orbit (cycle)
was recently discovered by \cite{Herault2011} in the shearing box
numerical framework (see \S\ref{setup}). Looking back at \S\ref{principles-hydro},
investigating the role of such invariant solutions in the MRI dynamo 
transition appears to be the next logical step to make progress on the
problem. This is the precise purpose of this work.

\subsection{Outline of the paper}
The framework of the study is introduced in
\S\ref{setup}. Section~\ref{turb} is devoted to a direct numerical
simulations exploration of the dynamo transition aiming at identifying
the imprint of invariant solutions. Section~\ref{cycles} describes the
numerical continuation and local stability analysis of several
nonlinear MRI dynamo cycles spotted in the transitional regime and
discusses a possible period-doubling route to chaos. A more direct
transition scenario is put forward in \S\ref{global}, where
a detailed characterization of chaos-generating global homoclinic and
heteroclinic bifurcations associated with dynamo cycles is
presented. A direct illustration of their
connections with the transition described in \S\ref{turb} is finally
provided in \S\ref{tanglechaos}. Section~\ref{discussion}
summarizes the main results and discusses their relevance to
astrophysics, dynamo theory and shear flow turbulence, as well as some
directions for future work.

\section{Equations and numerical framework\label{setup}}
The theoretical and numerical framework are the same as in earlier work by
our group \citep{Herault2011}. The problem geometry, equations, and
symmetry reductions are reproduced in this section for completeness,
together with a few new notations and several important
technical comments. We also discuss the numerical resolution and
choices of system parameters. The numerical methods are described in
detail in three appendices. 

\subsection{The shearing sheet}
\begin{figure}
  \centerline{\includegraphics[width=0.6\linewidth]{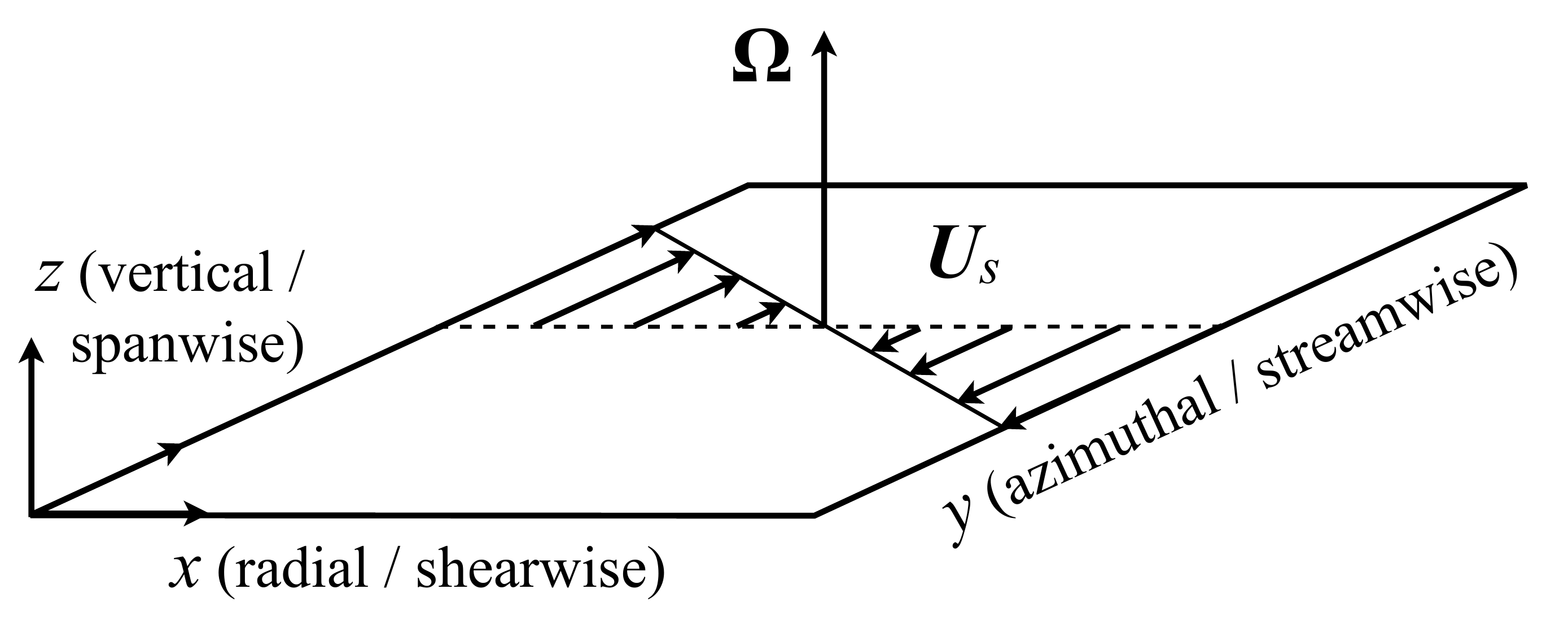}}
  \caption{The Keplerian shearing sheet.}
\label{shearing_sheet}
\end{figure}

We use the shearing sheet description of differentially rotating flows
\citep{goldreich65}, whereby an axisymmetric differential
rotation profile is approximated locally by a linear shear flow
$\vec{U}_s=-Sx\,\vec{e}_y$ and a uniform rotation rate
$\vec{\Omega}=\Omega\, \vec{e}_z$ (figure~\ref{shearing_sheet}). For a
Keplerian flow, $\Omega=(2/3)\, S$. Here, $(x,y,z)$ are respectively the
shearwise, streamwise and spanwise directions (radial, azimuthal 
and vertical in accretion disks). 
To comply with dynamo terminology, we refer to the $(x,z)$ projection
of vector fields as their poloidal component and to their $y$
projection as their toroidal component. ``Axisymmetric'' fields have
no $y$ dependence. Non-axisymmetric perturbations are usually referred
to as ``streamwise-dependent'' perturbations in plane Couette flow.
We consider incompressible velocity perturbations $\vec{u}$ and
a magnetic field $\vec{B}$ which evolve according to the
three-dimensional dissipative MHD equations in an unstratified
shearing sheet:
\begin{eqnarray}
\label{eq:NS}
\dpart{\,\vec{u}}{t}-Sx\,\dpart{\,\vec{u}}{y}+\vec{u}\cdot\grad{\,\vec{u}} &
=& -2\,\vec{\Omega}\times \vec{u} +S\,u_x\,\vec{e}_y -\grad{\,\Pi}
  \nonumber \\ & &
+\vec{B}\cdot\grad{\vec{B}}+\nu\Delta\,\vec{u}\,,
\end{eqnarray}
\begin{equation}
\label{eq:induc}
\dpart{\vec{B}}{t}-Sx\,\dpart{\vec{B}}{y}=-SB_x\,\vec{e}_y+\curl{\left(\vec{u}\times\vec{B}\right)}+\eta\Delta\vec{B}\,,
\end{equation}
\begin{equation}
\label{eq:div}
\div{\vec{u}}=0\,,\,\,\,\,\div{\vec{B}}=0\,.
\end{equation}
The kinetic and magnetic Reynolds numbers are defined by
$\Rey=SL^2/\nu$ and $\Reym=SL^2/\eta$, where $\nu$ and $\eta$ are
the constant kinematic viscosity and magnetic diffusivity, $L$ is
a typical scale of the spatial domain considered and
time is measured with respect to $S^{-1}$. We also introduce the
magnetic Prandtl number, $\Pm=\nu/\eta$.   $\Pi$ is the total pressure
(including magnetic pressure) divided by the uniform
density. $\vec{B}$ is expressed as an equivalent Alfv\'en
 velocity and both $\vec{u}$ and $\vec{B}$ are measured with 
 respect to $S L$. 

\subsection{Large-scale dynamo equation: the critical role of $\Reym$\label{setup-dynamo}}
The fundamentals of self-sustaining dynamo processes introduced qualitatively
in \S\ref{principles-SSPMHD} can be described in mathematical terms using a
``large-scale dynamo equation'' obtained by averaging
equation~(\ref{eq:induc}) over $y$:
\begin{equation}
\label{dynamoequation}
\frac{\partial \Ba }{\partial t}=-S\Bax\vec{e}_y+\overline{\vec{\nabla}\times\vec{\mathcal{E}}}+\eta\Delta\Ba~,
\end{equation}
where the overline denotes an average along the $y$ direction and
$\Ba=\Ba (x,z,t)$ only. The first term on the r.h.s. describes the
stretching of the axisymmetric (streamwise-independent) poloidal
magnetic field into a toroidal component. The second term is a
magnetic induction term which involves the axisymmetric projection
$\overline{\vec{\mathcal{E}}}$ of an electromotive force (EMF)
$\vec{\mathcal{E}}=\vec{u}\times\vec{B}$ generated by the
nonlinear couplings between velocity and magnetic
perturbations. In the MRI dynamo SSP, this EMF is the nonlinear
outcome of the transient development of a non-axisymmetric MRI and
drives large-scale field reversals \citep{Herault2011}. The third term
represents resistive damping. The presence of this term in the
dynamo equation makes the magnetic Reynolds number $\Reym$ a most
important parameter of the problem.
Equation~(\ref{dynamoequation}) is the equivalent in the MRI dynamo problem
of the equation for streamwise vortices and streaks in  the
hydrodynamic shear flow problem, with the lift-up term replaced by
the $\Omega$ effect term, the nonlinear advection term replaced by the
induction term, and the viscous term proportional to $1/\Rey$
replaced by the magnetic diffusion term proportional to $1/\Reym$.
Viscous dissipation is of course part of the full problem, but does
not appear explicitly in equation~(\ref{dynamoequation}).

\subsection{Symmetries\label{setup-sym}}
Enforcing symmetries is helpful to reduce the dynamics to
lower-dimensional subspaces and therefore makes it easier to
understand. \cite{nagata86} identified several symmetries for
three-dimensional nonlinear hydrodynamic solutions in Taylor-Couette
flow in the thin-gap limit, equivalent to a wall-bounded cartesian
plane Couette flow rotating about its spanwise $z$ axis. Similar
symmetries exist for MHD flows in the shearing
sheet. \cite{Herault2011} showed that
nonlinear states approaching a symmetry labelled $\mathcal{A}_1$ in
analogy with Nagata's work are regularly excited in the MRI dynamo
problem in the numerical configurations used in this paper. The
mathematical description of this symmetry is repeated in
appendix~\ref{appsym}. We will mostly investigate the dynamics in that
subspace and will show in \S\ref{turb} that this restriction does not
compromise the underlying fundamental complexity of the problem. 
Appropriate conjugation relations between Fourier modes
must be imposed throughout the numerical time integrations to enforce
the symmetry and eliminate the growth of non-symmetric perturbations
seeded by numerical noise. We will notably monitor the fundamental
Fourier mode in $z$ of $\Ba$, defined as
\begin{equation}
  \label{eq:axifieldfundamental}
\Bazero(z,t)=\Bazero(t)\cos{\left(\frac{2\pi}{L_z}\, z\right)}~,
\end{equation}
as this mode is always and by a large amount the dominant contribution
to the total axisymmetric magnetic field when the dynamics takes place
in the symmetric \Aone\ subspace.

\subsection{Numerical methods\label{setup-numerics}}
The numerics are based on a spectral version of the so-called
shearing box model, which is a numerical implementation of
equations~(\ref{eq:NS})-(\ref{eq:div}) in a finite numerical
domain. We use two complementary techniques to investigate
the dynamics. The first one, direct numerical simulation (DNS), is
used to integrate the equations in time starting from various initial
conditions. The second one, Newton iteration, is used to compute
accurate numerical representations of remarkable
solutions such as nonlinear cycles.

\subsubsection{Direct numerical simulations in the shearing box\label{setup-dns}}
DNS are carried out in the Keplerian shearing box with the SNOOPY code
\citep{lesur07} in a domain of size $(L_x,L_y,L_z)$, at numerical
resolution $(N_x,N_y,N_z$).  Let us summarize the important
qualitative features of the model (see appendix~\ref{appSB} for details).
The $y$ and $z$ directions are taken as periodic, while a
so-called shear periodicity is assumed in the $x$
direction. The latter amounts to assuming that a linear shear flow is
constantly imposed and that all quantities are periodic in all spatial
directions in a sheared Lagrangian frame. A discrete spectral basis
of ``shearing waves'' with constant $k_y$ and $k_z$ wavenumbers and
constant shearwise Lagrangian wavenumber $k_x'$ is used to represent
fields in the sheared Lagrangian frame. A pseudo-spectral method with
dealiasing is used to compute the nonlinear terms. The time
integration scheme is a first order operator splitting method coupling
an explicit third-order Runge-Kutta algorithm for the ideal MHD terms
and an implicit exact scheme for the dissipative terms.

Hydrodynamic turbulence and transport in shear flows have
been widely studied in such homogeneous shear flow simulations
\citep{pumir96,gualtieri02,casciola03}. The most notable difference
with wall-bounded plane Couette flow is the nonlinear feedback on the
background shear, which vanishes on average (in time) in the
shearing box. The absence of walls may also at first glance suggest
that important effects such as wave reflections are missing, but
a physical nonlinear scattering mechanism of non-axisymmetric 
shearing waves mediated by the radial (shearwise) modulation of
the axisymmetric (streamwise-independent) projections of the fields 
is actually present in the system \citep{Herault2011}.

\subsubsection{Newton-Krylov algorithm\label{setup-newton}}
Newton's method is a standard tool of nonlinear analysis 
which has proven most useful for computing simple nonlinear hydrodynamic
solutions in shear flows (see \S\ref{principles-hydro} and
\S\ref{principles-SSPMHD}). 
The most powerful and memory-saving methods to perform the algebra
in such high-dimensional systems (of typical size 10 000 or larger) involving
dense Jacobians are iterative methods such as Krylov
iteration. We developed our own Newton-Krylov solver PEANUTS
for the MRI dynamo problem along the lines described by
\cite{viswanath07}. The solver is based on the PETSc toolkit
\citep{petsc} and can be interfaced with different time-integrators 
to compute nonlinear equilibria, travelling waves and
periodic/relative periodic orbits for a variety of partial
differential equations.  The iterative GMRES algorithm is used to
solve the linear Jacobian system at each Newton iteration. For a
nonlinear cycle search, the code attempts to minimize
$||\vec{X}(T)-\vec{X}(0)||_2/||\vec{X}(0)||_2$, where
$\vec{X}(t)$ is a state vector containing all independent
field components at time $t$, and $T$ is a guess
for the period. The solver also performs pseudo-arclength
continuation of nonlinear solutions with respect to any relevant
system parameter and computes their local stability properties
with an iterative Arnoldi eigenvalue solver based on the SLEPc toolkit
\citep{slepc}. The code was tested against solutions of the
Kuramoto-Sivashinsky equation \citep{lan08}
before being implemented for the 3D MHD equations in the shearing box,
using SNOOPY as time integrator (appendix~\ref{appinterface}).

\subsubsection{Cyclic dynamics in the shearing box\label{setup-newton-SB}}
An important remark is that the implementation
of the shearing sheet equations in a discretized domain of
finite spatial extent introduces an internal time
$T_{SB}=L_y/(SL_x)$ in the system which breaks the continuous 
invariance of the shearing sheet equations under time translations
and turns it into a discrete one. Nonlinear periodic orbits (cycles)
are bound to have periods which are (\textit{a priori} arbitrary
large) integer multiples of $T_{SB}$. It must be emphasized that the
dynamical consequences of this property of the model are under control
\citep{Herault2011}. A qualitative dependence of
  the timescale of recurrent dynamics on the streamwise to 
  shearwise aspect ratio of the solutions is nevertheless
  expected in all shear flows in constrained geometries
  (including wall-bounded ones), as this ratio sets the typical
  timescale of shearing of streamwise-dependent structures that
  ultimately extract the energy from the base flow. Note however that
  this observation may not carry over to spatially extended and/or
  stratified configurations in which recurrent forms of MRI dynamo
  action are also observed \citep{simon12}; see \S\ref{astroturb}.

Appendix~\ref{appinterface-constraints} and \ref{traceWu} explain in
detail the restrictions that shear periodicity imposes on periodic
orbits and how this specificity of the shearing box can actually be
exploited to analyse some aspects of the dynamics in terms of
discrete-time maps.

\subsection{Parameters of the study: box size, $\Pm$ and numerical
  resolution\label{setup-parameters}}
We explore the MRI dynamo transition in an elongated box with
$L_x=0.7,L_y=20,L_z=2$. This choice, together with the reduction to a
symmetric subspace, is similar to that of a ``minimal flow unit'' in
wall-bounded shear flows and is motivated by the necessity to reduce the
nonlinear complexity of the problem \citep{Herault2011}. This setup
is undoubtedly restrictive, and it is currently not understood
which of the results obtained in such configurations should carry over to
very different setups (see again discussion in \S\ref{astroturb}). At
the current stage of understanding, this approach nevertheless remains
one of the most efficient to uncover some of the fundamental building
blocks of the transition.

We also restrict the analysis to the dependence of the
transition on the magnetic Reynolds number $\Reym$ which, as
equation~(\ref{dynamoequation}) indicates, is the
primary bifurcation parameter in the problem. A multidimensional
parametric study with respect to $\Rey$ or to the box
dimensions is unfortunately currently prohibitive in terms of both
human and computing time. MRI dynamo action has not been found in
  numerical simulations in periodic, unstratified shearing boxes at
  $\Pm < 1$ so far \citep{fromang07b}. The only option currently available 
  to address this problem is to study the detailed
  transition mechanisms in the $\Pm > 1$ regime in which the dynamo
  can easily be excited, hoping that the results will provide 
  new research directions to understand what is happening at low
  $\Pm$. We set $\Rey=70$ as in the work of
\cite{Herault2011} and focus on the transition in the typical range
$2<\Pm<10$. Possible ways to address the low $\Pm$ problem will be
discussed at the end of the paper in the light of the results.

The results of the paper are the outcome of several hundred
thousand DNS. Such an analysis could only be performed at a moderate
resolution $(N_x,N_y,N_z)=(24,12,36)$ (on top of which a 2/3
dealiasing rule was enforced) similar to that used in recent studies
of nonlinear cycles in hydrodynamic plane Couette flow
\citep{viswanath07,vanveen11}. The differences between
the results at this resolution and at double resolution
are very minor \citep{Herault2011} and do not affect any of the
conclusions on MRI dynamo cycles and their bifurcations. Analysing the
shape and convergence of energy spectra, we also found this resolution
to be largely acceptable in our box for most DNS simulations at $\Rey$
and $\Reym$ lower than 500. MRI dynamo cycles are
essentially large-scale coherent structures and can therefore actually
be safely followed at even higher $\Rey$ and $\Reym$ using this
resolution. The box size, $\Rey$, and numerical resolution 
remain the same throughout the paper and will not be repeated in the
figure captions.

\section{DNS cartography of the MRI dynamo transition\label{turb}}
The objective of this section is to offer a phenomenological geometric
and statistical perspective on the nature of the MRI dynamo
transition and to uncover its possible connections with nonlinear
invariant solutions. 

\subsection{Cartography procedure}
The approach is inspired by earlier work on hydrodynamic shear flows
\citep{Schmiegel1997,Moehlis2004a,Moehlis2004b,skufca06,schneider07a}
and consists in mapping the laminar-preturbulent MRI dynamo boundary in
the phase space of the corresponding dynamical system (see
appendix~\ref{appinterface-constraints}), as a function of the system
parameters. The high-dimensionality of this system  (almost 10 000
independent degrees of freedom) makes it impossible to probe this
border in the full phase space, as a function of all system
parameters. The cartography is restricted to a
two-dimensional plane, with $\Reym$ in the first dimension and the
amplitude of a given spatial form of initial conditions in the second
dimension. The initial conditions are generated as follows. For each
field component, we generate a random set of Fourier modes (white
noise in spectral space), symmetrize the resulting fields to enforce
reality and incompressibility, and normalize the total physical energy
  density to obtain a particular ``noise realization''. A given zero
net-flux initial condition is obtained by multiplying a given noise
realization by an amplitude factor $A$. We then perform a series of
DNS for each $\Reym$ and initial condition amplitude $A$ defined on a
two-dimensional parameter grid, and the typical dynamical lifetime
measured in each simulation is used to construct a two-dimensional
``transition map''. The lifetime definition is based on a somewhat arbitrary
small threshold for the total magnetic energy below which the system
is considered to have returned to the laminar state  ($10^{-4}$ in the units
used -- a typical dynamical state has $O(1)$ energy density). The general
aspect of the maps and their qualitative interpretation do not depend
critically on this definition and the threshold value used. Simulations in
which the dynamics lasts for times longer than the maximum time
indicated in each plot are coded with the same (dark/red) colour. The
results presented below required approximately 10 0000 DNS at
resolution $(N_x,N_y,N_z)=(24,12,36)$, each of them lasting between 
500 and 700 $S^{-1}$. 

\subsection{A fractal stability border?}
\begin{figure}
\centerline{\includegraphics{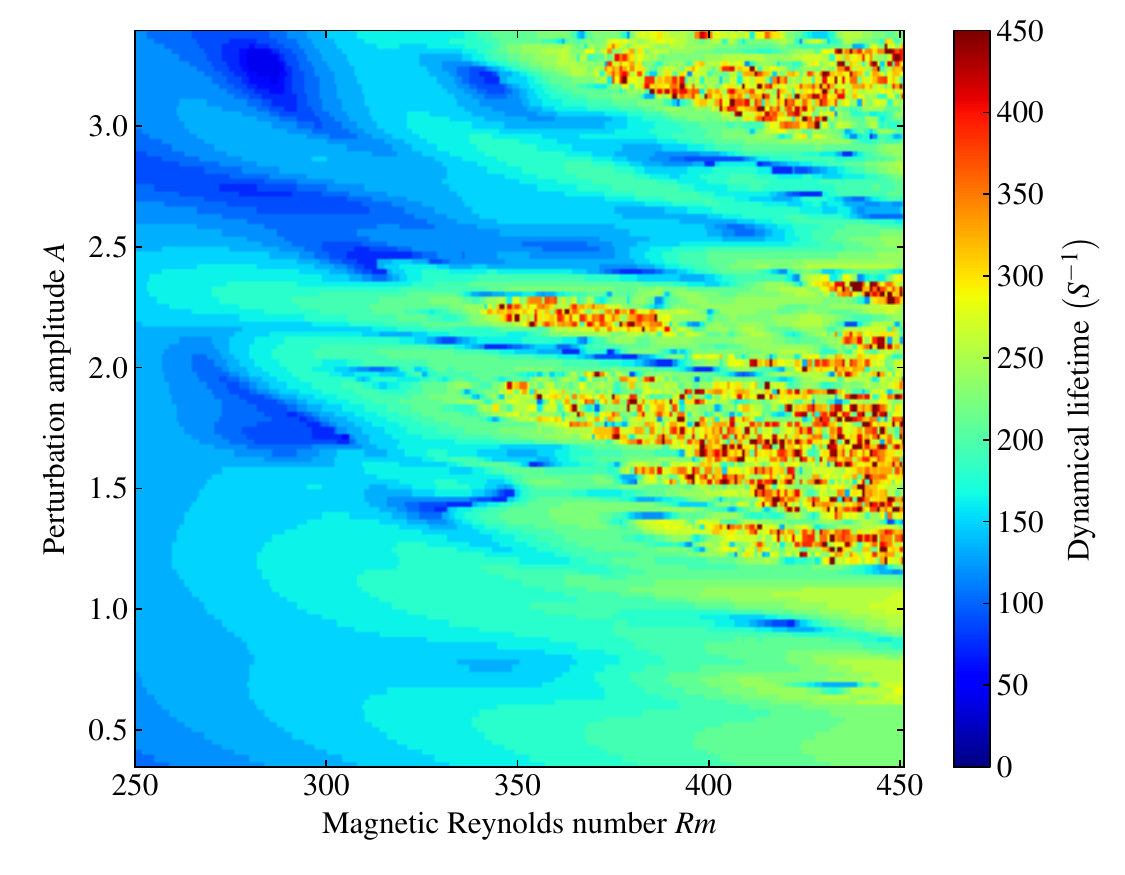}}
\caption{High-resolution transition map ($\delta\Reym =1,
  \delta A=0.023$) as a function of initial condition amplitude $A$
  and $\Reym$ for a given noise realization and no symmetry
  enforced.}
\label{map1}
\end{figure}

\begin{figure}
\centerline{\includegraphics[width=\linewidth]{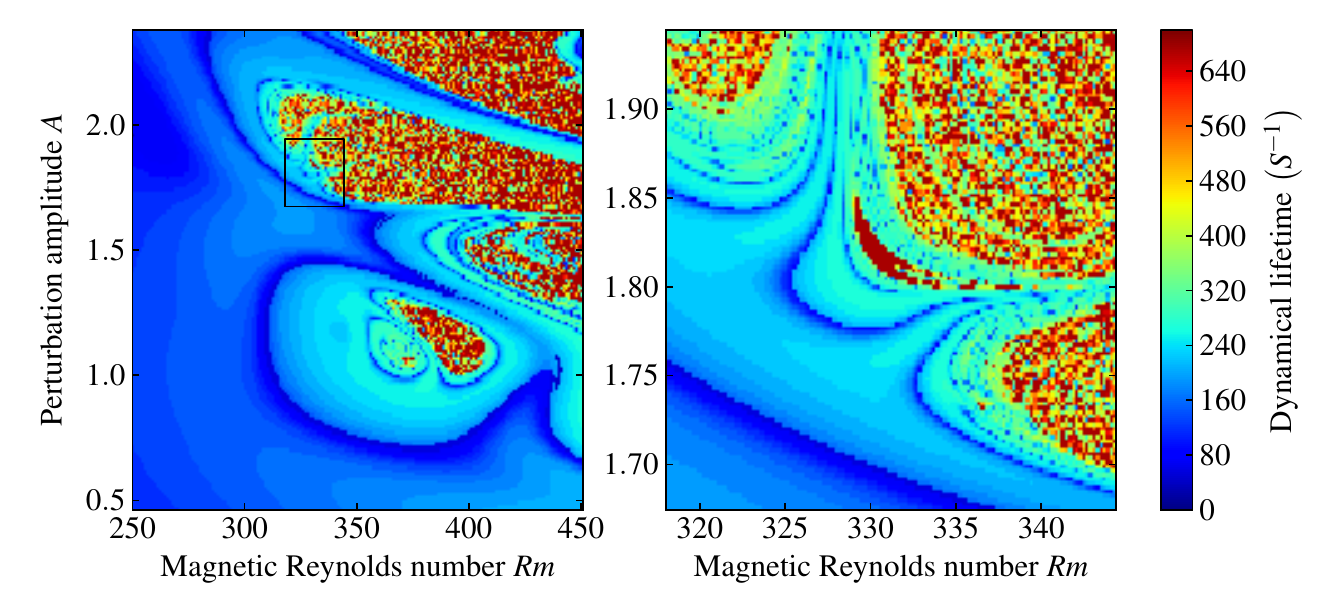}}
  \caption{High-resolution transition maps as a function 
    of initial condition amplitude $A$ and $\Reym$ for a given noise
    realization and \Aone\ symmetry enforced. Left: full map,
 resolution  $\delta\Reym=1, \delta A=0.0115$. Right:
 high-resolution map ($\delta\Reym=0.2, \delta A=0.0023$)
 computed for a restricted parameter range indicated by the black
 rectangle in the leftmost map.}
\label{map2}
\end{figure}

Figure~\ref{map1} displays a transition map for the full problem 
(no imposed symmetries), and a map resolution $\delta\Reym
=1, \delta A=0.023$. Figure~\ref{map2} presents another example
with the dynamics restricted to the \Aone\ subspace. 
Both maps in figures~\ref{map1} and \ref{map2} display a wide,
low-$\Reym$ region where return to the laminar state is quick (of the
order of a few tens of shearing times) and another region in which
much longer-lived three-dimensional dynamics is present. In this
region, the dynamics is highly sensitive on initial conditions,
i.e. a very small change in $A$ leads to very significant changes
in the history and lifetime of the dynamics.
These two regions are separated by a border with an intricate
geometry. Closer inspection reveals that the structure of the
map, and therefore the set of initial conditions leading to long-lived
transient chaos, is probably fractal. Figure~\ref{map2} shows a
high-resolution map (right) computed for a restricted parameter range
of the map on the left. Apparently random variations of turbulence
lifetime and the same colour patterns clearly replicate on smaller
scales. We note that the reduction of the dynamics to a symmetric
subspace does not affect the results at a qualitative level. This is
good news, because operating with symmetries makes the task of
understanding the transition much easier. 

\begin{figure}
\centerline{\includegraphics[width=\linewidth]{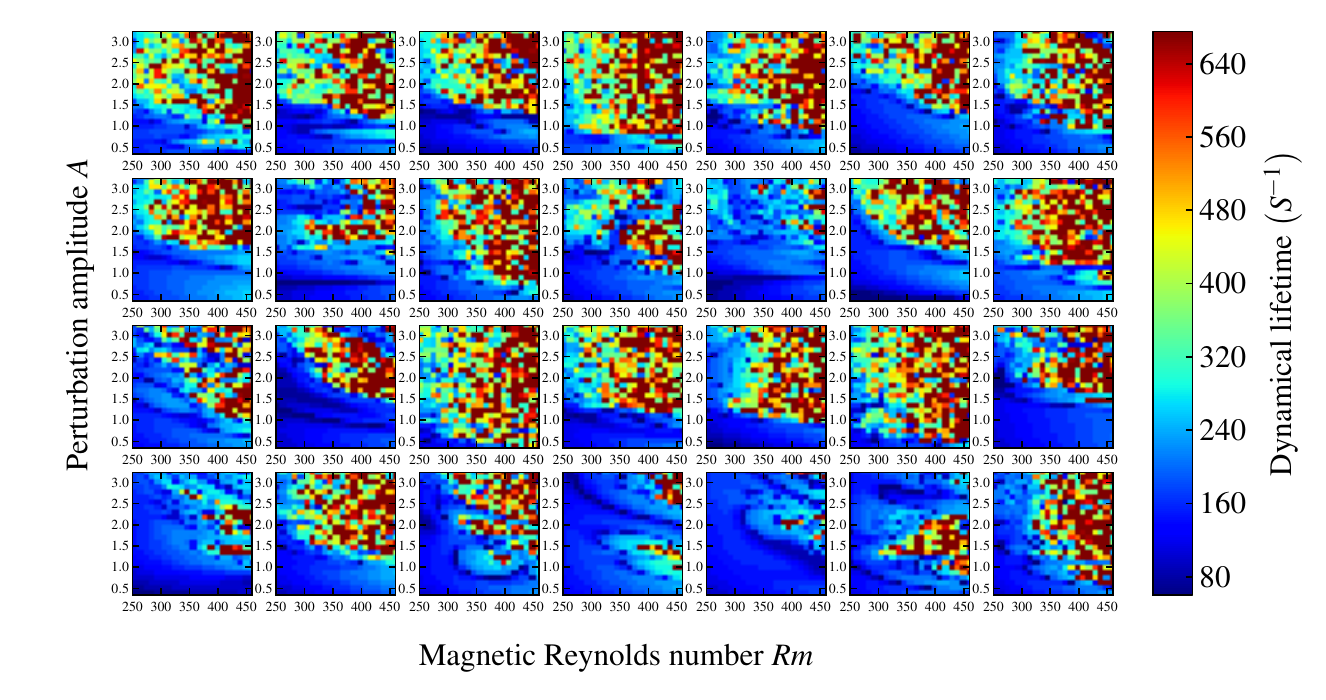}}
 \caption{Low resolution transition maps ($\delta\Reym =10,
   \delta A=0.115$) as a function of initial condition amplitude $A$
   and $\Reym$ for 28 different noise realizations.}
\label{multimap}
\end{figure}

In the previous examples, a single noise realization was used to
obtain a map of the laminar-turbulent transition border (the noise
realization in the symmetry-reduced case was different
from that in the full problem). Hence, these maps only explored
the laminar-turbulent border along a one-dimensional line in state
space. The direction of shooting was chosen randomly but once and for
all before the simulations. However, there is no
guarantee that results obtained this way are 
representative of the transition border as a
whole. To test this, we computed 28 different lower resolution maps
($\delta\Reym =10, \delta A=0.115$),
each of them based on a different noise realization
(figure~\ref{multimap}).  The results are interesting in two
respects. First, the fractal-like structure seems to be statistically
homogeneous in phase space. Except for a set of initial
directions of very small measure (possibly zero), it is quite
clear that shooting in any random direction will generate maps with
a similar aspect. Second, the critical $\Reym$ for the onset of 
transient chaotic dynamics depends both on
the amplitude of the initial perturbation and on the
noise realization (shooting direction). For instance, it is roughly
250 in map 1 but close to 380 in map 26 (the maps are
numbered from left to right and top to bottom). 

 The results, combined with the analysis of \cite{rempel10}, strongly
 suggest that the MRI dynamo transition involves a chaotic saddle and
 that chaotic trajectories with infinite lifetime live on a
 zero-measure set, the so-called ``edge of chaos'' in transitional
 hydrodynamic shear flows (see references above). 

\subsection{Islands of turbulence and transient appearance of dynamo
  cycles\label{turbislands}}
Much can be learned about the transition process by analysing the
contents of these maps in detail. 
From now on, we restrict the analysis to the symmetric case, which
makes it easier to identify potentially interesting dynamical
behaviour. In several maps, we observe that the first chaotic regions approached
as $\Reym$ is increased take the form of ``islands'' or elongated
``fingers''. Similarly, isolated zones in which the dynamics is
fast-decaying penetrate into chaotic regions. Such regions are
also clearly visible in similar studies of the hydrodynamic problem
\citep{Moehlis2004b}.

\begin{figure}
\centerline{\includegraphics[width=\linewidth]{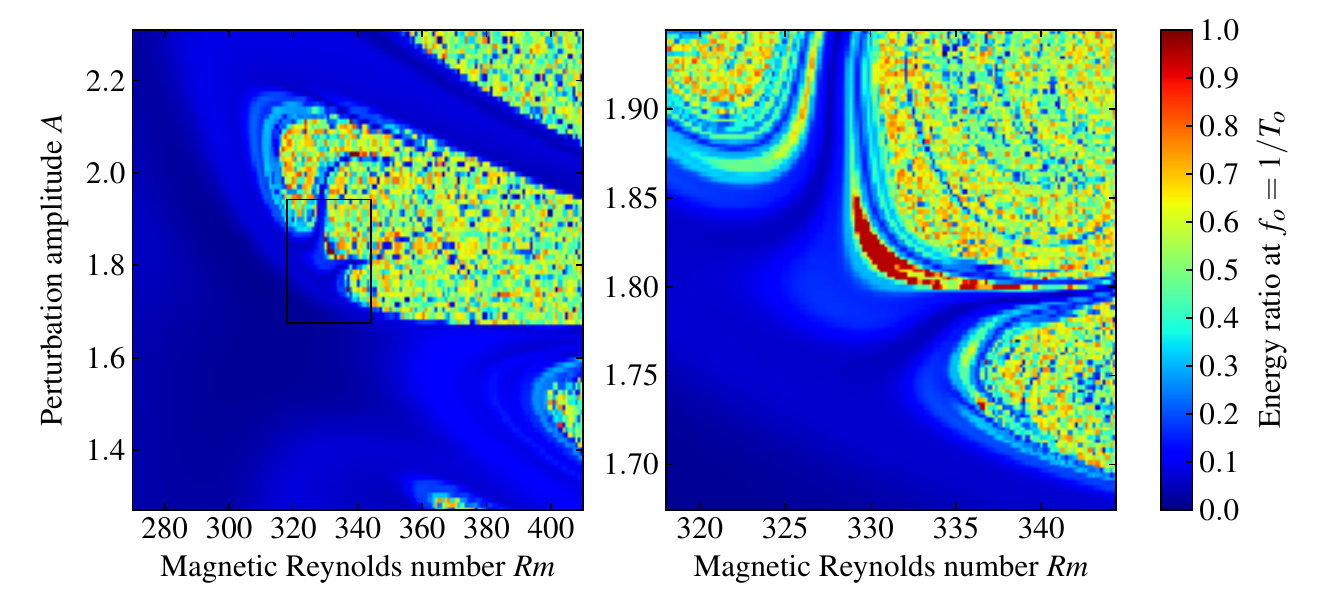}} 
 \caption{Normalized energy ratio of $\Bazeroy(t)$ at the fundamental
   frequency $f_o=1/T_o$ as a function of the perturbation amplitude
   $A$ and $\Reym$ for submaps of figure~\ref{map2}. The red/black
   spots indicate initial conditions lying in the basin of attraction
   of a stable MRI dynamo cycle.}
\label{mapratio}
\end{figure}

Most simulations in these island and finger regions are
characterized by a fairly extended period of recurrent dynamics with a
fundamental periodicity comparable to $T_o=2L_y/(SL_x)$
(see remarks in \S\ref{setup-newton-SB}), followed by a
decay to the laminar state. This observation can be made more
quantitative by computing the time Fourier transform of the toroidal
axisymmetric field $\Bazeroy(t)$, defined in
equation~(\ref{eq:axifieldfundamental}) for
each DNS in the map shown in figure~\ref{map2}, and by subsequently
plotting the energy ratio between the energy integrated over the peak
at the fundamental frequency $f_o=1/T_o$  and the total energy
(figure~\ref{mapratio}). Almost half of the large-scale magnetic field
energy is at the fundamental frequency on average in these
regions. Some areas even contain full sets of
simulations that ultimately converge
to truly periodic solutions. For instance, figure~\ref{mapratio}
(right) displays a zoom of the ``central'' structure of
figure~\ref{map2} (located around $A=1.82$ and $\Reym=330$). This
zoomed area contains a thin croissant-shaped region in which all 
simulations converge to a periodic orbit of period $T_o$
similar to the MRI dynamo cycle reported by \citet{Herault2011},
albeit more energetic. Other simulations in the same map converge to longer
period cycles (period-4, 6, 7, 8) or to recurrent but seemingly
aperiodic orbits. These long-period cycles correspond to modulated
$T_o$ oscillations with a modulation period larger than $T_o$. 
We will come back to this result in \S\ref{smale}.

\subsection{Other regimes and initial conditions}
 As cautioned in
\S\ref{setup-parameters}, it is as yet not possible to
conclude that the results described so far are representative of the dynamics
in all possible regimes. We checked that they were structurally
stable with respect to $\Rey$ for $\Rey\sim  100$ ($\Pm\sim 2-5$).
Small increases in $\Rey$ only seem to smooth out the fractal-like features
and to gently reduce the coherence between cycles and turbulent dynamics.
Preliminary results indicate that the situation becomes much more complex
and computationally expensive to analyse as $\Pm$ tends to unity. 
A parametric study of transition maps with respect to the aspect ratio
would also be useful, but any meaningful computational effort to span
a representative range of  box sizes with sufficiently small
increments so as to track down interesting features is currently
prohibitive. The results of \cite{rempel10} for much smaller
$L_y/L_x$ suggest that the phenomenology described above
pertains to different geometric configurations.

Finally, we point out that recurrent dynamics can be excited
  using either  zero net-flux random white noise seeds or various
  combinations of large-scale axisymmetric and non-axisymmetric
  perturbations \citep{Herault2011}. All simulations are ultimately
energetically dominated by the large-scale sinusoidal field
of equation~(\ref{eq:axifieldfundamental}), which is a key ingredient 
of the self-sustaining dynamo process.

\section{MRI dynamo cycles, saddle nodes and period-doublings\label{cycles}}

  Do MRI dynamo cycles play an active role in the transition
  process? Quantitative progress on this question first requires accurate
  computations of the defining features of such invariant solutions:
  how they are born, what their local stability properties are, etc.
We now present such an analysis for two seemingly important pairs of
cycles (labelled \SNone\ and \SNtwo) spotted in the simulations
presented in \S\ref{turb}.

\subsection{Restrictions of the study and numerical convergence}
The analysis is again restricted to the $\Reym$ dependence of the
problem and to the dynamics in the \Aone\ symmetric subspace
(both for the cycles and their eigenmodes). This facilitates the analysis
and the convergence of the Newton algorithm but does not impede the essential
basic mechanisms that we wish to investigate. It simply reduces the number
of active simple nonlinear invariant solutions and eliminates relative
periodic orbits, whose nature is fundamentally not different from that
of periodic orbits.

The results were obtained with the  Newton-Krylov PEANUTS solver. 
All of them are converged with a relative error
$||\vec{X}(T)-\vec{X}(0)||_2/||\vec{X}(0)||_2$  on the state vector
smaller than $10^{-5}$ for the same spatial resolution as that of the
DNS. Computing an accurate nonlinear MRI dynamo cycle with the solver
for a given set of parameters requires $\sim 10$ Newton
iterations, each of them involving $\sim 10$ Krylov
iterations. Hence, a single cycle computation typically requires 100
DNS. The results presented below are the aggregation of several
thousand such computations. In the few cases where continuation
appeared difficult, solutions were re-calculated with double
resolution, leading to only very minor differences. In
almost all ``hard'' cases, the difficulty to converge to
a cyclic solution was due to the presence of a local
bifurcation in the $\Reym$-neighbourhood of that cycle.

\subsection{Saddle node 1 (\SNone)}
\begin{figure}
  \centerline{\includegraphics[width=\linewidth]{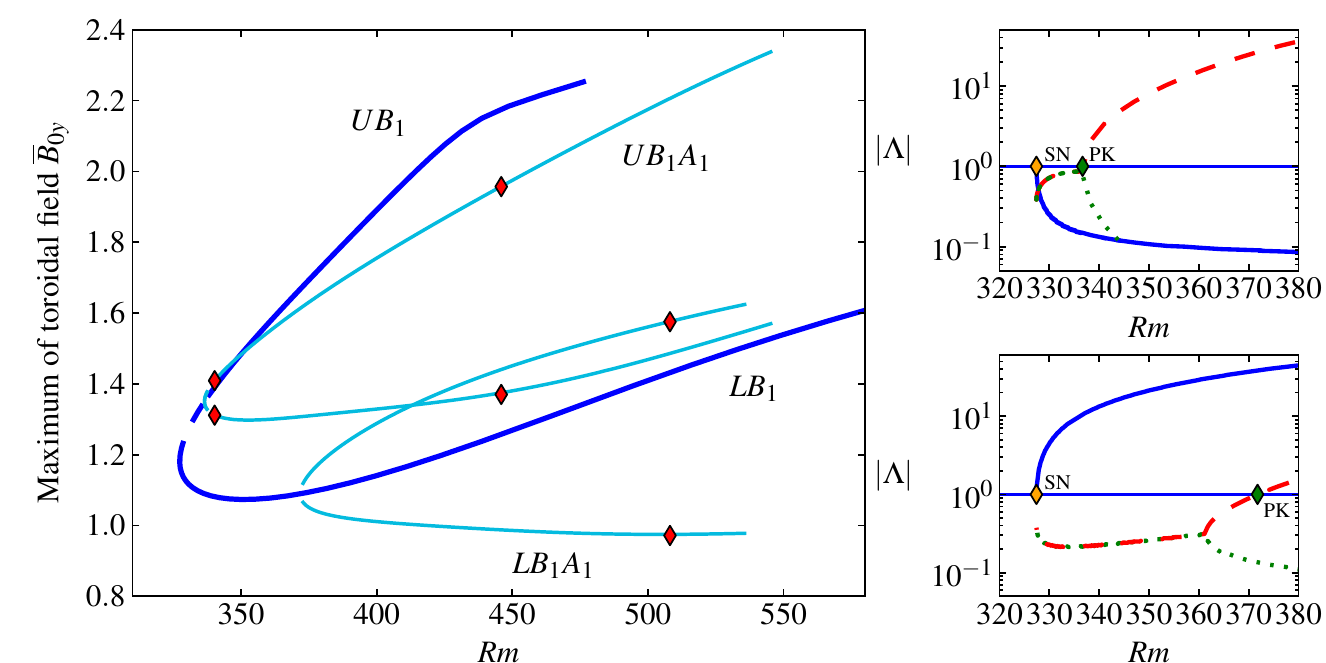}}
  \caption{Bifurcation diagram of the \SNone\ pair of cycles.
    Left: maximum amplitude of $\Bazeroy$
    over a period cycle $T_o$ as a function of $\Reym$. Solid and
    dashed lines correspond respectively to unstable and stable
    solutions. Lower branch ($LB_1$), upper branch ($UB_1$)
    are represented using thick/dark blue lines and their asymmetric
    bifurcations ($LB_1A_1$ and $UB_1A_1$) are represented with
    thin/light-blue lines. The locations of the first
    period-doubling bifurcations of asymmetric cycle branches are
    indicated by red diamonds.
    Right (upper branch on top, lower
    branch at the bottom): norm of the first three largest
    Floquet multipliers as a function of $\Reym$,
    with $\Lambda_1=1$ at the saddle node bifurcation (solid/blue
    line). Complex conjugate eigenvalues are superposed.}
\label{continuation_c1}
\end{figure}

We start by analysing the nonlinear MRI dynamo cycle found by
\cite{Herault2011}. We continued a low-resolution version of this
solution obtained at $\Reym=352$ at both lower and higher
$\Reym$. The continuation curve in figure~\ref{continuation_c1} (left)
shows the maximal amplitude of
$\Bazeroy$  over a cycle period $T_o$ as a function of $\Reym$ in
thick/dark blue line
($\Bazeroy$ is representative of the amplitude of the nonlinear solution).
There are actually two cycles born out of a saddle node bifurcation at
$\Reym=327.4$:  a lower branch $LB_1$ (the solution found by
\cite{Herault2011}), and an upper branch $UB_1$. $UB_1$ is
the cycle identified in figure~\ref{mapratio} in the croissant
region. Whatever quantity is used to construct the continuation curve
(maximum of $\Bazeroy$, of
nonlinear stresses, etc.), $UB_1$ is always observed to be more
vigorous than $LB_1$. The saddle node nature of the bifurcation is 
confirmed by looking at the $\Reym$-dependence of the largest
Floquet multipliers  of $LB_1$ and $UB_1$ (omitting neutral multipliers
corresponding to simple translations in $y$ and $z$), shown in the two
plots on the right of figure~\ref{continuation_c1}. At the bifurcation
point where the solutions merge, both have a neutral eigenvalue. Slightly
beyond the bifurcation point, $UB_1$ is actually stable, explaining the
convergence of simulations in the croissant region. $LB_1$ 
has a single unstable eigenvalue near the bifurcation and is
unstable for all $\Reym$. Both cycles can be continued to much
larger $\Reym$ well beyond the saddle node with no difficulty. The
upper branch curve has a knee around $\Reym=430$,
beyond which an asymptotic regime seems to be reached.

\subsubsection{Detailed stability analysis of $LB_1$ and $UB_1$}
The full stability analysis of $LB_1$ and $UB_1$ is summarized in table~\ref{tab_c1}.
$LB_1$ is originally unstable through its first
Floquet multiplier $\Lambda_1$. Near the saddle node, $\Lambda_2$ and
$\Lambda_3$ form a stable pair of complex conjugate
multipliers. Increasing $\Reym$, $\Lambda_2$ and $\Lambda_3$ collide
and turn into a real stable pair of multipliers. Just after this collision, at
$\Reym=371.24$, $\Lambda_2$ becomes larger than one. This pitchfork
bifurcation results in two new branches of periodic
solutions with period $T_o$, whose most notable feature is to be
asymmetric in time at larger $\Reym$. For these two branches,
the maximum field amplitude in the first $T_o/2$  is different from
that during the second half of the cycle. Closer inspection reveals
that the two branches are physically identical and symmetric
under a simultaneous space and time transformation which, as far as
the large-scale components of the magnetic field is concerned,
corresponds to a half-period phase shift combined with a change
in the field polarity\footnote{This is a well-known effect in
  highly symmetrical dynamical systems \citep{swift84}, reported 
  for instance in double-diffusive convection
  \citep{knobloch81,moore83,knobloch86} and in simple shear
  flow models \citep{Moehlis2004b}.}. We will henceforth only
focus on one of these branches, called $LB_1A_1$.

The stability domain of $UB_1$ is limited. Increasing
$\Reym$ from the saddle node, the first real stable Floquet
multiplier $\Lambda_1$ decreases. Originally $\Lambda_2$ and
$\Lambda_3$ are stable and complex conjugate. As $\Reym$ increases,
they turn into two stable real multipliers, the
largest of which is responsible for a supercritical pitchfork bifurcation at
$\Reym=336.54$ (dashed/red line on figure~\ref{continuation_c1} (top
right)). This bifurcation creates two physically identical 
asymmetric in time periodic solutions of period $T_o$. We call one of
them $UB_1A_1$. 

\begin{table}
  \begin{center}
  \begin{tabular}{ccccccccccccc}
      $\Reym$  & Bifurcation type  & Branch &\hspace{0.5cm} & $\Real$ &  $\Imag $ &\hspace{0.5cm} & $\Real$ & $\Imag$ &\hspace{0.5cm} & $\Real $ & $\Imag  $\\[3pt]
             
       327.40   & Saddle node & & & 1 & 0 & & -0.30  & 0.23 & & -0.30 & -0.23\\
       336.55   & Pitchfork & $UB_1$  & & 1 & 0 & &  0.75  & 0 & & 0.14 & 0\\
       371.24  & Pitchfork & $LB_1$ & & 37.9 & 0 & &  1  & 0 & & 0.14 & 0\\
               
       340.28   & P-doubling & $UB_1A_1$ & & -1 & 0 & & -0.66  & 0 & & 0.16 & 0\\
       439.03  & Torus? & $UB_1$ &  & 45.1 & 0 & & -0.15  & 0.99 & & -0.15 & -0.99\\
       445.92   & P-doubling & $UB_1A_1$ & & -935 & 0 & & -1  & 0 & & 0.078 & 0\\
        
       508.12 & P-doubling & $LB_1A_1$ & & 5279 &  0 & & -1 & 0 & & -0.15 & 0\\
       $\sim$634 & P-doubling & $LB_1A_1$ & & -2.63 &  0 & & -1 & 0 & & N/A & N/A \\     
  \end{tabular}
  \caption{Bifurcations and first three Floquet multipliers 
    (excluding neutral multipliers) of the \SNone\ pair of cycles
    $LB_1$ and $UB_1$ and their asymmetric child cycles $LB_1A_1$ and
    $UB_1A_1$ for increasing $\Reym$. When
    possible, two significant decimal digits have been kept. Approximate
    zeroes are used to denote strongly stable multipliers of very
    small magnitude. The multipliers are simply
    sorted in descending order of magnitude
    here and do not necessarily correspond to the
    $\Lambda_1$, $\Lambda_2$ and $\Lambda_3$ terminology used in the
    text. $\Lambda_1$, $\Lambda_2$ and $\Lambda_3$ are only used to
    tag the first three Floquet multipliers at the saddle node
    bifurcation to follow them in $\Reym$.}
  \label{tab_c1}
  \end{center}
\end{table}

\subsubsection{Period-doublings of $LB_1A_1$ and $UB_1A_1$} 
We now describe the stability analysis of the asymmetric branches
$LB_1A_1$ and $UB_1A_1$, whose maximum $\Bazeroy$ amplitudes during each
half-cycle are used to represent their continuation in
figure~\ref{continuation_c1}. $LB_1A_1$ (thin solid/light-blue line) is
originally unstable as it inherits $\Lambda_1>1$ from its parent
$LB_1$, while $UB_1A_1$ has a small stability domain
($336.5<\Reym\leq 340$), 
thin dashed/light-blue line). Increasing $\Reym$ from $371.24$, $LB_1A_1$
undergoes two successive period-doubling bifurcations as two of its
real Floquet multipliers become smaller than -1. The first
period-doubling  (red diamonds in figure~\ref{continuation_c1})
takes place at $\Reym=508.1$ and the second one at
$\Reym=634$. We obtain two well-converged cycles of period $2T_o$
whose typical amplitudes remain close to that
of $LB_1A_1$. A similar result is obtained for $UB_1A_1$ as two of its
real Floquet multipliers cross -1 at $\Reym=340.28$ and
$\Reym=445.92$. The first period-doubling is a primary loss of
stability of $UB_1A_1$. 

The results hint at a period-doubling cascade of MRI dynamo
cycles. However, even though we managed to
continue several period-doubled cycles in $\Reym$, we found it much
harder to obtain accurate results for their
stability. We obtained a converged Floquet multiplier of -1 for one
of the period-doubled solutions branching from $UB_1A_1$ but were
unable to converge on the corresponding
period-4 orbits. The confusion of the Newton solver may be due to the
packing on the $\Reym$ line of successive period-doubling bifurcations
in a \cite{feigenbaum78} sequence, to the high instability of the
cycles, to the close resemblance between them near bifurcation
points, or to the dynamical complexity associated with  global
bifurcations taking place in parallel (see \S\ref{global}).

\subsection{Saddle node 2 (\SNtwo)}
\begin{figure}
  \centerline{\includegraphics[width=\linewidth]{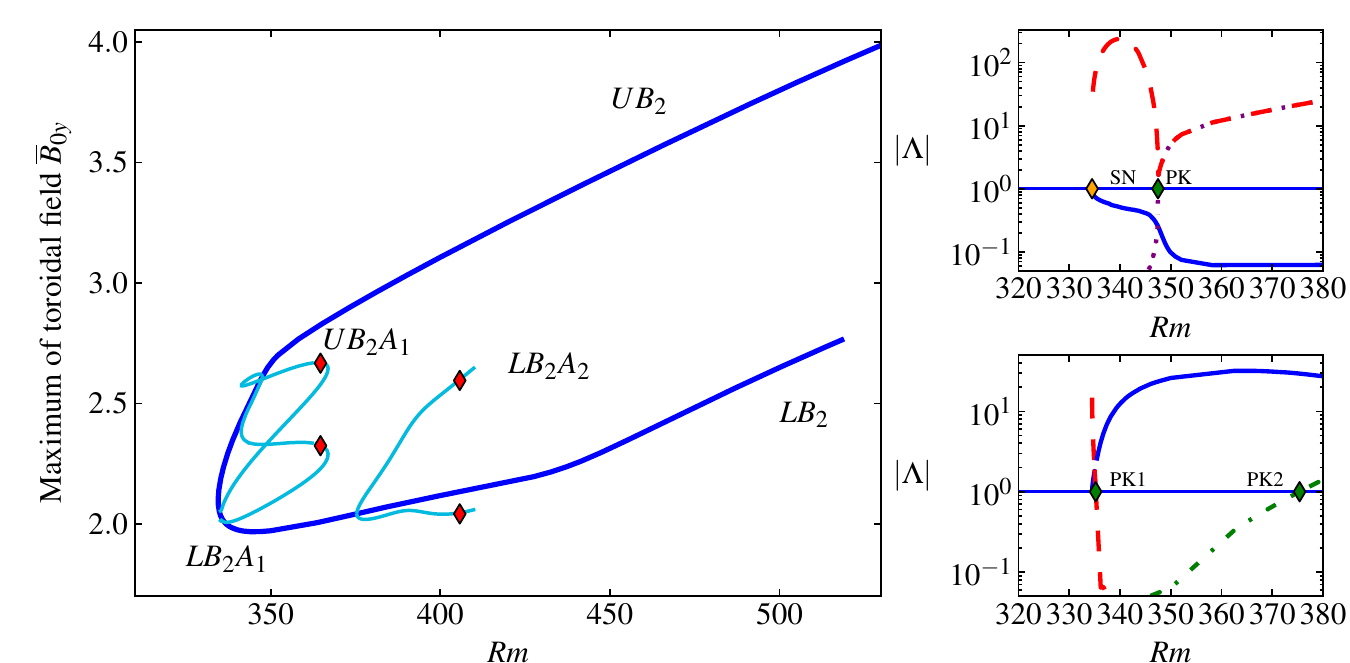}}
  \caption{Bifurcation diagram of the \SNtwo\ pair of cycles.
    Same legend as  figure~\ref{continuation_c1}. An additional
    Floquet multiplier (purple dotted line)
    has been added in the top right plot 
    (stability analysis of $UB_2$).  The stability analysis of $LB_2$
    is shown in the bottom right plot.}
\label{continuation_c2}
\end{figure}

We learned in \S\ref{turb} that several families of cycles probably
take an active part in the transition, but there is no guarantee that
the previous analysis of \SNone\ cycles is representative of all such cycles.
We therefore attempted to capture other cycles by feeding the Newton solver
with well-chosen DNS snapshots hopefully close enough to periodic
trajectories. Another simple cycle of period $T_o$ at $\Reym=352$ that
did not correspond to either $LB_1$ or $UB_1$ was found. Its
continuation, displayed in figure~\ref{continuation_c2} (left), shows
that it is also born out
of a saddle node bifurcation and is a lower branch saddle, which we
henceforth call $LB_2$ (the upper branch is called $UB_2$). The saddle
node nature of the bifurcation is  confirmed by analysing the
$\Reym$-dependence of the Floquet multipliers (omitting once again the
neutral multipliers) of $LB_2$ and $UB_2$, shown in the two plots on
figure~\ref{continuation_c2} (right). Just as in the previous case,
both solutions have a neutral eigenvalue where they merge at the
bifurcation point, located at $\Reym=334.52$. Unlike  \SNone\ though,
the neutral eigenvalue at
the saddle node is not the largest of all eigenvalues but the second
largest. Hence $UB_2$ is born unstable and the saddle $LB_2$ is born
with two unstable eigenvalues. Both cycles can be continued to much
larger $\Reym$, well beyond the saddle node, without any major
difficulty (except again in the neighbourhood of local
bifurcations described below). At large $\Reym$, the maximum
amplitude of $\Bazeroy$ and nonlinear stresses over a period both
increase monotonically for increasing $\Reym$.

\subsubsection{Detailed stability analysis of $LB_2$ and $UB_2$} 
The stability analysis of $LB_2$ and $UB_2$ is summarized in
table~\ref{tab_c2}. They undergo the same kind of primary bifurcations
as $LB_1$ and $UB_1$ but the detailed situation is slightly more
difficult to analyse. Near the saddle node, both
$\Lambda_1$ and $\Lambda_2$ of $LB_2$ are larger than one. 
As $\Reym$ increases, $\Lambda_1$ decreases,
resulting in a subcritical reverse pitchfork bifurcation at
$\Reym=335.25$ (dashed/red line in figure~\ref{continuation_c2}
(bottom right)).  This bifurcation generates two branches
of physically identical, asymmetric in time periodic orbits of period $T_o$.
We only keep track of one of these branches which we hereafter call
$LB_2A_1$. The maximum of $\Bazeroy$ over each half-cycle is
represented by a pair of thin solid/light-blue lines in 
figure~\ref{continuation_c2} (left). One notable difference with the
\SNone\ case is that instead of staying close to $LB_2$, the
asymmetric branch $LB_2A_1$ turns at $\Reym=366.9$ and goes back to
lower $\Reym$ before folding again at $\Reym=341.32$ to join $UB_2$ at
$\Reym= 347.52$. A computation of the stability of $UB_2$ at this
point reveals that $UB_2$ undergoes a
subcritical pitchfork bifurcation associated with a fourth Floquet
multiplier (purple dotted line in the top right plot of
figure~\ref{continuation_c2}) passing through 1. Thus, $LB_2A_1$ is
connected to both $LB_2$ and $UB_2$ (hence the second label
$UB_2A_1$). $LB_2$ undergoes another pitchfork bifurcation at
$\Reym=375.41$ to asymmetric in time cycles of period $T_o$
($LB_2A_2$) disconnected from $UB_2$.

\begin{table}
  \begin{center}
  \begin{tabular}{ccccccccccccc}
      $\Reym$  & Bifurcation type  & Branch &\hspace{0.5cm}& $\Real$   &  $\Imag$ &\hspace{0.5cm}& $\Real$ & $\Imag $ &\hspace{0.5cm}&  $\Real$ & $\Imag  $\\[3pt]             
       334.52   & Saddle node &  & & 14.8 & 0 & & 1  & 0 & & 0 & 0.02\\
       335.25   & Rev. Pitchfork & $LB_2$ & & 2.18 & 0 & & 1  & 0 & & 0 & 0.03\\
       347.52  & Pitchfork & $UB_2$ & & 2.40 & 0 & & 1  & 0 & & 0.25    & 0\\
       375.41  & Pitchfork & $LB_2$ & &  29.5 & 0 & & 1  & 0 & & 0.09 & 0\\
       364.67   & P-doubling & $UB_2A_1$ & & 62.9 & 0 & &  -1  & 0 & &
       -0.06 & 0\\
       405.73   & P-doubling & $LB_2A_1$ & & 619 &  0 & & -1 & 0 & & -0.36 & 0\\
  \end{tabular}
  \caption{Same as table~\ref{tab_c1} for the \SNtwo\ pair of cycles
 $LB_2$ and $UB_2$  and their asymmetric child cycles $LB_2A_1$ and
    $UB_2A_1$.}
  \label{tab_c2}
  \end{center}
\end{table}

\subsubsection{Period-doublings of $LB_2A_1$ and $UB_2A_1$} 
The asymmetric cycles $LB_2A_1$/$UB_2A_1$
and $LB_2A_2$ are always unstable and
undergo several period-doubling bifurcations. The first
period-doubling for each branch is indicated with red diamonds in
figure~\ref{continuation_c2}. We found it impossible in that case to
converge cleanly on the period-2 solutions with the Newton solver. 
Just as in the previous example, this lack of convergence probably
indicates that the local phase space landscape is packed
with many very similar nonlinear invariant solutions.

\subsection{Other periodic structures}
\SNone\ and \SNtwo\ are the most frequent nonlinear MRI dynamo cycles
underlying the dynamics in the symmetric subspace \Aone, but other
recurrent patterns were also spotted and some transition maps indicated
that preturbulent dynamics could occur at $\Reym$ as low as 250, well
below the critical saddle node $\Reym$ of \SNone\ and
\SNtwo. Searching for other types of recurrences, we discovered that
at least two other cycles of period $T_o$ are involved in transitional
dynamics at $\Reym<327$. The Newton solver was able to converge to the
first one. This cycle is even more
energetic than $UB_2$ and also appears in a saddle node
bifurcation. It is confined to a very small range
$264\lesssim\Reym\lesssim300$, which suggests that it takes part in
the dynamics in some of the ``island'' regions spotted in
\S\ref{turb}. The second type of recurrence is quite different from
the cycles described before, as the large-scale field component appears to
reverse several times per $T_{SB}$. Unfortunately, we were not able to
converge to the underlying nonlinear periodic orbit with reasonable accuracy.

\subsection{Relevance of period-doublings to the chaotic transition}
The results described so far indicate that supercritical period-doubling
  cascades \citep{feigenbaum78} of nonlinear cycles participate in the
  transition, but they do not actually demonstrate that such cascades
  are the crux of the matter. In particular, the transient nature of the dynamics
  suggests that it is structured around a chaotic saddle, not an
  attractor. It is well known that global boundary crises of
  preexisting attractors provide a possible route to transient chaos
  \citep[see][and discussion in
  \S\ref{hydrodiscussion}]{grebogi82,kreilos12}, but in our simulations
  transient chaos is almost always found at $\Reym$ smaller than
  that a which even the first period-doublings occur. This suggests
  that other bifurcations take place in parallel to these
  period-doubling cascades. As we are about to show, the transition to
  transient chaos is most likely triggered by global homoclinic and
  heteroclinic bifurcations of unstable MRI dynamo cycles.

\section{Chaos-generating global bifurcations of MRI dynamo
  cycles\label{global}}
   Global homoclinic (respectively heteroclinic) bifurcations 
  of nonlinear invariant solutions of a dynamical system are
  associated with the transverse collision in phase space between 
  their unstable manifold $\Wu$ and their stable manifold $\Ws$
  (respectively that of another invariant solution). These
  bifurcations are a defining signature of the dynamical stretching
  and folding in the vicinity of these solutions, and the
  Birkhoff-Smale theorem  \citep{birkhoff35,smale67} states that they
  are in fact associated with the emergence of chaotic behaviour in the
  system. In this section, we provide detailed numerical evidence that
  the unstable MRI dynamo cycles described in \S\ref{cycles} provide
  the dynamical skeleton for many such bifurcations at $\Reym$ barely
  larger than their critical saddle node values.

   The problem will be investigated in two complementary ways. At a
   homoclinic bifurcation, the unstable manifold of an unstable
   invariant solution folds back towards the solution and oscillates
   indefinitely around it, crossing its stable manifold an infinite
   number of times to form a Poincar\'e tangle \citep{ott02}. This
   complex behaviour can be detected using specific techniques to
   visualize the geometry of the unstable  manifold. The second method
   consists in searching for signatures of the many dynamical
   consequences of such bifurcations, which include the formation of
   Smale horseshoes, the existence of homoclinic trajectories, and
   that of stable cycles of arbitrarily long periods.

\subsection{Numerical procedures: Poincar\'e sections and
  low-dimensional projections} 
Several dedicated numerical procedures must be used to perform this 
analysis. They are described step by step in
appendix~\ref{appglobalnumerics}. We find it useful to briefly
describe two key points of the analysis here. As explained in detail
in appendix~\ref{traceWu}, the shear periodicity makes it possible to
analyse the MRI dynamo flow in the shearing box as a discrete-time map
(denoted by $\Phi_o$ in the following), with an iteration of the map
corresponding to a time integration of the fluid equations during
$T_o$. Looking at the state of the system stroboscopically every $T_o$
is equivalent to introducing a Poincar\'e section in the flow and
makes it possible to visualize the unstable dynamics around MRI cycles
more easily. Although looking at full snapshots of numerical
simulations every $T_o$ is not very informative, a low-dimensional
projection of the dynamics can be introduced to
uncover interesting behaviour. We use a two-dimensional projection on
the large-scale axisymmetric magnetic field plane ($\Bazerox$,$\Bazeroy$),
in which MRI dynamo cycles appear as points and their unstable
manifolds as lines, if the cycles have a single unstable
eigenvalue (this is always the case in the examples below). The
following analysis required several 10 000 DNS at resolution
$(N_x,N_y,N_z)=(24,12,36)$, each simulation running for several hundred shearing times.

\subsection{Homoclinic and heteroclinic tangles of \SNone\ cycles}
\subsubsection{Tangles involving $LB_1$}

\begin{figure}
  \centerline{\includegraphics[width=\linewidth]{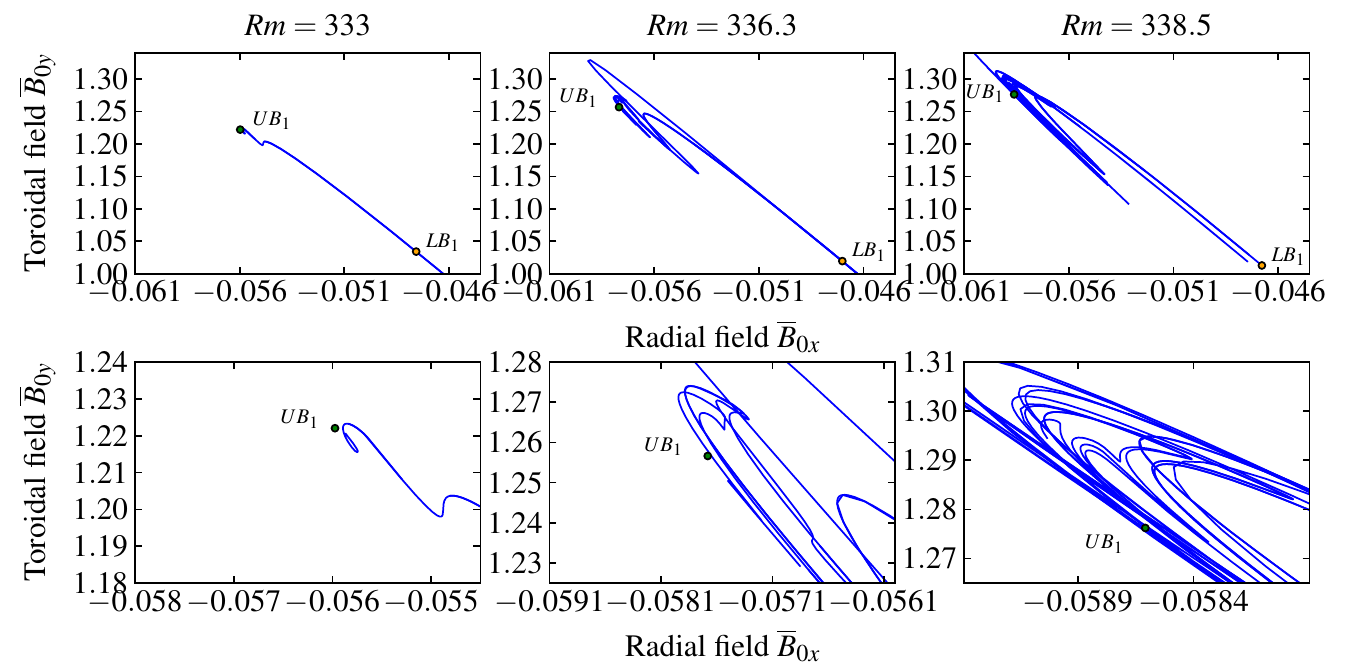}}
  \caption{Projections of the unstable manifold of the saddle cycle
    $LB_1$ as a function of $\Reym$ for $333<\Reym<338.5$. The plot
    reveals the increasingly folded geometry of $\Wu(LB_1)$ resulting from
    the parallel formation of a homoclinic tangle of $LB_1$ and of 
    a heteroclinic tangle between $LB_1$ and $UB_1$.
    Top: projections of $\Wu(LB_1)$ in the
    ($\Bazerox$,$\Bazeroy$) plane. Bottom:
    zooms on the neighbourhood of $UB_1$ showing the heteroclinic
    folding of $\Wu(LB_1)$ around $UB_1$.}
\label{foldLB1}
\end{figure}

\begin{figure}
\centerline{\includegraphics[width=\linewidth]{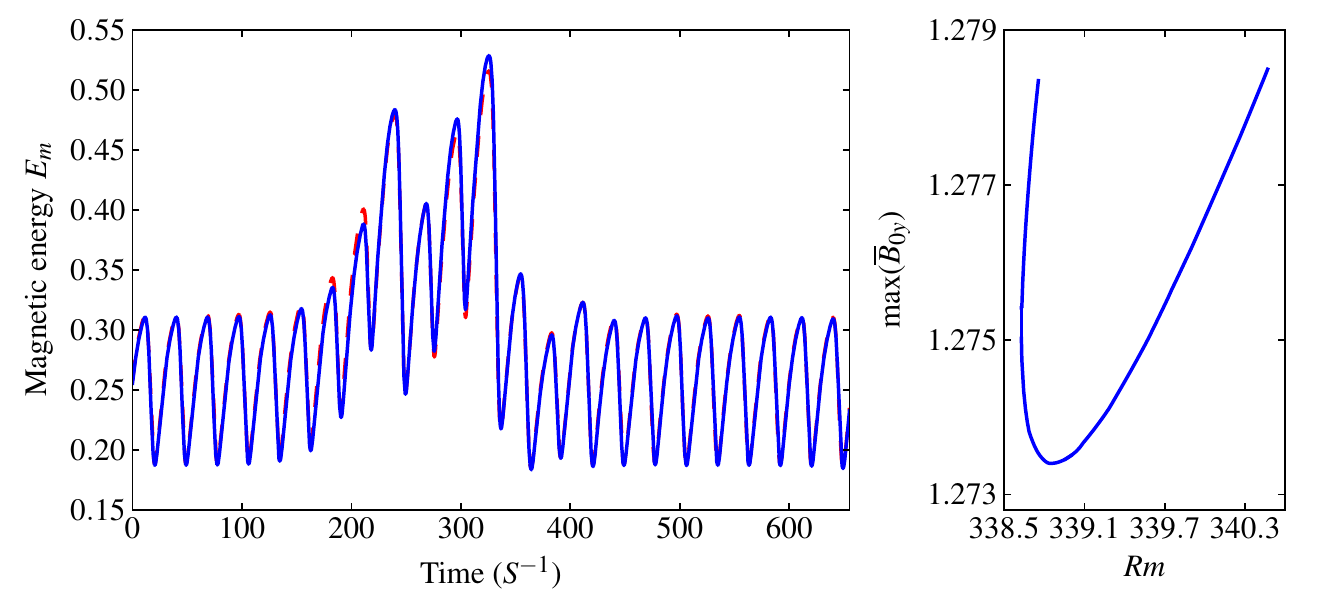}}
\caption{Left: time-evolution of the total magnetic energy along the
  least energetic homoclinic orbit of $LB_1$ at $\Reym=340.5$ (solid/blue line) and
  $\Reym=338.7$  (dashed/red line). Right: continuation curve of
  the homoclinic orbit representing the maximum of $\Bazeroy$ over
  successive snapshots taken every $T_o$ along the homoclinic orbit,
  as a function of $\Reym$.}
\label{homotrajLB1}
\end{figure}

We first studied the geometry of the unstable manifold of $LB_1$ between
$\Reym=333$ and $\Reym=338.5$. Figure~\ref{foldLB1} reveals the formation of a 
heteroclinic tangle between $\Wu(LB_1)$ and  $\Ws(UB_1)$, and to a
lesser extent of a homoclinic tangle of $LB_1$. The formation of the
heteroclinic tangle is apparent in the lower zoomed
plots, as $\Wu(LB_1)$ starts to fold repeatedly around
$UB_1$ as $\Reym$ increases. It is confirmed by a computation of a
heteroclinic orbit between $LB_1$ and $UB_1$ (not shown). The
  formation of the homoclinic tangle is not obvious because
  $\Wu(LB_1)$ first approaches $UB_1$ and folds
around it in a very complicated way before making its first
return to $LB_1$. Indirect evidence for this tangle
was found by computing a orbit homoclinic to $LB_1$
 whose intrinsic core length (or transition time)
seems to be $3T_o$ \citep[difficult to show rigorously in the
absence of a dedicated technique; see][]{sterling99}. 
Figure~\ref{homotrajLB1} shows the
time-evolution of magnetic energy along this orbit. 
A continuation of this orbit with respect to $\Reym$
was performed with a dedicated continuation
procedure (figure~\ref{homotrajLB1} (right)).
The bifurcation to this particular homoclinic orbit occurs at
$\Reym=338.62$, smaller than the critical $\Reym$ at which the first 
period-doubling of either $LB_1A_1$ or $UB_1A_1$ occurs. Note that within the same
tangle, homoclinic orbits with different cores are created at slightly
different $\Reym$ values. Figure~\ref{homotrajLB1}
shows that two branches of distinct homoclinic solutions exist and
merge smoothly at a homoclinic bifurcation of the saddle node type. 

\subsubsection{Homoclinic tangle of $UB_1$\label{homoUB1}}
\begin{figure}
  \centerline{\includegraphics[width=\linewidth]{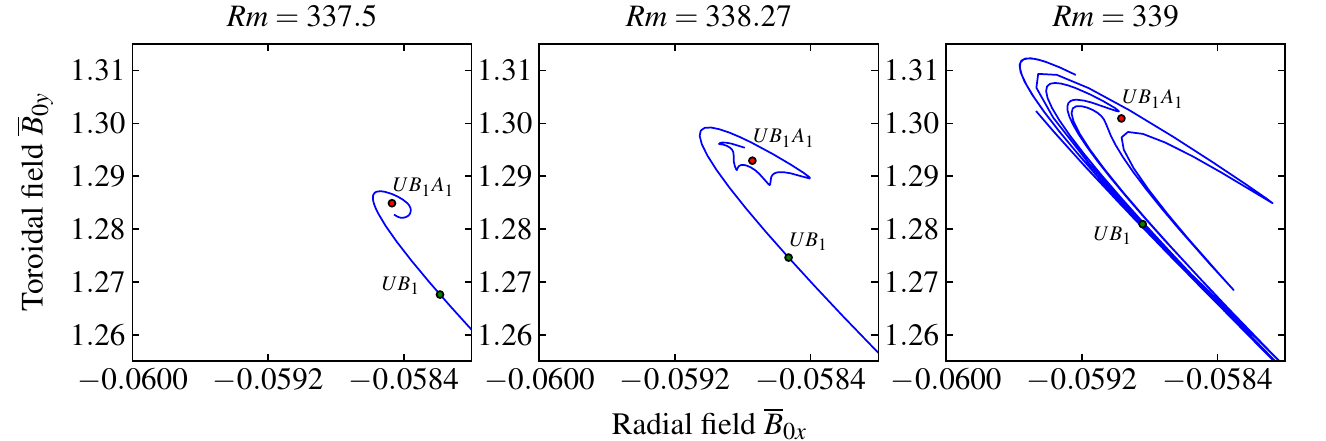}}
  \caption{Projections of the unstable manifold of $UB_1$ in the
    ($\Bazerox$,$\Bazeroy$) plane as a function of  $\Reym$ for
    $337.5<\Reym<339$. The homoclinic tangle is clearly visible at
    $\Reym=339$.}
\label{foldUB1}
\end{figure}

\begin{figure}
\centerline{\includegraphics[width=\linewidth]{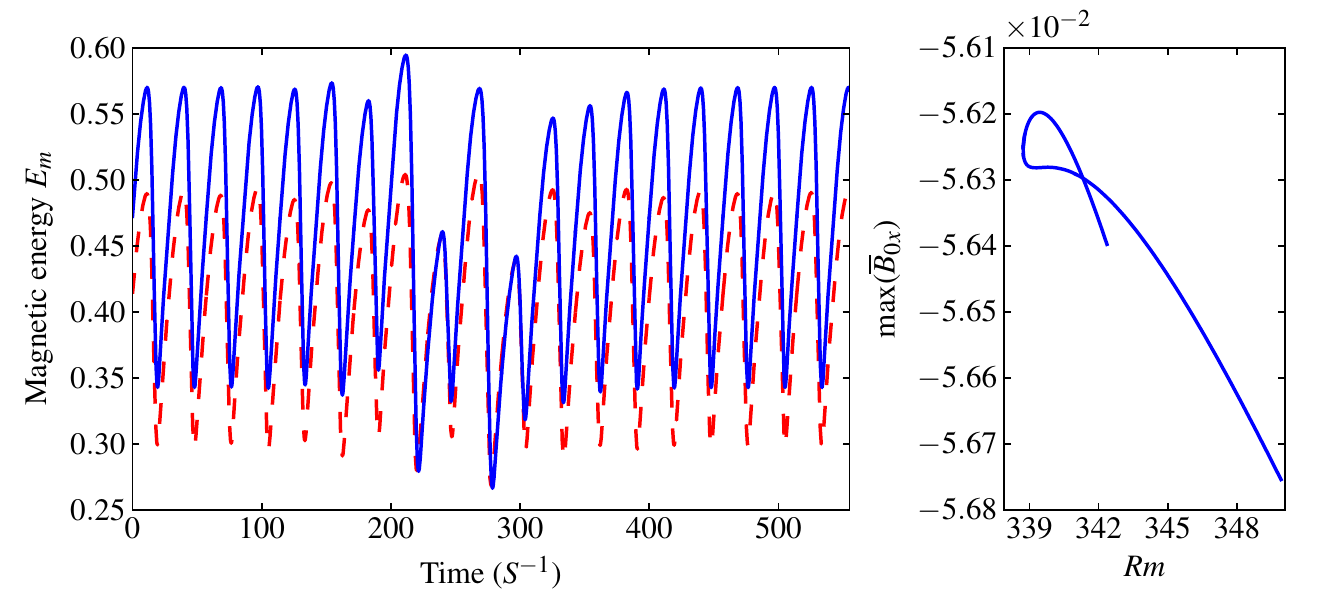}} 
\caption{Left: time-evolution of the total magnetic energy along the
  least energetic homoclinic orbit of $UB_1$ at $\Reym=349.9$
  (solid/blue line) and
  $\Reym=338.7$  (dashed/red line). Right: continuation curve of
  the homoclinic orbit representing the maximum of $\Bazerox$ over
  successive snapshots taken every $T_o$ along the homoclinic orbit,
  as a function of $\Reym$.}
\label{homotrajUB1}
\end{figure}

Is this behaviour generic? We performed a similar
study of the unstable manifold of $UB_1$ for $337.5<\Reym<339$. 
Figure~\ref{foldUB1} shows a projection of $\Wu(UB_1)$ at
three different $\Reym$. For $\Reym=337.5$, $\Wu(UB_1)$
approaches the asymmetric solution $UB_1A_1$ and loops
around it. At this $\Reym$, $UB_1A_1$ is a stable focus (one
stable real eigenvalue and two stable complex conjugate ones) and
$\Wu(UB_1)$ clearly feels its local attraction. At slightly larger
$\Reym=338.27$,  $UB_1A_1$ is less focusing and
$\Wu(UB_1)$ somehow ``escapes'' from its
local neighbourhood. We observe the formation of spikes 
which tend to be attracted back to the neighbourhood of $UB_1$. 
At $\Reym=339$, the spikes have clearly been
stretched along the unstable direction of $UB_1$ and a homoclinic
tangle is born: $\Wu(UB_1)$ oscillates indefinitely in front of $UB_1$
at ever smaller distances of $UB_1$ in the direction perpendicular to its
unstable eigendirection, hinting at an infinite number of
transverse intersections between its stable and unstable manifolds. 
The conclusion is supported by the computation a homoclinic orbit
of $UB_1$ (figure~\ref{homotrajUB1}). The continuation of this
orbit with respect to $\Reym$ (figure~\ref{homotrajUB1}
(right)) shows that two branches of distinct homoclinic solutions appear
at $\Reym=338.72$ in a homoclinic saddle node bifurcation. 

\subsection{Other heteroclinic tangles}
\begin{figure}
\centerline{\includegraphics[width=\linewidth]{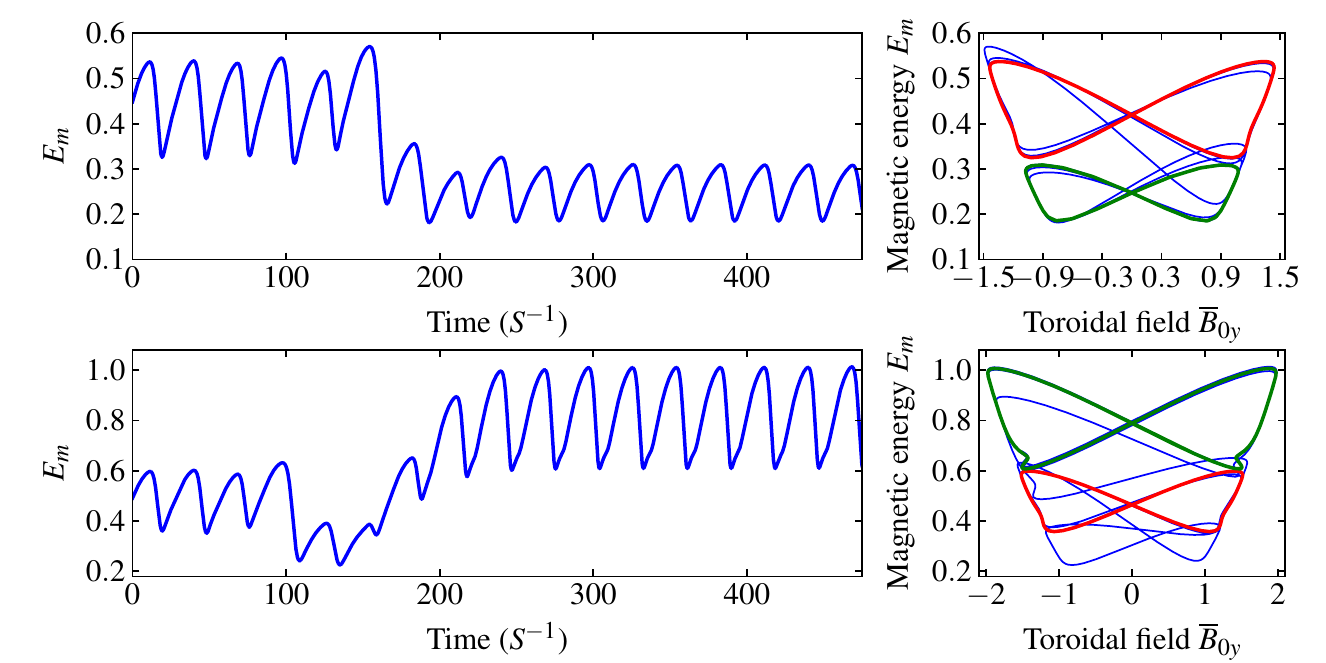}}
\caption{Heteroclinic orbits leaving $UB_1$. Top: heteroclinic
  orbit from $UB_1$ to $LB_1$ ($\Reym=345$). Bottom:
  heteroclinic orbit from $UB_1$ to $LB_2$ ($\Reym=352$). Left: evolution of
  magnetic energy over 500 $S^{-1}$. Right: phase portraits representing
  magnetic energy as a function of $\Bazeroy$. In both cases, the
  thin solid/blue line represents the heteroclinic orbit, the thick
  solid/red line represents the MRI dynamo cycle from which the heteroclinic
  trajectory departs, and the thick solid/green line represents the
  MRI dynamo  cycle to which the trajectory asymptotes.}
\label{heteroUB1LB1}
\end{figure}

We found evidence for two other heteroclinic tangles. At $\Reym$
larger than 343,  $\Wu(UB_1)$ visits the neighbourhood of $LB_1$,
making folds in its
immediate vicinity. A heteroclinic orbit from $UB_1$ to $LB_1$ for
$\Reym=345$ is shown in figure~\ref{heteroUB1LB1} (top).
Another notable result is the existence of heteroclinic
tangles  between cycles born out of different saddle node
bifurcations. At $\Reym>350$, $\Wu(UB_1)$ approaches $LB_2$ and makes
folds in its vicinity. A heteroclinic orbit between $UB_1$
and $LB_2$ at $\Reym=352$ is shown in figure~\ref{heteroUB1LB1} (bottom).

\subsection{Smale horseshoe and period-doublings of stable node cycles\label{smale}}
We now present several supplementary numerical results which
further confirm the occurrence of global bifurcations and make the
connection between the tangles and the emergence of chaos in the
system explicit. This paragraph is rather technical, and
readers only interested in the transition phenomenology may want to
jump directly to \S\ref{tanglechaos}. The analysis is restricted to
the homoclinic tangle of $UB_1$, but similar results can be obtained
for the other identified tangles.

We first verify the occurrence of horseshoe-type dynamics \citep{ott02}
predicted by the Birk\-hoff-Smale theorem in the tangle of $UB_1$.
The theorem basically states that if a fixed point of a
discrete-time map (such as $\Phi_o$) has a homoclinic tangle, then a
sufficiently high number of iterates of the map behaves as a
horseshoe map in the neighbourhood of the fixed point. 
This means that if we consider a small initial subset $D$ of state
space close to $UB_1$ and aligned with $\Wu(UB_1)$, there should exist
an integer $N\geq1$ such that the intersection between $D$ and
$\Phi_o^N(D)$  is not empty. Besides, this intersection should contain
a fundamental periodic orbit of period $NT_o$ (in the sense that it
cannot be constructed from shorter-period orbits).
The set of points defined as the infinite intersection
$\bigcap_{k=0}^{\infty} \Phi_o^{kN}(D)$  should in fact contain a
countable infinity of periodic orbits.

\begin{figure}
\centerline{\includegraphics[width=\linewidth]{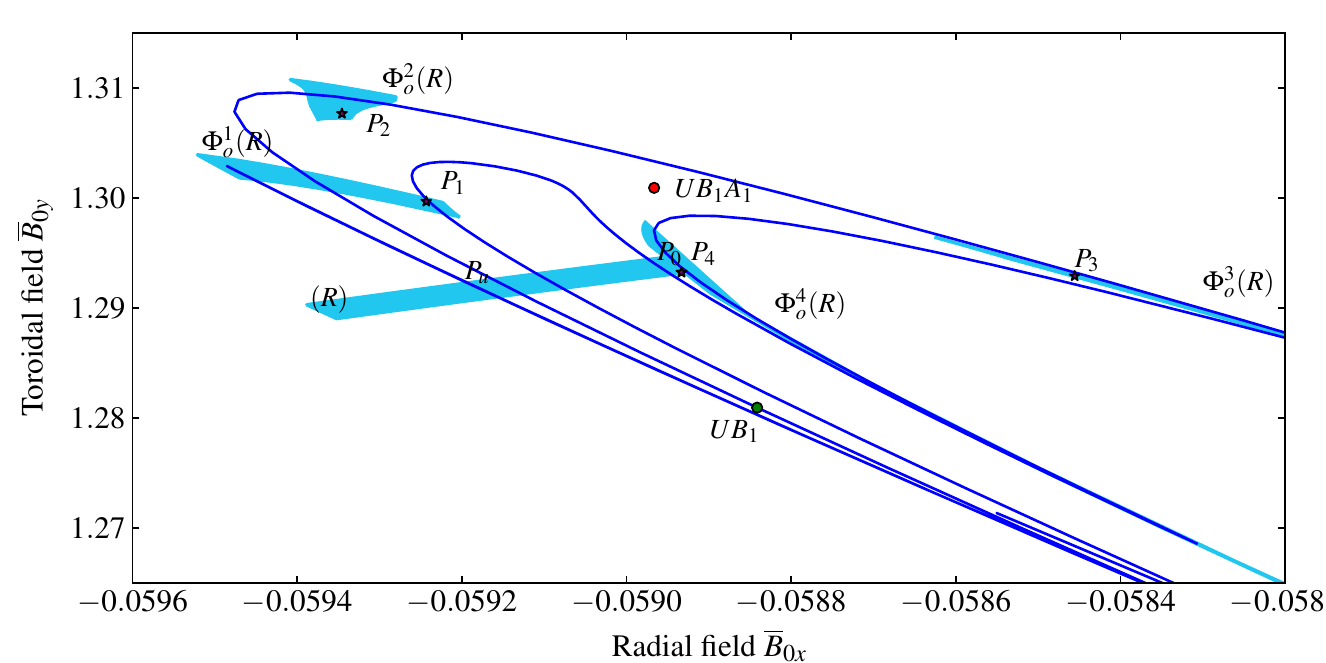}}
\caption{Stroboscopic dynamics in the homoclinic tangle of $UB_1$
  projected in the ($\Bazerox,\Bazeroy$) plane at $\Reym=339$. $UB_1$ and
  $UB_1A_1$ are represented with filled circles. $\Wu(UB_1)$ is
  represented by a solid/blue line. An initial patch $R$
  of phase space overlapping  $\Wu(UB_1)$  and its successive images
  under $\Phi_o$ are shown in grey/light-blue. The plot
  shows the intersection between $R$ and $\Phi_o^4(R)$. The 
  projected trajectory of the period-4 cycle is shown as a succession
  of stars $P_0$ to $P_3$.}
\label{horseshoe}
\end{figure}

To confirm this behaviour in the vicinity of $UB_1$ at $\Reym=339$, 
we tried to establish the existence of long-period orbits associated
with horseshoe intersections. Consider the return map representing the
$j^{\mathrm{th}}$ $\Phi_o$ iterate of an initial segment of
$\Wu(UB_1)$ versus its $i^{\mathrm{th}}$ iterate, with $j>i$. The fixed
points of the curves constructed that way should be periodic orbits of
period $j-i$. Plotting the return maps of $\Bazeroy$ between the first iterate
of $\Phi_o$ and its fifth iterate, we discovered such an
approximate fixed point, which was fed into the Newton solver
instructed to look for a cycle of period $4T_o$. The solver easily 
converged on a genuine period-4 cycle which is not a repetition of
shorter cycles.

\begin{figure}
\centerline{\includegraphics[width=\linewidth]{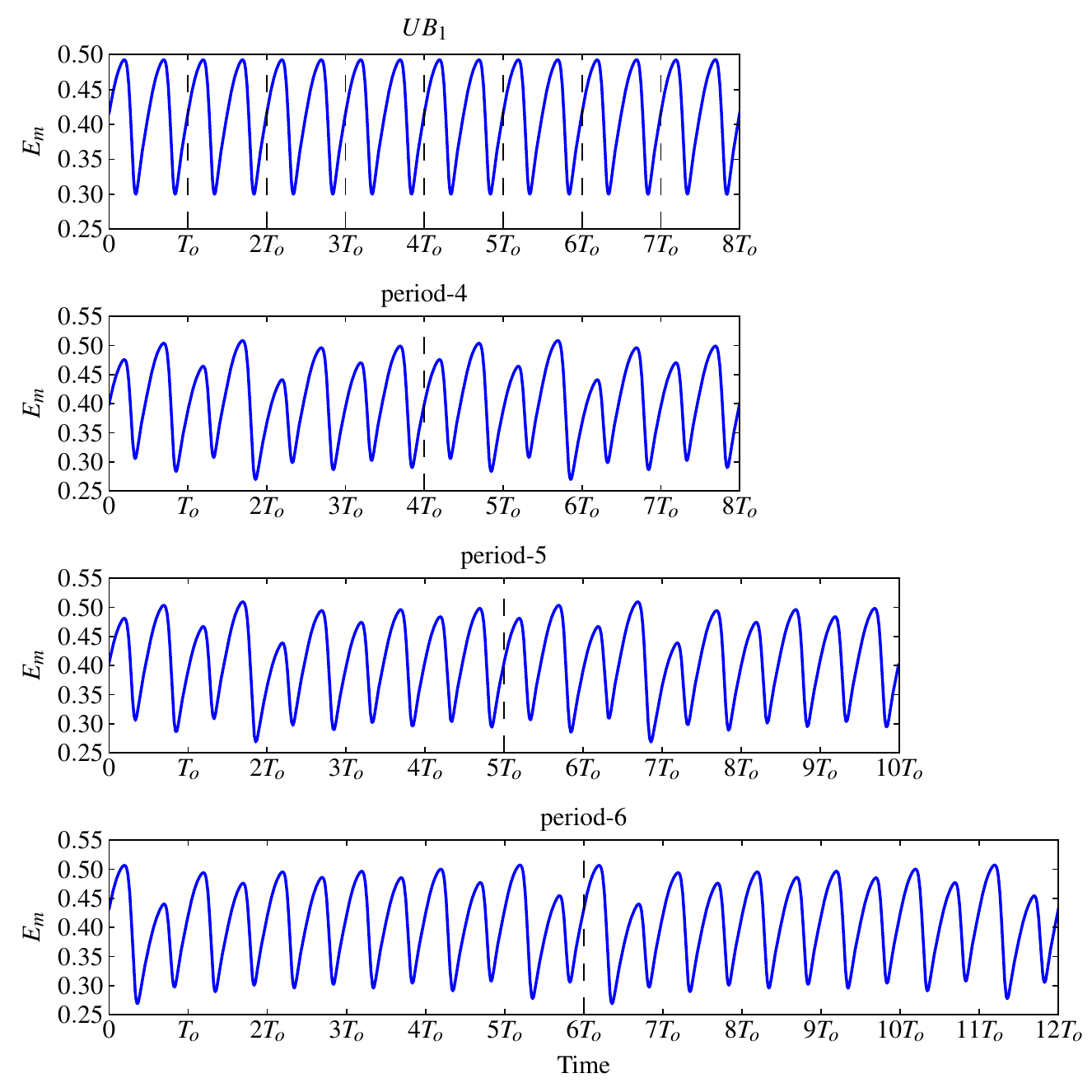}}
\caption{Several long-period saddle node cycles in the homoclinic tangle of
  $UB_1$. The time-evolution of $UB_1$ is also shown and repeated for
  $8T_o$ for comparison.}
\label{longperiod}
\end{figure}

The period-4 cycle can be used to ``image'' the horseshoe map. 
The dynamical state of this cycle after each $T_o$ can be visualized as
a set of four points $P_0$, $P_1$, $P_2$, and $P_3$ in the projection plane,
 shown with stars in figure~\ref{horseshoe} ($P_4$, the image of $P_3$,
coincides with $P_0$). Let us consider the point $P_u$ 
 on the initial segment of $\Wu(UB_1)$ closest to $P_0$ and
 construct a small patch $R$ of phase space enclosing that point
 (grey/light-blue shade) by computing a set of states
 consisting in joint perturbations of $UB_1$ (whose state
 vector we denote by $\vec{X}_e$) along its unstable and first two
 stable eigenvectors $\vec{X}_u$, $\vec{X}_{s1}$, $\vec{X}_{s2}$ :
\begin{equation}
\vec{X} (\varepsilon_u,\varepsilon_{s1},\varepsilon_{s2})=
\vec{X}_e+\varepsilon_u \vec{X}_u+\varepsilon_{s1} \vec{X}_{s1}+ \varepsilon_{s2} \vec{X}_{s2}~.
\end{equation}
The values of $\varepsilon_u$ are taken very small compared to those
of $\varepsilon_{s1}$ and $\varepsilon_{s2}$, so that $R$ is strongly elongated along the
stable directions of $UB_1$. Each of these states is integrated
by successive DNS of period $T_o$ ($\Phi_o$ iterations) to obtain a
collection of images of $R$ under increasing powers of $\Phi_o$
(grey/light-blue shades in figure~\ref{horseshoe}). The map stretches
$R$ along $\Wu(UB_1)$ and contracts it in the directions perpendicular
to it. After four iterations, the image $\Phi_o^4(R)$ of the
original patch $R$ intersects it in the projected plane, suggesting
that $\Phi_o^4$ is a horseshoe map. Note that $P_o$ is both in $R$ and 
$\Phi_o^4(R)$, so $\Phi_o^4(R)$ and $R$ do indeed intersect in the full
phase space. Using the return map technique with different $j-i$, we found
 several fundamental cycles of period 3, 4, 5, 6, 7, 8
 and 9$T_o$, indicating that the homoclinic tangle hosts
 several horseshoes, each of them corresponding to 
a different number of iterates of $\Phi_o$.
The evolution of magnetic energy for several such cycles, some of
which were actually spotted in DNS in \S\ref{turbislands},
is shown in figure~\ref{longperiod}. Being close to $UB_1$ in phase
space, all the cycles display $UB_1$-like oscillations which appear to
be modulated on the longer fundamental period of the cycles.

The previous results finally make it possible to check various 
mathematical predictions for two-dimensional dissipative maps relating
the formation of horseshoes to that of stable long-period upper branch
cycles appearing in saddle node
bifurcations \citep{gavrilov72,newhouse79,robinson83,palis93}. 
Continuation of the long-period cycles discovered in the tangle indeed
reveals that they all arise from saddle node bifurcations at slightly
different $\Reym$. The first ones to appear are a pair of period-6 cycles
at $\Reym=338.72$, followed by period-5 ($\Reym=338.75$), period-7
($\Reym=338.89$), period-4 ($\Reym=338.91$) and finally period-3
($\Reym=339.39$). A local stability analysis shows that all upper
branch solutions have a small domain of stability
near the bifurcation, perhaps explaining why some of them
can be spotted in DNS (\S\ref{turbislands}), and that all of them
undergo a period-doubling bifurcation at slightly larger
$\Reym$, consistent with another prediction of \cite{yorke83}.
Further similarities with bifurcations of the H\'enon map 
 will be briefly discussed in \S\ref{hydrodiscussion}.

\section{Chaos in the \SNone\ tangles and transition maps\label{tanglechaos}}
\begin{figure}
\centerline{\includegraphics[width=\linewidth]{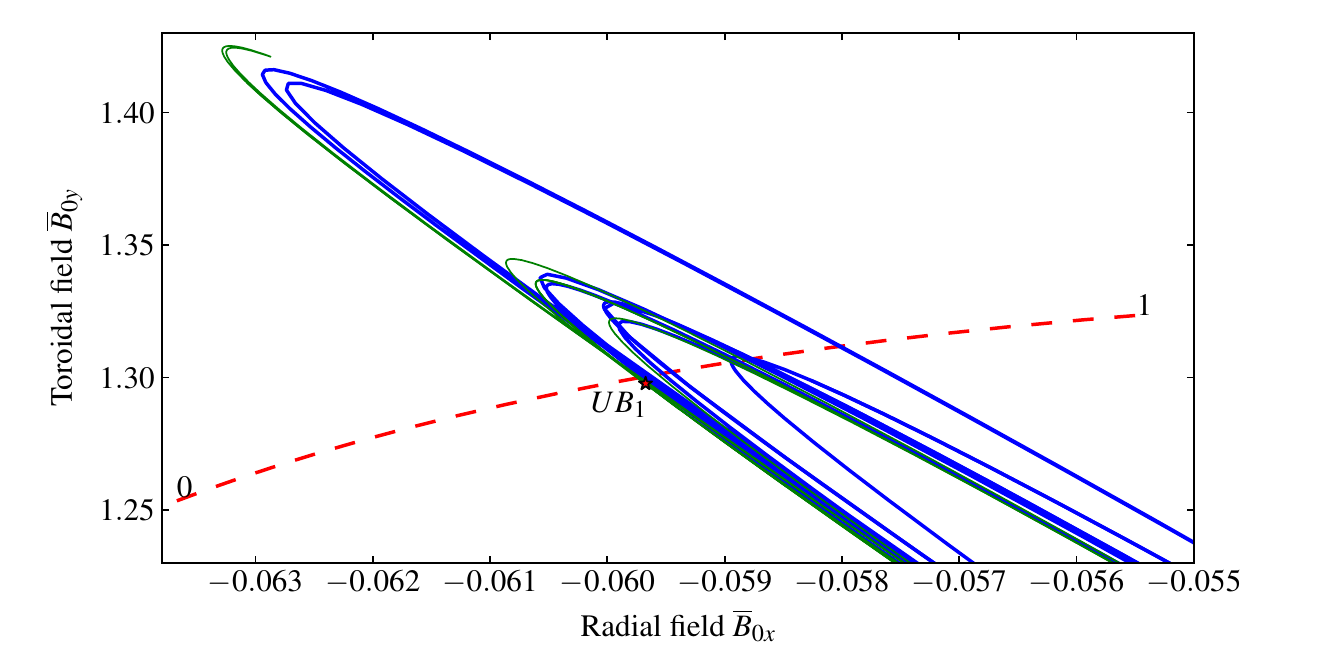}}
\centerline{\includegraphics[width=\linewidth]{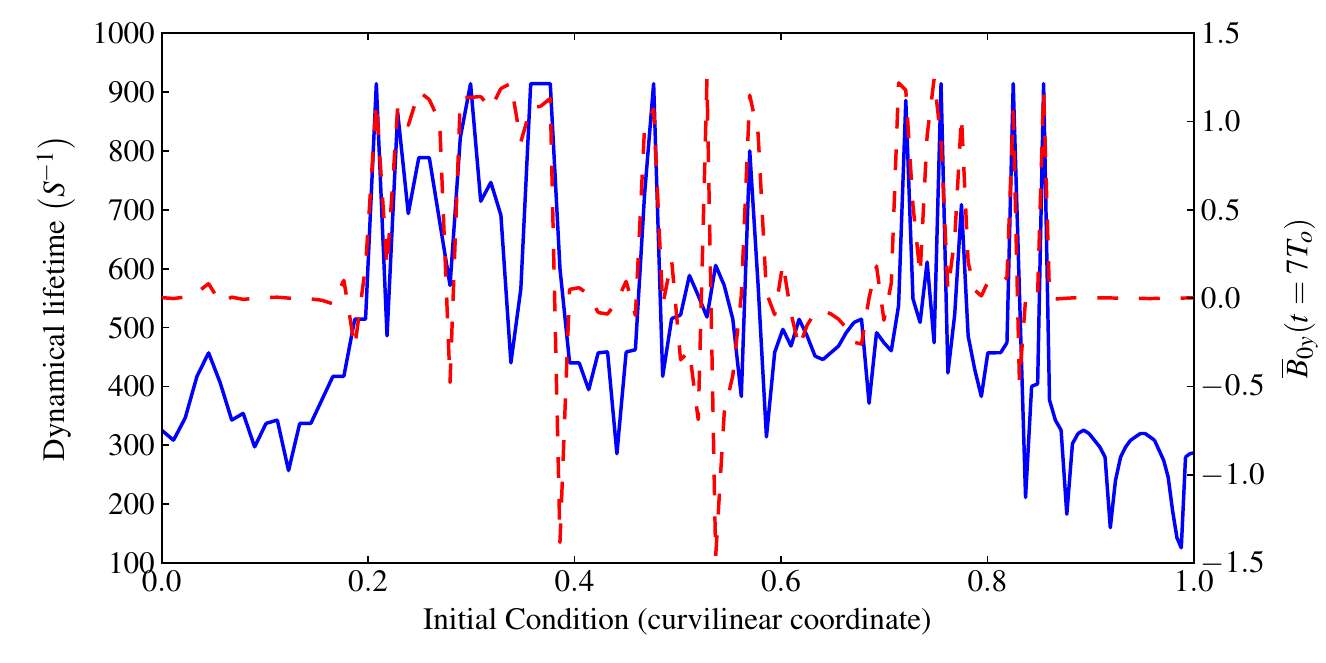}}
\caption{Dynamics in the vicinity of $UB_1$ at $\Reym=341$ as
  probed along a line of initial conditions 
  across the homoclinic tangle of $UB_1$. Top:
  $(\Bazerox,\Bazeroy)$ plane projection of 
  $\Wu(UB_1)$ (thin solid/green line), $\Wu(LB_1)$ (thick solid/blue
  line) and of the phase space line along which initial conditions are
  calculated (dashed/red line). Bottom: dynamical lifetime
  (solid/blue line) and amplitude of $\Bazeroy$ after $t=7T_o$
  (dashed/red line) for each simulation, as a
  function of the corresponding initial condition labelled by a
  normalized curvilinear coordinate along the line.}
\label{chaos}
\end{figure}

The objective of this short section is to offer a more intuitive
  illustration of the dynamical consequences of these global bifurcations
  and to show that they provide a natural explanation for the 
  results of \S\ref{turb}. We present a numerical experiment
  which probes the sensitivity of the system on a set of
  initial conditions taken along a line in phase space crossing
the homoclinic tangle of $UB_1$ (figure~\ref{foldUB1}). The idea is to
mimic the experiments of \S\ref{turb} which shooted along random
directions in phase space. We restrict the analysis to the dynamics in
the vicinity of $UB_1$ and to a single $\Reym=341$. The same procedure
could easily be repeated for other tangles. The set of initial
conditions crossing the tangle is computed along the same lines as the
patch $R$ in \S\ref{smale}. Projections of this locus of initial
states and of $\Wu(UB_1)$ and $\Wu(LB_1$) are shown in figure~\ref{chaos}.

Figure~\ref{chaos} (bottom) displays the dynamical lifetime
(solid/blue line) and the amplitude of $\Bazeroy$  after $t=7T_o$
(dashed/red line), measured in each simulation as a function of a
normalized curvilinear coordinate along the initial condition line. 
Sensitive dependence on
  initial conditions is observed as soon as the dynamics is
  initialized close enough to $UB_1$, producing a typical pattern
  of lifetimes reminiscent of those obtained by
slicing vertically the transition maps of \S\ref{turb}. This
sensitivity is lost for continuous sets of initial conditions
sufficiently far away from the tangle. In that case, the flow
relaminarizes much more quickly, just like in the smooth blue/dark
regions separating turbulent islands and fingers in the maps.
The natural explanation for this chaotic behaviour is the stretching
and folding of the dynamics in the phase space neighbourhood of $UB_1$.

\section{Discussion\label{discussion}}
\subsection{Summary of the main results}
In this paper, we identified possible transition
mechanisms to chaotic, ``preturbulent'' dynamics in the three-dimensional,
incompressible nonlinear MRI dynamo problem in Keplerian shear flow, 
using the numerical shearing box framework. We focused on bifurcations
occurring as the magnetic Reynolds number $\Reym$ of the system is
varied, for fixed ``minimal'' box size $(L_x,L_y,L_z)=(0.7,20,2)$,
fixed kinetic Reynolds number $\Rey=70$ and for a range of magnetic Prandtl
  number $2\lesssim\Pm\lesssim 8$. In most parts of the work, the dynamics was
restricted to an invariant symmetric subspace to simplify the analysis.

In \S\ref{turb}, we presented a phenomenological description of the
MRI dynamo transition based on a numerical cartography of the
laminar-chaotic boundary in the phase space of the system. 
It was shown that this transition does not occur at a well-defined
critical $\Reym$ and that the stability border has a complex,
fractal-like shape. The transitional dynamics  is always transient but
appears to be fairly recurrent and long-lived in islands or
finger-shaped regions, hinting at the presence of nonlinear periodic
solutions (cycles).

In \S\ref{cycles}, we described accurate computations,
continuations and stability analyses of two families of nonlinear MRI
dynamo cycles whose dynamics were transiently spotted in
\S\ref{turb}. These cycles, including that computed by
\cite{Herault2011}, appear in pairs in saddle node bifurcations
at $\Reym$ typical of the transition as a whole. All cycles appear
either to be unstable straight from the saddle node bifurcation or to
lose stability quickly in pitchfork bifurcations, resulting in the
creation of new
cycles whose main characteristic is to be asymmetric in time. These
cycles themselves undergo period-doubling bifurcations. In some cases,
period-doublings of the period-doubled cycles were also found, indicating a
possible route to chaos via a cascade of supercritical period-doublings.
However, it was pointed out that the first period-doublings take
place at $\Reym$ larger than those at which transient chaos was
typically found in \S\ref{turb}.

This problem was addressed in \S\ref{global}. Studying the geometry
of the unstable manifolds of several cycles studied in \S\ref{cycles},
we found that chaos-generating global homoclinic and heteroclinic
bifurcations occur in the system at $\Reym$ barely larger than that of
the saddle node bifurcations themselves. Quantitative verifications of
several expected dynamical consequences  of homoclinic bifurcations  
(horseshoe-type dynamics, existence of homoclinic and heteroclinic
orbits and of many long-period cycles) were provided to support this
claim. In \S\ref{tanglechaos}, the sensitivity of
the dynamics on initial conditions was confirmed in the vicinity of one of the
cycles after a homoclinic bifurcation, and significant similarities
with the transition maps described in \S\ref{turb} were pointed out.

Based on these results, we argue that the first germs of
chaotic MRI dynamo action in this system are provided
by global bifurcations of a few energy-injecting nonlinear
MRI dynamo cycles whose dynamics takes place at scales comparable to
the system scale at mild $\Reym$ and $\Rey$. The complex-shaped chaotic
regions observed in transition maps are interpreted as signatures of
the homoclinic and heteroclinic tangles surrounding these cycles in
phase space. Heteroclinic connections between lower
and upper branch cycles born out of different saddle nodes indicate
that the system can easily jump from low energy states to higher
energy states (and conversely). An increasing number of cycles
involving smaller spatial scales are expected to appear in successive
saddle node bifurcations at increasingly large $\Rey$ and $\Reym$ and
to complexify the dynamics even further.

\subsection{Connections with astrophysical accretion flows\label{astroturb}}
The highly idealized nature of the numerical experiments
  presented in this work (constrained geometry, incompressible flow,
  fairly low $\Rey$) does not make it possible to 
  extrapolate the results directly to astrophysical accretion disks,
  but it is worth discussing which conclusions are likely to be at least
  qualitatively relevant or useful for understanding MHD turbulence
  and dynamo action in disks. An important first remark is
  that the results are the direct outcome of simulations of the
  full three-dimensional incompressible dissipative MHD equations, not
  of a mean-field model or of a very low-dimensional toy dynamical
  system. Besides, \cite{Herault2011} showed that the MRI dynamo
  cycles studied in both papers rely on several physical mechanisms
  which are all highly relevant to disks (and experiments): the
  $\Omega$ effect, the non-axisymmetric
  magnetorotational instability (which should take the form of a
  sheared spiral pattern in a cylindrical flow), nonlinear advection
  and induction mechanisms, and wave scattering processes. 
  Several studies \citep{lesur08, rempel10} also suggest that chaotic
  saddles and periodic dynamics carry over to different
  geometric configurations in the incompressible framework. 
  These arguments strongly suggest that the self-sustaining MRI dynamo
  process is a robut physical nonlinear phenomenon
  which should be present in a variety of Keplerian-like shear flows
  of conducting fluids (except perhaps at low $\Pm$; see below).

  Many simulations in vertically stratified Keplerian shearing boxes
  \citep{branden95,stone96,gressel10,davis10}, some of them
  of large horizontal extent \citep{simon12}, also exhibit
  recurrent dynamics. However, several aspects of this dynamics
  appear to be more difficult to describe than the periodic solutions
  identified in this paper. In particular,  magnetic reversals
  reported in these simulations probably result
  from the combined nonlinear action of many successive
  MRI-unstable shearing wave packets, while
  the cycles studied in \S\ref{cycles} require just one simple
  MRI-unstable shearing wave per large-scale field reversal
  \citep{Herault2011}. The latter cycles therefore illustrate the closure
  of the toroidal to poloidal part of the MRI dynamo loop in its
  purest form and are in that sense the most simple MRI dynamo cycles
  that can be constructed from the physics. Much remains to be
  understood to connect these simple nonlinear solutions and their
  bifurcations to the more statistical phenomena observed in other
  configurations. Finally, we have not discussed in this paper the
  effects of vertical stratification and vertical boundary conditions,
  which add yet another layer of complexity to the problem. Both are
  notably likely to have an important impact on the typical timescale
  and $\Pm$-dependence of the dynamo through the combined effects of
  magnetic buoyancy, vertical symmetry breaking and magnetic helicity
  expulsion \citep{davis10,kapyla11,simon11,oishi11}.

The main conclusion of this discussion is that the incompressible
shearing box model does not include all the possible
effects relevant to accretion disks physics but incorporates enough
ingredients of the full problem for the MRI dynamo to be active at
$\Pm >1$ in its simplest possible form. The results therefore seem to
provide a tangible starting point to address the question of the
existence of an MRI dynamo mechanism  at low $\Pm$ from a quantitative
nonlinear dynamics perspective. In particular, they make it possible
to ask computationally challenging but well-defined questions such as
whether a self-sustaining MRI dynamo process, MRI dynamo cycles (not
necessarily those described in this paper) and chaos-generating
bifurcations are present in this regime. Multi-parameter explorations
along the lines of the numerical study presented in \S\ref{turb}
represent a complementary avenue of research in this respect.

The results may to a lesser extent also be useful for studying how (MHD)
turbulent angular momentum transport varies with the
various dimensionless parameters of the problem. All the cycles found in the
range of magnetic Prandtl numbers investigated transport a significant
amount of angular momentum, comparable to or larger than the typical
transport rates estimated in most past DNS studies of the problem\footnote{Quite
  remarkably, in the configuration studied, the Maxwell
  stress throughout the periodic dynamics is dominated by the
  correlation between the large-scale axisymmetric ``dynamo'' field
  components $\Bazerox$ and $\Bazeroy$, not by the self-correlations
  of non-axisymmetric MRI disturbances.}. However, whether a few
large-scale cycles, most notably the more energetic upper branch
solutions, are representative of fully turbulent states remains to be
assessed.  Making a representative census of all the organizing cycles
present at a given $\Reym$ and $\Rey$ is probably required to obtain
reliable estimates of the magnitude of turbulent transport using periodic orbit
theory \citep{cvit92} but appears difficult in such a high-dimensional
system in turbulent regimes. A similar problem is currently
investigated in hydrodynamic turbulence \citep{cvit10,chandler13}.

\subsection{Connections with dynamo theory and experiments}
The results also confirm the subcritical, non-kinematic nature of
the MRI dynamo. In particular, MRI dynamo cycles do not
bifurcate linearly from a preexisting turbulent hydrodynamic
state. The electromotive force involved in the
self-sustaining regeneration process of the large-scale axisymmetric
field is also always observed to be a complicated nonlinear time-delayed
function of the field in the configuration studied, indicating that
the coherent, large-scale periodic dynamics described by \cite{Herault2011}
cannot be simply described in terms of classical mean-field theory.
It may be possible to fit numerically the $\overline{\vec{\mathcal{E}}}$ versus
$\overline{\vec{B}}$ relation using various methods but a rigorous,
physically motivated statistical closure theory valid for
generic physical configurations remains to be developed in connection
with the observations made in \S\ref{astroturb}.

A second significant result in this context is the
very explicit identification of global homoclinic and heteroclinic
bifurcations in a three-dimensional MHD system. Similar bifurcations
have long been invoked in the context of the geodynamo and
experimental dynamos \citep[see
e.g.][]{gallet12,gissinger12} and there have actually been reports of
heteroclinic dynamo orbits in the VKS experiment \citep{monchaux09}. 
A compelling observation very likely related to the
occurrence of  such bifurcations is the remarkable dynamical
complexity offered  by the MRI dynamo and other instability-driven
dynamos \citep{cline03,tobias11} even in fairly low-resolution
three-dimensional numerical experiments. Not only are fully
three-dimensional, large-scale (comparable to the size of the domain)
MRI dynamo cycles easy to excite,  the problem also contains
all the germs of chaotic dynamo action, 
a hallmark of many astrophysical dynamos. It is therefore tempting to
conjecture that instability-driven dynamos relying on  self-sustaining
processes play an important role in the generation of large-scale,
time-dependent magnetic fields in differentially rotating astrophysical flows
such as those encountered in stellar interiors. Global simulations
might be able to test this in the future.

This work may finally turn out to be useful to future laboratory
experiments on dynamos and on the MRI. The low $\Pm$ issue discussed earlier
is also important in this context, as all experiments to date
have been conducted in liquid metals for which \hbox{$\Pm\ll 1$}, but
upcoming plasma dynamo experiments may make it possible to study the
MRI dynamo in Keplerian-like flows in $\Pm >1$
regimes \citep{collins12}. Another possible caveat relative to the
connection with experiments is that the shearing box has no walls and  
no curvature. Similar work in wall-bounded
  curved geometries is needed to assess the
  existence of MRI dynamo cycles in this kind of configuration and to
  make quantitative predictions for experimental MRI dynamo
  thresholds and periods. As explained by \cite{Herault2011}, the
  essential triad couplings characteristic of all self-sustaining
  processes in shear  flows do not require walls. The existence of
  time-periodic nonlinear hydrodynamic solutions
  in \textit{wall-bounded} shear flows
  \citep{kawahara01,viswanath07,kreilos12} also strongly suggests that
  the time-periodic nature of the solutions computed in this paper
  is not a shearing box artefact. Of course, MRI dynamo cycle periods in
  wall-bounded geometry should not be quantized.

\subsection{Connections with other chaotic systems and hydrodynamic
  transition of shear flows\label{hydrodiscussion}}
  The dynamical behaviour described in the previous sections has much
  in common with several well-known simpler dynamical systems, such as
  chaotic scattering by three potential hills \citep{ding90} and,
  perhaps more importantly, with the bifurcations of the two-dimensional
  H\'enon map \citep{henon76,grebogi83,sterling99}. The sequence of 
  global bifurcations uncovered in \S\ref{global} is actually almost
  identical to that following the saddle node bifurcation to period-1
  cycles in that map \citep{grebogi87,goswami02}. We did not 
  investigate the question but it cannot be ruled out the closure of
  $\Wu(UB_1)$ is a H\'enon-type attractor in a limited range of $\Reym$.

  The most interesting connection is with recent advances on the problem
  of hydrodynamic transition of shear flows, in particular with the work
  of \cite{kreilos12} on plane Couette flow. They describe 
  a basin boundary crisis \citep{grebogi82,ott02} which results from 
  the collision between a chaotic attractor and the stable manifold of
  a lower branch saddle  equilibrium. Indirect evidence for a
  homoclinic tangle of a saddle solution has also been reported in
  Couette flow by \cite{vanveen11}.  The results altogether suggest
  that various homoclinic and heteroclinic crises
  involving different types of repelling and attracting structures are
  possible in these problems. Their occurrence probably depends
  on the flow geometry, aspect ratio and detailed properties of
  the nonlinear saddle node solutions taking part in the transition
  process.

  Note finally that further bifurcations may occur at larger
    $\Rey$ or $\Reym$ in these problems and change the nature of the
    dynamics, as exemplified by the Lorenz system
    \citep{sparrow82}. Multiple crises scenarios could perhaps be
    investigated with the techniques used in this paper, but the task
    appears difficult in view of the dynamical complexity already
    present at mild parameter values. Another interesting research
    direction in this respect is the possible route to spatio\-temporal
    chaos in spatially extended systems proposed by \cite{pomeau86}
    and recently explored by
    \cite{manneville09,willis2009,moxey2010,barkley2011,avila11}.

\medskip

We thank S\'ebastien Fromang, Michael Proctor, Erico
Rempel, Bertrand Georgeot and Alexander Schekochihin for many useful
discussions. This research was supported by the University Paul
Sabatier of Toulouse under an AO3 grant, by the
Midi-Pyr\'en\'ees region, by the French National Program for Stellar
Physics (PNPS), by the Leverhulme Trust Network for Magnetized Plasma
Turbulence and by the National Science Foundation under Grant
No. PHY05-51164. Numerical calculations were carried out on the CALMIP
platform (CICT, University of Toulouse), whose assistance is
gratefully acknowledged.

\appendix
\section{DNS techniques - SNOOPY\label{appSB} and the shearing box}
\subsection{Shear periodicity}
The SNOOPY code is a pseudo-spectral code \citep{canuto}
for the direct numerical simulation of the three-dimensional,
incompressible dissipative MHD equations. It provides an
implementation of the shearing box model assuming simple spatial
periodicity in the $y$ and $z$ directions. Due to the presence of
shear, the $x$ direction cannot be taken periodic. To see this,
consider the left-hand side of the Navier-Stokes and induction
equations~(\ref{eq:NS})-(\ref{eq:induc}) in the presence of a linear
shear flow $\vec{U}_S=-Sx\vec{e}_y$:
\begin{equation}
\label{eq:shear1}
\dpart{\,\vec{\Psi}}{t}-Sx\,\dpart{\,\vec{\Psi}}{y} =\cdots ~,
\end{equation}
where $\vec{\Psi}$ is either $\vec{u}$ or $\vec{B}$. This
equation has an explicit linear $x$-dependence which prevents any
periodic decomposition of solutions. In more physical
terms, assuming that a non-axisymmetric ($y$-dependent)
physical structure is strictly periodic in $x$ at some time $t$, then
it is clear that the periodicity is destroyed as the
structure is sheared. In order to use a spectral
representation in the simulations, the solution is to go to a sheared
Lagrangian frame of reference \citep[see][]{umurhan04} defined
according to
\begin{equation}
x'=x\;,\quad y'=y+Sxt\;, \quad   z'=z\;,\quad t'=t.\label{transformshear}
\end{equation}
In that frame, equation~(\ref{eq:shear1}) transforms into
\begin{equation}
\label{eq:shear2}
\dpart{\,\vec{\Psi}}{t'}= \cdots~,
\end{equation}
which does not have any explicit dependence on $x'$. Periodic boundary
condition in $x'$ can be assumed, hence the term shear periodicity. 

\subsection{Spectral representation: shearing waves}
The previous discussion makes it clear that any field $\vec{\Psi}$ in the
discretized sheared Lagrangian frame $(x_p',y_q',z_r')$ can be
transformed to Fourier space according to 
\begin{equation}
\label{fourier}
\hat{\vec{\Psi}}_{l,m,n}=\!\sum_{-N_x/2}^{N_x/2}\sum_{-N_y/2}^{N_y/2}\sum_{-N_z/2}^{N_z/2}\!\!\vec{\Psi}(x_p',y_q',z_r')\exp\left[-i(k_{x,l}'x_p'+k_{y,m}'y_q'+k_{z,n}'z_r')\right]\;,
\end{equation}
where the $k_{x,l}'=2\pi l/L_x$ are the Lagrangian Fourier
wavenumbers in the periodic $x'$ direction and the $k_{y,m}'=2\pi
m/L_y$ and $k_{z,n}'=2\pi n/L_z$ are the Fourier
wavenumbers in the periodic $y'$ and $z'$ directions. Using
equation~(\ref{transformshear}) in equation~(\ref{fourier}) to look
at the evolution in the Eulerian frame of reference
$(x,y,z)$, the shearing of a non-axisymmetric
plane wave with given wavenumber $k_{y,m}'$ in the $y'=y$ direction
and Lagrangian wavenumber $k_{x,l}'$ in the $x'$ direction is
described  by a time-dependent Eulerian wavevector
\begin{equation}
\label{SWapp}
k_x(t)=k_{x,l}'+Sk_{y,m}'t\;,\quad k_y=k_{y,m}'\;,\quad k_z=k_{z,n}'.
\end{equation}
This equation translates the phy\-sical obser\-vation that the ty\-pical
scale of non-axisym\-metric structures along the $x$ direction becomes
increasingly smaller as the structure gets sheared. Waves whose
wavevectors follow this kind of evolution are called shearing
waves \citep{kelvin1887,orr1907}. As a result, the right hand side of
equation~(\ref{eq:shear2}) (the kind of equation integrated in
practice) is explicitly dependent on time, as any term
involving spatial derivatives with respect to the shearwise direction
(for instance the dissipative terms) now involves expressions of the
form~(\ref{SWapp}) in spectral space.

\subsection{Remapping procedure\label{appSB-remap}}
\begin{figure}
\centering
\includegraphics[width=\linewidth]{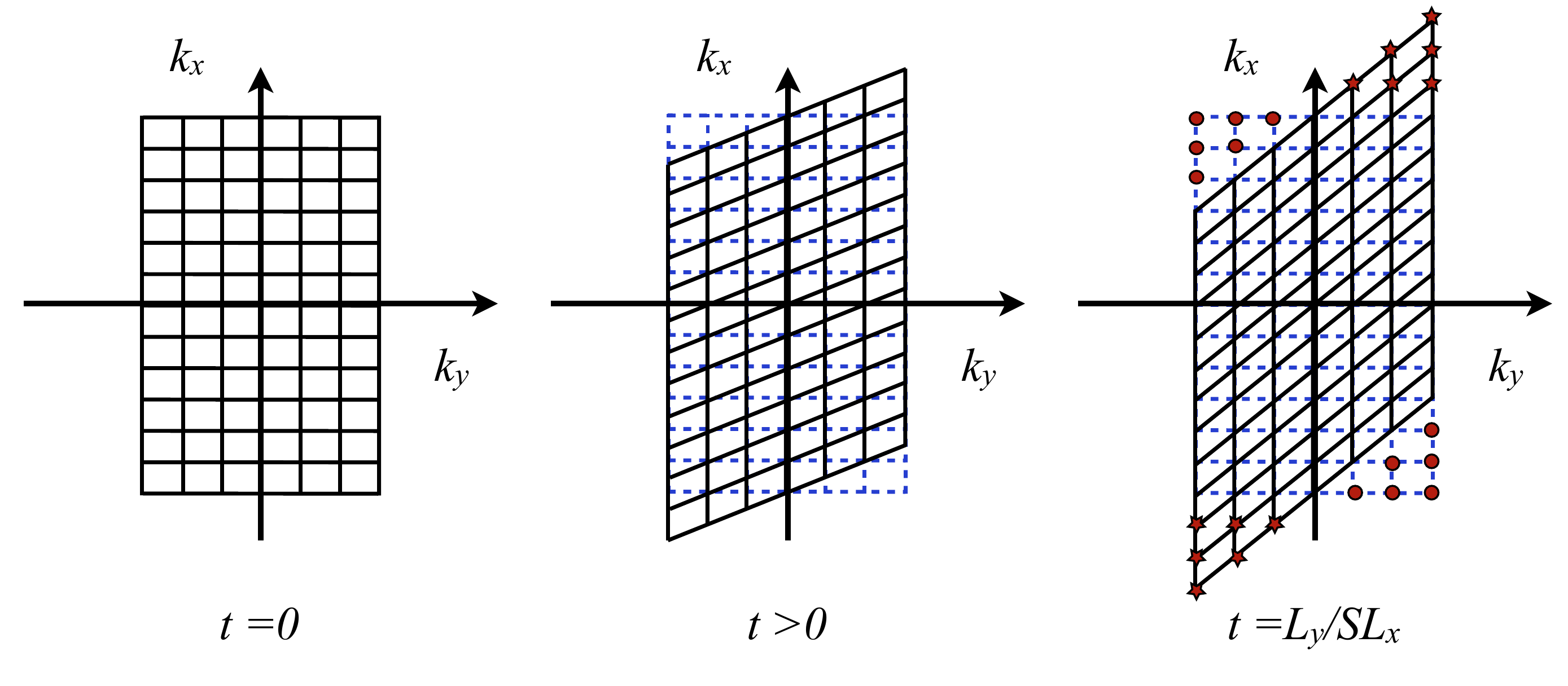}
\caption{Time-evolution of the wavenumber grid. The
  sheared grid is shown in solid/black line, the reference cartesian
  grid in dashed/blue line. The two grids coincide after $t=T_{SB}$,
  except for the points shown with symbols. Red stars indicate 
  strongly trailing wavevectors ($k_yk_x>0$) with $|k_x|>\pi N_x/L_x$. The
corresponding Fourier basis vectors are pruned and replaced with new
leading basis vectors ($k_yk_x<0$), shown as full/red circles on the
rightmost sketch.}
\label{sheargrid}
\end{figure}

Equation~(\ref{SWapp}) shows that the ``effective'' wavenumber grid
used to compute spatial derivatives in spectral space is
sheared. Now, if we simply consider the nonlinear evolution of a given
initial set of shearing waves defined on such a grid, some physical
nonlinearities and therefore some physics will be missed. Shearing
waves rapidly evolve into strongly trailing
($k_x(t)k_y>0$) sheared structures with ever smaller scale in $x$ and
are therefore dissipated extremely efficiently
\citep{knobloch85,korycansky92}. This kind of
dynamics is not self-sustaining. Even worse, all the Fourier
modes available to describe the dynamics end up strongly
trailing after some time, after which there are no modes left
to describe new leading waves ($k_x(t)k_y<0$) that could potentially
be \textit{physically excited}. This problem is solved by introducing a
remapping procedure. We show in figure~\ref{sheargrid} a cartesian
wavenumber grid $(k_{x,l}',k_{y,m}',k_{z,n}')$ (dashed/blue lines)
coinciding with the simulation grid at $t=0$. The
simulation grid (solid/black lines) is sheared according to
equation~(\ref{SWapp}) as time increases but coincides with the
reference grid every $T_{SB}=L_y/SL_x$. Indeed, after $T_{SB}$, the
effective wavenumber grid is
\begin{equation}
k_x(T_{SB})=k_{x,l}'+k_{y,m}'S\,T_{SB}= k_{x,l+m}'~,\quad
k_y=k_{y,m}',\quad k_z=k_{z,n}'~.
\end{equation} 
In other words, a shearing wave with wavenumber $k_{y,m}'$
in the $y$ direction and initial wavenumber $k_{x,l}'$ in the
$x$ direction can be described at $t=T_{SB}$ as a shearing wave
with the same wavenumber $k_{y,m}'$ in the $y$ direction but a
different wavenumber $k_{x,l+m}'$ in the $x$ direction. This makes it
possible to reproject the fields on the reference wavenumber grid at
this particular time. For instance, the Fourier coefficient
$\hat{\vec{\Psi}}_{l,m,n}$ during the first $T_{SB}$ of the simulation 
becomes $\hat{\vec{\Psi}}_{l+m,m,n}$ between $T_{SB}$ and $2\,T_{SB}$,
and so on.  Beside this simple reallocation of Fourier space, the
Fourier coefficients of trailing shearing waves whose absolute value
of the radial wavenumber $|k_x|$ exceeds $\pi N_x/L_x$ (the Nyquist
wavenumber) at $T_{SB}$  are set to zero at the corresponding
time. The corresponding basis vectors (red stars in the
figure) are pruned and replaced by new shearing wave basis vectors
with strongly leading wavenumber in the wavenumber grid (full/red circles).

Because we only update the basis of representation of the fields
during the remapping procedure and not the fields themselves, the 
procedure does not affect artificially the physical evolution as long
as the simulations are well-resolved. The pruned wavevectors represent
strongly trailing waves which have already been heavily damped by
dissipative processes when they are pruned. We always find that
pruning losses are negligible in SNOOPY compared to the physical
dissipation for well-resolved simulations
\citep{lesur05}. The remapping procedure does not by itself inject
energy into new leading waves either, but merely provides room for
them in the wavenumber grid. A pseudo-spectral method with dealiasing
is used to compute all nonlinear terms of
equations~(\ref{eq:NS})-(\ref{eq:div}) at each time
step. These terms describe the physical nonlinearities of the system
and are the only ones that can seed energy into leading waves when the
corresponding basis wavevectors are introduced, as
the energy content of these new waves is strictly enforced to be zero
when the remapping is performed \citep{Herault2011}.

\subsection{Symmetries\label{appsym}}
\cite{nagata86} identified several possible symmetries for
three-dimensional nonlinear hydrodynamic solutions in centrifugally
unstable (Rayleigh-unstable) Taylor-Couette flow in the thin-gap limit,
which corresponds to a cartesian wall-bounded plane Couette flow
rotating along its spanwise $z$ axis. Several similar symmetries can be
derived for MHD flows in the Keplerian shearing box. We consider a
physical domain of size $(L_x,L_y,L_z)$ and introduce a reference
frame comoving with the fluid at $L_x/2$:
\begin{equation}
  \label{eq:xprime}
  \tilde{y}= y-\f{L_x}{2}  St~.
\end{equation}
Most simulations and nonlinear solutions described in the paper are
restricted to a symmetric $\mathcal{A}_1$ subspace in which the fields
are described according to :\smallskip
\newcommand{\rme}{\mathrm{e}}
\newcommand{\rmo}{\mathrm{o}}
\begin{equation}
\vec{u}:\left\{
\begin{array}{lllr}
u_x & = & &
u_{x,\rme\rme}(t)\cos\left(k_{z\rme}z\right)\sin\left(k_{y\rme}\tilde{y}+k_x(t)x\right)\\
& & + &u_{x,\rmo\rmo}(t)\cos\left(k_{z\rmo}z\right)\cos\left(k_{y\rmo}\tilde{y}+k_x(t)x\right)\smallskip\\
u_y & = & &
u_{y,\rme\rme}(t)\cos\left(k_{z\rme}z\right)\sin\left(k_{y\rme}\tilde{y}+k_x(t)x\right)\\
& & + & u_{y,\rmo\rmo}(t)\cos\left(k_{z\rmo}z\right)\cos\left(k_{y\rmo}\tilde{y}+k_x(t)x\right)\smallskip\\
u_z  & = &  &
u_{z,\rme\rme}(t)\sin\left(k_{z\rme}z\right)\cos\left(k_{y\rme}\tilde{y}+k_x(t)x\right)\\
& & + & u_{z,\rmo\rmo}(t)\sin\left(k_{z\rmo}z\right)\sin\left(k_{y\rmo}\tilde{y}+k_x(t)x\right)
\end{array}\right\}~,
\end{equation}

\begin{equation}
\vec{B}:\left\{
\begin{array}{lllr}
B_x & = &  &
B_{x,\rme\rmo}(t)\cos\left(k_{z\rme}z\right)\sin\left(k_{y\rmo}\tilde{y}+k_x(t)x\right) \\
&&  + &
B_{x,\rmo\rme}(t)\cos\left(k_{z\rmo}z\right)\cos\left(k_{y\rme}\tilde{y}+k_x(t)x\right)\smallskip\\
B_y & = &  &
B_{y,\rme\rmo}(t)\cos\left(k_{z\rme}z\right)\sin\left(k_{y\rmo}\tilde{y}+k_x(t)x\right) \\
&&  + &   B_{y,\rmo\rme}(t)\cos\left(k_{z\rmo}z\right)\cos\left(k_{y\rme}\tilde{y}+k_x(t)x\right)\smallskip \\
B_z & = &  &
 B_{z,\rme\rmo}(t)\sin\left(k_{z\rme}z\right)\cos\left(k_{y\rmo}\tilde{y}+k_x(t)x\right)\\
&& + & B_{z,\rmo\rme}(t)\sin\left(k_{z\rmo}z\right)\sin\left(k_{y\rme}\tilde{y}+k_x(t)x\right)
\end{array}\right\}~,\bigskip
\end{equation}
where the $\rme$ and $\rmo$ subscripts indicate that the associated discrete 
wavenumbers are based on even and odd relative integers, respectively. $k_x(t)$ 
in each individual trigonometric expression is defined implicitly by 
equation~(\ref{SWapp}) using the $k_y$ wavenumber of the same expression. 
This symmetry is conserved by the MHD equations~(\ref{eq:NS})-(\ref{eq:div}). 

\section{Newton solver - PEANUTS \label{appinterface}}
The main difficulty in interfacing a Newton solver with a
high-dimensional PDE integrator comes from the construction of a
mapping between the full multidimensional representation of fields
most often used in DNS and a reduced one-dimensional state
vector representation required for the Newton solver. The problem is
particularly tedious for incompressible fluid dynamics
and/or when spectral dealiasing is used, as 
one must analyse carefully which modes of the full
representation must be kept in the state vector representation in that
case to avoid numerical trouble associated with singular algebra. 
This appendix describes how to switch from the full three-dimensional
representation of MHD fields in SNOOPY to a reduced
one-dimensional state vector representation
in the Newton solver PEANUTS. We also discuss the important issue
of enforcing spatial and temporal phase constraints in the shearing
box to compute simple invariant solutions.

\subsection{Mode counting}
We wish to count the number of independent real variables of the state
vector to obtain an invertible problem, starting from $N_x N_y
N_z$ real values in physical space for each field component. 
 The divergence-free constraint on both the velocity and
magnetic field reduces the number of independent field components to
4. We also restrict the dynamics to a subspace in which the
mean velocity is zero, which reduces the number of independent
 variables for each velocity component by 1. Finally, since the
 shearing box equations conserve magnetic flux, all box-averaged
 magnetic field components should vanish, reducing the number of
 independent variables for each magnetic component by 1.
The number of independent variables for the MHD equations in the
shearing box in three dimensions is therefore  $N_v = 4(N_x N_y
 N_z -1)$. If we also reject the Nyquist modes, we
find 
\begin{equation}\label{Nv2}
  N_v = 4(N_x N_y N_z -2)~.
\end{equation}

If dealiasing is used, $N_i$ ($i=x,y,z$) is always taken a multiple of
$3$ and $2$, so that  $n_i=2N_i/3$ are integers. The dealiasing
procedure also zeroes the Fourier components in the Nyquist planes
$n_i/2$, leaving $n_i-1$ variables in each direction and a total count
of $(n_x-1) (n_y-1) (n_z-1)$ per field component. The same number of
constraints as before must be subtracted, leading  to a number of
independent variables equal to
\begin{equation}\label{Nv3}
  N_v = 4\left[(n_x-1) (n_y-1) (n_z-1) -1\right]~.
\end{equation}

\subsection{Reduction of field variables}
The procedure to switch back and forth between the full three-dimensional
representation of complex Fourier-transformed fields $\vec{\tilde \Psi}$ (with
$\vec{\Psi}=\vec{u}$ or $\vec{B}$) and the reduced state
vector representation of size $N_v$ is summarized in
table~\ref{reducing} for the dealiased problem (the only one
of practical importance). In the forward mapping
(DNS$\rightarrow$Newton), only two field components are kept for each
wavenumber to satisfy the incompressibility constraint, while
$\vec{k}=\vec{0}$ modes are suppressed to ensure zero mean velocity
and magnetic flux conservation. In the reverse mapping
(Newton$\rightarrow$DNS), the missing component is recovered
thanks to the incompressibility condition expressed in Fourier
space. The wavenumber range indicated in the table results from the
real nature of the physical fields (the complex Fourier amplitudes for
the missing wavenumbers are recovered by complex conjugation).
\begin{table}
  \begin{center}
  \begin{tabular}{lccccc}
          & & & Hold variables &  wavenumber range & Missing component  \\[3pt]
             
            $ k_z\ne 0$ & & &  ${\tilde \Psi}_x $ and ${\tilde \Psi}_y $ &  $k_z=1$ to $(n_z-2)/2$    & ${\tilde \Psi}_{z}=-[k_x{\tilde \Psi}_x+k_x{\tilde \Psi}_y]/k_z$\\
            
           $k_z=0$   &  $k_y\ne 0$ & &   ${\tilde \Psi}_z $ and  ${\tilde \Psi}_x $ & $k_y=1$ to $(n_y-2)/2$ & ${\tilde \Psi}_{y}=-k_x{\tilde \Psi}_x/k_y$\\

         & $k_y=0$& $k_x\ne 0$ &${\tilde \Psi}_y$ and  ${\tilde \Psi}_z $ & $k_x=1$
to $(n_x-2)/2$  &  ${\tilde \Psi}_{x}= 0$\\ 
        &  &$k_x= 0$  & removed  \\                    
  \end{tabular}
  \caption{State vector mappings. For each Fourier mode, hold variables
    denote variables kept during the forward mapping between the DNS
    and Newton codes. The missing components are those recovered in the
    reverse mapping from the Newton code to the DNS codes.}
  \label{reducing}
  \end{center}
\end{table}

\subsection{Newton algorithm, phase constraints and shear periodicity\label{appinterface-constraints}}
A general method for imposing constraints in the Newton algorithm 
to compute relative periodic orbits of wall-bounded shear flows is
described by \cite{viswanath07}. The shearing box case
is a bit special and should be discussed in some detail.
We denote a full state vector in the Newton solver representation by
$\vec{X}(t)$. This vector of size $N_v$ contains all the independent
complex Fourier modes of all the independent components of the
physical fields $(\vec{u},\vec{B}$). Our dynamical
system is formally described by
\begin{equation}\label{sysdyn}
  \frac{d\vec{X}}{dt}=\vec{F}(\vec{X})~,
\end{equation}
where $\vec{F}$ is a nonlinear ``MRI dynamo operator''. This system can be
integrated by DNS, and a Newton solver is used to compute relative
periodic orbits of period $T$ obeying
\begin{equation}
  \label{eq:orbits}
  \vec{X}(\vec{X}_0,T)=\vec{\tau}(\vec{X}_0)~,
\end{equation}
where $\vec{X}_0$ is the state of the system at $t=0$  and $\vec{\tau}$ is a
translation operator in the $y$ and $z$ directions, which are the
two directions of invariance of the shearing box (recall that spatial
periodicity is assumed along these directions). In principle,  the
problem has $N_t=N_v+3$ unknowns: the $N_v$ independent
variables of the state vector, the period $T$ of the orbit and two
phase shifts in $y$ and $z$, which can be recast in the form
of two phase velocities $C_y$ and $C_z$ using the period $T$. 
These extra variables are needed because the original
problem of size $N_v$ is under-determined due to translational
space invariances (any phase-shifted solution is also a
solution) and time invariance (any state $\vec{X}_0$ on a periodic
trajectory is a solution of equation~(\ref{eq:orbits})). An equal
number of constraints must be imposed to lift the degeneracies and
to make the numerical problem non-singular.

The Newton method solves equation~(\ref{eq:orbits}) iteratively. Starting
from an initial guess $\vec{X}_{g,0}$ of a point on a cycle, a succession
of corrections $\delta\vec{X}_i$ to that guess are calculated from
successive Jacobian inversions, producing a sequence of states
$\vec{X}_{g,i}$ that should approach the desired solution if $\vec{X}_{g,0}$ is 
not too far from the solution. The extra constraints required to lift
the translational and time degeneracies described above are
\begin{eqnarray}
\left\langle \delta\vec{X}_i,\frac{\partial \vec{X}_{g,i}}{\partial y}\right\rangle & = & 0~,\label{C1}\\
\left\langle \delta\vec{X}_i,\frac{\partial \vec{X}_{g,i}}{\partial z}\right\rangle & = & 0~,\label{C2}\\
\left\langle \delta\vec{X}_i,\vec{F}(\vec{X}_{g,i})\right\rangle &= & 0~,\label{C3}
\end{eqnarray}
where the brackets denote a simple scalar product between state
vectors in complex space. These constraints basically impose that the
corrections performed by the solver be orthogonal to the directions of
phase space along which the flow is invariant. However, the 
shear periodicity described in appendix~\ref{appSB} introduces
an internal time $T_{SB}=L_y/SL_x$ in the system that breaks the
continuous translational time invariance of the equations and turns it
into a discrete one. To see this, consider the state $\vec{X}$
of the system at time $t=0$. Because of the shearing of the basis of
wavevectors on which the various fields are projected,
it does not make sense to compare the Fourier coefficients for that
state to those for different state at an arbitrary time $t$, because
in general the wavevector basis at time $t$ will be different from
that at $t=0$. The only times at which such a comparison is meaningful
are the multiples of $T_{SB}$ because for these times the sheared Fourier
grid coincides with that at $t=0$. In less technical
terms, the time-periodicity of shear periodic boundary conditions
in the $x$ direction only makes it possible for periodic orbits of
period that are multiples of $T_{SB}$ to be present in the shearing box.
Actually, because cycles require at least two reversals of the
large-scale axisymmetric components of the fields triggered by
successive nonlinear shearing wave interactions, they can only have
periods multiple of $T_o=2T_{SB}$ \citep{Herault2011}. In practice, a 
Newton search for a cycle whose estimated period is $NT_o$ can be
performed without imposing (\ref{C3}) by simply specifying the number
$N$ of successive $T_o$ integrations. 

Finally, enforcing the \Aone\ symmetry described in
appendix~\ref{appsym} in the time integrations breaks the continuous
translational invariances in space and notably eliminates relative
periodic orbits. In that situation, the solver can converge to periodic
orbits without imposing the phase constraints~(\ref{C1})-(\ref{C2}). We
checked that convergence to the same periodic orbits is
obtained whether the phase constraints are imposed or not.
We also checked in several important cases that convergence to a given
cycle with the Newton solver could be obtained independently of
whether symmetries are imposed or not.

\section{Numerical methods for homoclinic bifurcations\label{appglobalnumerics}}
\subsection{Imaging unstable manifolds in the MRI dynamo problem\label{traceWu}}
A standard technique to analyse the dynamics in the neighbourhood
of nonlinear invariant solutions is to look at the geometry of their
unstable manifolds $\Wu$ using  Poincar\'e return sections
\citep{simo89}. The method has for instance been used to analyse the
nonlinear dynamics of the
Kuramoto-Sivashinsky equation \citep{christiansen97,lan08}, and was 
recently introduced in the context of hydrodynamic plane Couette flow
and pipe flow \citep{gibson08,gibson09,halcrow09,vanveen11,willis13}.
Tracing an unstable manifold of a nonlinear periodic orbit 
is easier when the orbit has a single unstable
eigenvalue, as happens to be the case in transitional ranges of
$\Reym$ for the MRI dynamo cycles born out of
\SNone\ and \SNtwo, when the dynamics is limited to the \Aone\
symmetric subspace. Assume we have a periodic orbit whose full
phase space coordinates are denoted by $\vec{X}_e$. Close to
$\vec{X}_e$, a good approximation of a point in the unstable manifold
is given by the local tangent
\begin{equation}
\label{tangent}
 \vec{X}_0=\vec{X}_e+\varepsilon \vec{X}_u~,
\end{equation}
where $\vec{X}_u$ is the eigenvector associated with the
unstable eigenvalue and $\varepsilon$ is small enough 
for the tangent approximation of $\Wu$ at
$\vec{X}_e$ to be valid (up to some reasonable accuracy). A full local
segment of $\Wu$ can be approximated by sprinkling a large number of
points with different $\varepsilon$ on this tangent line. This set of
points is then used  as a set of initial conditions which can be
integrated along the flow for as much time as desired. As $\Wu$ is 
invariant under the flow by definition, all 
image points resulting from the integration of the initial conditions
in the initial segment also lie in $\Wu$ and can therefore be
used to construct arbitrary long pieces of $\Wu$ further away from
$\vec{X}_e$. Using successive crossings of dynamical trajectories with
appropriate Poincar\'e sections in flows, or discrete iterations in the
case of maps, and further projecting the dynamics on a low-dimensional
subspace of the full phase space finally makes it possible to
visualize $\Wu$ as a simple collection of points lying on a
one-dimensional curve.

The numerical formulation of the problem makes it possible to analyse
the nonlinear dynamics in terms of discrete-time map $\Phi_o$ which
consists in integrating the fluid equations during $T_o$. Looking at
the state of the system stroboscopically every $T_o$ is equivalent
to introducing a Poincar\'e section.  In that stroboscopic
representation, MRI dynamo cycles simply appear as nonlinear
equilibrium points (hence the notation $\vec{X}_e$). Although looking
at full snapshots of numerical simulations every $T_o$ is not very
informative, a low-dimensional projection of the dynamics can be
introduced to uncover interesting behaviour. We use a two-dimensional
projection on the large-scale axisymmetric magnetic field plane
($\Bazerox$,$\Bazeroy$), equivalent to a streamwise vortex / streak
projection in the hydrodynamic problem. Possible projection
degeneracies can be lifted by introducing a third variable
(e. g. another Fourier component of the spectral representation of the fields).

Let us summarize this technical description in order to make it
a practical, useful recipe. To uncover the geometry of 
the unstable manifold $\Wu$ of a MRI dynamo cycle $\vec{X}_e$ with a
single unstable eigenvalue, a preliminary requirement is to calculate
the unstable eigenvector $\vec{X}_u$ of $\vec{X}_e$. This data is obtained
from the SLEPc eigenvalue problem solver during the local stability
analysis presented in \S\ref{cycles}. Using this vector and applying 
equation~(\ref{tangent}) for a large set of $\varepsilon$ in a given range
of small enough values, we generate a large set of initial conditions,
or``initial segment'' tangent to $\Wu$ at $\vec{X}_e$. Each initial
condition in that set is then integrated in time with the SNOOPY code
for a succession of periods $T_o$ (typically 8 in the paper). After
each $T_o$ time integration (each corresponding to one $\Phi_o$ iteration),
the results of all the simulations are projected on the
two-dimensional plane ($\Bazerox$,$\Bazeroy$) to obtain successive
``image segments'' of the initial one. As $\Wu$ is invariant under
time integrations, these successive curved segments represent
increasingly larger portions of the projection of $\Wu$ in the plane.

\subsection{Computation and continuation of homoclinic orbits\label{apphomo}}
Homoclinic orbits of fixed points of discrete-time maps are
characterized by a unique symbolic sequence (``core'') of finite
intrinsic length (or transition time $N_t$). This sequence is preceded
by an infinite repetition of the symbol associated with the fixed
point and is followed by another infinite repetition of the same
symbol. These head and tail sequences represent
exponential escape from the fixed point and
exponential convergence back to it towards infinite times, while
the core sequence of finite length represents the nonlinear excursion
in the tangle, far away from the fixed point. A well-known map for
which the zoology of homoclinic bifurcations and associated symbolic
dynamics has been thoroughly studied is the H\'enon map
\citep{henon76,grassberger89,sterling99}.

\subsubsection{Capturing homoclinic orbits}
We only describe the homoclinic case, as the generalization to heteroclinic
orbits is straightforward. To investigate homoclinic orbits of a MRI
dynamo cycle (which we recall is a fixed point of the $\Phi_o$ map) in
the associated tangle, we first construct the projection in the
$(\Bazerox,\Bazeroy)$ plane of the unstable manifold $\Wu$ of the cycle
using the technique described in \S\ref{traceWu}. 
We typically construct 5 to 20 image segments in that case, depending
on the desired approximation of the orbit. We then consider the fold
of $\Wu$ approaching closest to the fixed point and identify
the image point in the fold closest to it.
We then track the pre-images of this point, back to its
parent initial condition in the initial tangent
segment. If the value of $\varepsilon$ used to construct this
particular initial condition according to equation~(\ref{tangent}) is
sufficiently small so that the point can be considered to be part of
the ``head'' sequence of a homoclinic orbit, the sequence of points
resulting from successive $T_o$ integrations of the initial condition
already provides a good approximation of that orbit. However, as $\Wu$ is
strongly stretched along itself close to the fixed point, a little
change in $\varepsilon$ can result in an important divergence of
trajectories after just a few iterations. To obtain a
reasonably accurate approximation of a specific homoclinic
orbit, it is in general desirable to refine the initial condition.
We use a bisection method on $\varepsilon$ to minimize the
distance between the last point of integration and the
fixed point. Approximate orbits with an arbitrarily long head can be
computed by using arbitrarily small initial segments. Approximate
orbits with an arbitrarily long tail are obtained by increasing the
target number of $T_o$ integrations of the algorithm.

\subsubsection{Continuing homoclinic orbits in parameter space}
Assume that a homoclinic
orbit $\mathcal{H}$ of an equilibrium point $\vec{X}_e$ of the map
$\Phi_o$ has been approximately computed at a given
$\Reym$ as a set of $N$ successive states $\vec{H}_{i}$ 
($i=0,\cdots,N-1$) using the technique described above. $\vec{H}_{0}$
is the initial state closest to $\vec{X}_e$, $\vec{H}_{1}$ its
image under the action of $\Phi_o$, and so on. We also know from the
calculation of $\mathcal{H}$ the parameter $\varepsilon_0$ from which
$\vec{H}_{0}$ was computed.

To follow this orbit at a slightly different $\Reym+\Delta
\Reym$, we first recompute $\vec{X}_e$ and its unstable eigenvector
$\vec{X}_u$ at the target $\Reym$ using a standard continuation
algorithm and stability solver, and construct a new perturbed state
$\vec{H}_{0}^0$ using the same $\varepsilon_0$ as before. 
From this state, we perform $N$ iterations of $\Phi_o$ of the map at
the target $\Reym$ and obtain a new set of states
$\vec{H}_i^0$. This set is not in general a good
approximation of the homoclinic orbit at the new  $\Reym$. It
usually increasingly diverges from $\mathcal{H}$ as the number of
iterations of $\Phi_o$ increases. But we can use the
difference between the two sets of states $\vec{H}_i^0$ and $\vec{H}_i$ to
guess a new $\varepsilon$ and better approximate the first point
computed along the homoclinic orbit at the target $\Reym$.
We denote by $F_N$ the function of $\varepsilon$ which, for a given
equilibrium $\vec{X}_e$ and corresponding unstable eigenvector
$\vec{X}_u$ of $\Phi_o$, gives the projection of the image of
$\vec{X}_e+\varepsilon\vec{X}_u$ under the action of $\Phi_o^N$ on 
the ($\Bazerox,\Bazeroy$) plane:
\begin{eqnarray}
F_N : \mathbb{R} & \to & \mathbb{R}^2 \\
\varepsilon & \mapsto &\mbox{P}\left[\Phi_o^N(\vec{X}_e+\varepsilon \vec{X}_u)\right]~.
\end{eqnarray}
The two-dimensional projection P is not mandatory but makes
the algorithm much faster than if we were to use the full
state vector. We use a simple Newton algorithm to perform the
desired $\varepsilon$ correction.  Two possibilities must be
considered. If the difference between $\vec{H}_i^0$ and $\vec{H}_i$
only becomes larger than a given threshold value in the tail sequence of the
homoclinic orbit (i.e. after the ``bursting'' transition sequence of
length $N_t$ has been passed), say at iteration $N_d$, then a first Newton
step is set up to converge to $\vec{H}_{N_d}=\vec{X}_e$ at the
target $\Reym$. In that case, the first Newton step reads
\begin{equation}
\varepsilon_{1} = \varepsilon_0 - \tilde{\mathsfbi{J}_0}\,\mbox{P}\left[\vec{H}_{N_d}^0 - \vec{X}_e\right]\,,
\end{equation}
where $\tilde{\mathsfbi{J}_0}=((\mathsfbi{J}_0^T\mathsfbi{J}_0)^{-1})\mathsfbi{J}_0^T$ is the
pseudo-inverse of
$\displaystyle{\mathsfbi{J}_0=\left.\frac{d{F_{N_d}}}{d\varepsilon}\right|_{\varepsilon=\varepsilon_0}}$\!\!\!\!.\smallskip \\
$\varepsilon_1$ is used in a new prediction step to recompute an
updated initial state $\vec{H}^1_0$. This state is integrated as before,
producing a full new sequence $\vec{H}^1_i$. In
general, one Newton step will not be enough to converge to
the homoclinic solution at the target $\Reym$, but the iteration of
$\Phi_o$ at which the new sequence will diverge from the homoclinic
orbit at the original $\Reym$ will be larger (and therefore still in the
tail sequence). To take advantage of this, we 
increment $N_d$ by one at each Newton iteration so that points
ever further in the tail of the approximate orbit close in on
$\vec{X}_e$. At step $n$, the correction step will read:
\begin{equation}
\varepsilon_{n} = \varepsilon_{n-1} - \tilde{\mathsfbi{J}}_{n-1}\,\mbox{P}\left[\vec{H}_{N_d+n-1}^{n-1} - \vec{X}_e\right]\,,\medskip
\end{equation}
with
$\tilde{\mathsfbi{J}}_{n-1}=((\mathsfbi{J}_{n-1}^T\mathsfbi{J}_{n-1})^{-1})\mathsfbi{J}_{n-1}^T$
the pseudo-inverse of
$\displaystyle{\mathsfbi{J}_{n-1}=\left.\frac{d F_{N_d+n-1}}{d
      \varepsilon}\right|_{\varepsilon=\varepsilon_{n-1}}}$\!\!\!\!.\smallskip\\
In practice, the $\mathsfbi{J}_i$ Jacobians (which for the projection
used are just two-dimensional vectors) are computed approximately
using a first-order finite difference formula.

The algorithm must be slightly modified in the case of divergence in the
head sequence (i.e. before the transition sequence of length
$N_t$). In that case, we would like that  all the points of the
approximate orbit at the target $\Reym$ whose index is between $N_d$
and the end of the transition sequence stay close to the corresponding
points of the orbit $\mathcal{H}$ at the previous $\Reym$. The first steps
of the correction algorithm are modified according to:
\begin{eqnarray}
\varepsilon_{1} & =  & \varepsilon_0 - \tilde{\mathsfbi{J}}_{0}\,\mbox{P}\left[\vec{H}^0_{N_d} - \vec{H}_{N_d}\right]\,,\\
\varepsilon_{2} & = & \varepsilon_1 - \tilde{\mathsfbi{J}}_{1}\,\mbox{P}\left[\vec{H}^1_{N_d+1} - \vec{H}_{N_d+1}\right]\,,
\end{eqnarray}
and so on until the tail sequence of the orbit is reached. Once this
is achieved, $\vec{X}_e$ becomes the target of the Newton
algorithm and the same procedure as before is used. 

This algorithm is somewhat empirical (there is no mathematical
guarantee of convergence) but nevertheless give full
satisfaction.  Obviously, for such a strategy to work,
$\Delta \Reym$ must be small enough so that $\mathcal{H}$ at the
target $\Reym$ differs only weakly from $\mathcal{H}$ at the original
$\Reym$. A detailed examination of the results is necessary 
to check that the orbit found at the target $\Reym$ is a
good numerical approximation of a homoclinic orbit, and that it is
indeed the continuation of that obtained at the preceding $\Reym$.

\bibliographystyle{jfm}
\bibliography{refs}

\begin{thebibliography}{164}
\expandafter\ifx\csname natexlab\endcsname\relax\def\natexlab#1{#1}\fi

\bibitem[{Armitage}(2010)]{armitage10}
{\sc {Armitage}, P.~J.} 2010 {\em {Astrophysics of Planet Formation}\/}.
  Cambridge University Press.

\bibitem[{Avila} {\em et~al.\/}(2011){Avila}, {Moxey}, {de Lozar}, {Avila},
  {Barkley} \& {Hof}]{avila11}
{\sc {Avila}, K., {Moxey}, D., {de Lozar}, A., {Avila}, M., {Barkley}, D. \&
  {Hof}, B.} 2011 {The onset of turbulence in pipe flow}. {\em Science\/} {\bf
  333}, 192.

\bibitem[Balay {\em et~al.\/}(2011)Balay, Brown, Buschelman, Gropp, Kaushik,
  Knepley, McInnes, Smith \& Zhang]{petsc}
{\sc Balay, Satish, Brown, Jed, Buschelman, Kris, Gropp, William~D., Kaushik,
  Dinesh, Knepley, Matthew~G., McInnes, Lois~Curfman, Smith, Barry~F. \& Zhang,
  Hong} 2011 {PETSc} {W}eb page. \texttt{http://www.mcs.anl.gov/petsc}.

\bibitem[{Balbus}(2003)]{balbus03}
{\sc {Balbus}, S.~A.} 2003 {Enhanced angular momentum transport in accretion
  disks}. {\em Annu. Rev. Astron. Astrophys.\/} {\bf 41}, 555.

\bibitem[{Balbus} \& {Hawley}(1991)]{balbus91}
{\sc {Balbus}, S.~A. \& {Hawley}, J.~F.} 1991 {A powerful local shear
  instability in weakly magnetized disks. I. Linear analysis}. {\em Astrophys.
  J.\/} {\bf 376}, 214.

\bibitem[{Balbus} \& {Hawley}(1992)]{balbus92}
{\sc {Balbus}, S.~A. \& {Hawley}, J.~F.} 1992 {A powerful local shear
  instability in weakly magnetized disks. IV. Nonaxisymmetric perturbations}.
  {\em Astrophys. J.\/} {\bf 400}, 610.

\bibitem[{Balbus} \& {Hawley}(1998)]{balbus98}
{\sc {Balbus}, S.~A. \& {Hawley}, J.~F.} 1998 {Instability, turbulence, and
  enhanced transport in accretion disks}. {\em Rev. Mod. Phys.\/} {\bf 70}, 1.

\bibitem[{Balbus} \& {Henri}(2008)]{balbus08}
{\sc {Balbus}, S.~A. \& {Henri}, P.} 2008 {On the magnetic Prandtl number
  behavior of accretion disks}. {\em Astrophys. J.\/} {\bf 674}, 408.

\bibitem[{Barkley}(2011)]{barkley2011}
{\sc {Barkley}, D.} 2011 {Simplifying the complexity of pipe flow}. {\em Phys.
  Rev. E\/} {\bf 84}, 016309.

\bibitem[{Birkhoff}(1935)]{birkhoff35}
{\sc {Birkhoff}, G.~D.} 1935 Nouvelles recherches sur les syst\`emes
  dynamiques. {\em Mem. Pont. Acad. Sci. Nov. Lyn.\/} {\bf 1}, 85.

\bibitem[{Bodo} {\em et~al.\/}(2011){Bodo}, {Cattaneo}, {Ferrari}, {Mignone} \&
  {Rossi}]{bodo11}
{\sc {Bodo}, G., {Cattaneo}, F., {Ferrari}, A., {Mignone}, A. \& {Rossi}, P.}
  2011 {Symmetries, scaling laws, and convergence in shearing-box simulations
  of magneto-rotational instability driven turbulence}. {\em Astrophys. J.\/}
  {\bf 739}, 82.

\bibitem[{Brandenburg} \& {Dintrans}(2006)]{branden06}
{\sc {Brandenburg}, A. \& {Dintrans}, B.} 2006 {Nonaxisymmetric stability in
  the shearing sheet approximation}. {\em Astron. Astrophys.\/} {\bf 450}, 437.

\bibitem[{Brandenburg} {\em et~al.\/}(1995){Brandenburg}, {Nordlund}, {Stein}
  \& {Torkelsson}]{branden95}
{\sc {Brandenburg}, A., {Nordlund}, A., {Stein}, R.~F. \& {Torkelsson}, U.}
  1995 {Dynamo-generated turbulence and large-scale magnetic fields in a
  Keplerian shear flow}. {\em Astrophys. J.\/} {\bf 446}, 741.

\bibitem[{Canuto} {\em et~al.\/}(1988){Canuto}, {Hussaini}, {Quarteroni} \&
  {Zang}]{canuto}
{\sc {Canuto}, C., {Hussaini}, M.~Y., {Quarteroni}, A. \& {Zang}, T.~A.} 1988
  {\em Spectral Methods in Fluid Dynamics\/}. Springer.

\bibitem[Casciola {\em et~al.\/}(2003)Casciola, Gualtieri, Benzi \&
  Piva]{casciola03}
{\sc Casciola, C.~M., Gualtieri, P., Benzi, R. \& Piva, R.} 2003 Scale-by-scale
  budget and similarity laws for shear turbulence. {\em J. Fluid Mech.\/} {\bf
  476}, 105.

\bibitem[Chandler \& {Kerswell}(2013)]{chandler13}
{\sc Chandler, G.~J. \& {Kerswell}, R.~R.} 2013 {Invariant recurrent solutions
  embedded in a turbulent two-dimensional {K}olmogorov flow}. {\em J. Fluid
  Mech.\/} {\bf 722}, 554.

\bibitem[{C}handrasekhar(1960)]{chandra60}
{\sc {C}handrasekhar, {S}.} 1960 {\em {Proc. Natl. Acad. Sci.}\/} {\bf 46},
  253.

\bibitem[{Childress} \& {Gilbert}(1995)]{childress95}
{\sc {Childress}, S. \& {Gilbert}, A.~D.} 1995 {\em {Stretch, Twist, Fold. The
  Fast Dynamo}\/}. Springer-Verlag.

\bibitem[{Christiansen} {\em et~al.\/}(1997){Christiansen}, {Cvitanovi\'c} \&
  {Putkaradze}]{christiansen97}
{\sc {Christiansen}, F., {Cvitanovi\'c}, P. \& {Putkaradze}, V.} 1997
  {Spatiotemporal chaos in terms of unstable recurrent patterns}. {\em
  Nonlinearity\/} {\bf 10}, 55.

\bibitem[{Cline} {\em et~al.\/}(2003){Cline}, {Brummell} \&
  {Cattaneo}]{cline03}
{\sc {Cline}, K.~S., {Brummell}, N.~H. \& {Cattaneo}, F.} 2003 {Dynamo action
  driven by shear and magnetic buoyancy}. {\em Astrophys. J.\/} {\bf 599},
  1449.

\bibitem[Collins {\em et~al.\/}(2012)Collins, Katz, Wallace, Jara-Almonte,
  Reese, Zweibel \& Forest]{collins12}
{\sc Collins, C., Katz, N., Wallace, J., Jara-Almonte, J., Reese, I., Zweibel,
  E. \& Forest, C.~B.} 2012 Stirring unmagnetized plasma. {\em Phys. Rev.
  Lett.\/} {\bf 108}, 115001.

\bibitem[{Cvitanovi{\'c}}(1992)]{cvit92}
{\sc {Cvitanovi{\'c}}, P.} 1992 {Periodic orbit theory in classical and quantum
  mechanics}. {\em Chaos\/} {\bf 2}, 1.

\bibitem[{Cvitanovi{\'c}} \& {Gibson}(2010)]{cvit10}
{\sc {Cvitanovi{\'c}}, P. \& {Gibson}, J.~F.} 2010 {Geometry of the turbulence
  in wall-bounded shear flows: periodic orbits}. {\em Physica Scripta\/} {\bf
  142}, 014007.

\bibitem[{Darbyshire} \& {Mullin}(1995)]{darbyshire95}
{\sc {Darbyshire}, A.~G. \& {Mullin}, T.} 1995 {Transition to turbulence in
  constant-mass-flux pipe flow}. {\em J. Fluid Mech.\/} {\bf 289}, 83.

\bibitem[{Dauchot} \& {Daviaud}(1995{\natexlab{{\em a\/}}})]{dauchot95a}
{\sc {Dauchot}, O. \& {Daviaud}, F.} 1995{\natexlab{{\em a\/}}} {Finite
  amplitude perturbation and spots growth mechanism in plane Couette flow}.
  {\em Phys. Fluids\/} {\bf 7}, 335.

\bibitem[{Dauchot} \& {Daviaud}(1995{\natexlab{{\em b\/}}})]{dauchot95b}
{\sc {Dauchot}, O. \& {Daviaud}, F.} 1995{\natexlab{{\em b\/}}} {Streamwise
  vortices in plane Couette flow}. {\em Phys. Fluids\/} {\bf 7}, 901.

\bibitem[{Davies} \& {Hughes}(2011)]{davies10}
{\sc {Davies}, C.~R. \& {Hughes}, D.~W.} 2011 {The mean electromotive force
  resulting from magnetic buoyancy instability}. {\em Astrophys. J.\/} {\bf
  727}, 112.

\bibitem[{Davis} {\em et~al.\/}(2010){Davis}, {Stone} \& {Pessah}]{davis10}
{\sc {Davis}, S.~W., {Stone}, J.~M. \& {Pessah}, M.~E.} 2010 {Sustained
  magnetorotational turbulence in local simulations of stratified disks with
  zero net magnetic flux}. {\em Astrophys. J.\/} {\bf 713}, 52.

\bibitem[{de Lozar} {\em et~al.\/}(2012){de Lozar}, {Mellibovsky}, {Avila} \&
  {Hof}]{lozar12}
{\sc {de Lozar}, A., {Mellibovsky}, F., {Avila}, M. \& {Hof}, B.} 2012 {Edge
  state in pipe flow experiments}. {\em Phys. Rev. Lett.\/} {\bf 108}, 214502.

\bibitem[{Ding} {\em et~al.\/}(1990){Ding}, {Grebogi}, {Ott} \&
  {Yorke}]{ding90}
{\sc {Ding}, M., {Grebogi}, C., {Ott}, E. \& {Yorke}, J.~A.} 1990 {Transition
  to chaotic scattering}. {\em Phys. Rev. A\/} {\bf 42}, 7025.

\bibitem[{Duguet} {\em et~al.\/}(2008{\natexlab{{\em a\/}}}){Duguet}, {Pringle}
  \& {Kerswell}]{duguet08a}
{\sc {Duguet}, Y., {Pringle}, C.~C.~T. \& {Kerswell}, R.~R.}
  2008{\natexlab{{\em a\/}}} {Relative periodic orbits in transitional pipe
  flow}. {\em Phys. Fluids\/} {\bf 20}, 114102.

\bibitem[{Duguet} {\em et~al.\/}(2008{\natexlab{{\em b\/}}}){Duguet}, {Willis}
  \& {Kerswell}]{duguet08b}
{\sc {Duguet}, Y., {Willis}, A.~P. \& {Kerswell}, R.~R.} 2008{\natexlab{{\em
  b\/}}} {Transition in pipe flow: the saddle structure on the boundary of
  turbulence}. {\em J. Fluid Mech.\/} {\bf 613}, 255.

\bibitem[{Eckhardt}(2009)]{eckhardt09}
{\sc {Eckhardt}, B.} 2009 {Introduction. Turbulence transition in pipe flow:
  125th anniversary of the publication of Reynolds' paper}. {\em Proc. R. Soc.
  Lond. Series A\/} {\bf 367}, 449.

\bibitem[{Eckhardt} {\em et~al.\/}(2007){Eckhardt}, {Schneider}, {Hof} \&
  {Westerweel}]{eckhardt07}
{\sc {Eckhardt}, B., {Schneider}, T.~M., {Hof}, B. \& {Westerweel}, J.} 2007
  {Turbulence transition in pipe flow}. {\em Annu. Rev. Fluid Mech.\/} {\bf
  39}, 447.

\bibitem[{Faisst} \& {Eckhardt}(2003)]{faisst03}
{\sc {Faisst}, H. \& {Eckhardt}, B.} 2003 {Traveling waves in pipe flow}. {\em
  Phys. Rev. Lett.\/} {\bf 91}, 224502.

\bibitem[{Faisst} \& {Eckhardt}(2004)]{faisst04}
{\sc {Faisst}, H. \& {Eckhardt}, B.} 2004 {Sensitive dependence on initial
  conditions in transition to turbulence in pipe flow}. {\em J. Fluid Mech.\/}
  {\bf 504}, 343.

\bibitem[{Feigenbaum}(1978)]{feigenbaum78}
{\sc {Feigenbaum}, M.~J.} 1978 {Quantitative universality for a class of
  nonlinear transformations}. {\em J. Stat. Phys.\/} {\bf 19}, 25.

\bibitem[{Frank} {\em et~al.\/}(2002){Frank}, {King} \& {Raine}]{frank02}
{\sc {Frank}, J., {King}, A. \& {Raine}, D.~J.} 2002 {\em {Accretion Power in
  Astrophysics: Third Edition}\/}. Cambridge University Press.

\bibitem[{Fromang} \& {Papaloizou}(2007)]{fromang07a}
{\sc {Fromang}, S. \& {Papaloizou}, J. C.~B.} 2007 {MHD simulations of the
  magnetorotational instability in a shearing box with zero net flux. I. The
  issue of convergence}. {\em Astron. Astrophys.\/} {\bf 476}, 1113.

\bibitem[{Fromang} {\em et~al.\/}(2007){Fromang}, {Papaloizou}, {Lesur} \&
  {Heinemann}]{fromang07b}
{\sc {Fromang}, S., {Papaloizou}, J. C.~B., {Lesur}, G. \& {Heinemann}, T.}
  2007 {MHD simulations of the magnetorotational instability in a shearing box
  with zero net flux. II. The effect of transport coefficients}. {\em Astron.
  Astrophys.\/} {\bf 476}, 1123.

\bibitem[{Gallet} {\em et~al.\/}(2012){Gallet}, {Herault}, {Laroche},
  {P\'etr\'elis} \& {Fauve}]{gallet12}
{\sc {Gallet}, B., {Herault}, J., {Laroche}, C., {P\'etr\'elis}, F \& {Fauve},
  S.} 2012 Reversals of large-scale fields generated over a turbulent
  background. {\em Geophys. Astrophys. Fluid Dyn.\/} {\bf 106}, 468.

\bibitem[{Gavrilov} \& {Shil'nikov}(1972)]{gavrilov72}
{\sc {Gavrilov}, N. \& {Shil'nikov}, L.} 1972 On the three dimensional
  dynamical systems close to a system with a structurally unstable homoclinic
  curve, i. {\em Math. USSR Sbornik\/} {\bf 17}, 467.

\bibitem[{Gibson} {\em et~al.\/}(2008){Gibson}, {Halcrow} \&
  {Cvitanovi{\'c}}]{gibson08}
{\sc {Gibson}, J.~F., {Halcrow}, J. \& {Cvitanovi{\'c}}, P.} 2008 {Visualizing
  the geometry of state space in plane Couette flow}. {\em J. Fluid Mech.\/}
  {\bf 611}, 107.

\bibitem[{Gibson} {\em et~al.\/}(2009){Gibson}, {Halcrow} \&
  {Cvitanovi{\'c}}]{gibson09}
{\sc {Gibson}, J.~F., {Halcrow}, J. \& {Cvitanovi{\'c}}, P.} 2009 {Equilibrium
  and travelling-wave solutions of plane Couette flow}. {\em J. Fluid Mech.\/}
  {\bf 638}, 243.

\bibitem[{Gissinger}(2012)]{gissinger12}
{\sc {Gissinger}, C.} 2012 {A new deterministic model for chaotic reversals}.
  {\em Eur. Phys. J. B\/} {\bf 85}, 137.

\bibitem[{Goldreich} \& {Lynden-Bell}(1965)]{goldreich65}
{\sc {Goldreich}, P. \& {Lynden-Bell}, D.} 1965 {II. Spiral arms as sheared
  gravitational instabilities}. {\em Mon. Not. R. Astron. Soc.\/} {\bf 130},
  125.

\bibitem[{Goswami} \& {Basu}(2002)]{goswami02}
{\sc {Goswami}, B.~K. \& {Basu}, S.} 2002 {Self-similar organization of
  Gavrilov-Silnikov-Newhouse sinks}. {\em Phys. Rev. E\/} {\bf 65}, 036210.

\bibitem[{Grassberger} {\em et~al.\/}(1989){Grassberger}, {Kantz} \&
  {Moenig}]{grassberger89}
{\sc {Grassberger}, P., {Kantz}, H. \& {Moenig}, U.} 1989 {On the symbolic
  dynamics of the {H}\'enon map}. {\em J. Phys. A\/} {\bf 22}, 5217.

\bibitem[{Grebogi} {\em et~al.\/}(1982){Grebogi}, {Ott} \& {Yorke}]{grebogi82}
{\sc {Grebogi}, C., {Ott}, E. \& {Yorke}, J.~A.} 1982 {Chaotic attractors in
  crisis}. {\em Phys. Rev. Lett.\/} {\bf 48}, 1507.

\bibitem[{Grebogi} {\em et~al.\/}(1983){Grebogi}, {Ott} \& {Yorke}]{grebogi83}
{\sc {Grebogi}, C., {Ott}, E. \& {Yorke}, J.~A.} 1983 {Crises, sudden changes
  in chaotic attractors, and transient chaos}. {\em Physica D\/} {\bf 7}, 181.

\bibitem[{Grebogi} {\em et~al.\/}(1987){Grebogi}, {Ott} \& {Yorke}]{grebogi87}
{\sc {Grebogi}, C., {Ott}, E. \& {Yorke}, J.~A.} 1987 {Basin boundary
  metamorphoses: Changes in accessible boundary orbits}. {\em Physica D\/} {\bf
  24}, 243.

\bibitem[{Gressel}(2010)]{gressel10}
{\sc {Gressel}, O.} 2010 {A mean-field approach to the propagation of field
  patterns in stratified magnetorotational turbulence}. {\em Mon. Not. R.
  Astron. Soc.\/} {\bf 405}, 41.

\bibitem[{Grossmann}(2000)]{grossmann2000}
{\sc {Grossmann}, S.} 2000 {The onset of shear flow turbulence}. {\em Rev. Mod.
  Phys.\/} {\bf 72}, 603.

\bibitem[{Gualtieri} {\em et~al.\/}(2002){Gualtieri}, {Casciola}, {Benzi},
  {Amati} \& {Piva}]{gualtieri02}
{\sc {Gualtieri}, P., {Casciola}, C.~M., {Benzi}, R., {Amati}, G. \& {Piva},
  R.} 2002 {Scaling laws and intermittency in homogeneous shear flow}. {\em
  Phys. Fluids\/} {\bf 14}, 583.

\bibitem[{Halcrow} {\em et~al.\/}(2009){Halcrow}, {Gibson}, {Cvitanovi{\'c}} \&
  {Viswanath}]{halcrow09}
{\sc {Halcrow}, J., {Gibson}, J.~F., {Cvitanovi{\'c}}, P. \& {Viswanath}, D.}
  2009 {Heteroclinic connections in plane Couette flow}. {\em J. Fluid Mech.\/}
  {\bf 621}, 365.

\bibitem[{Hamilton} {\em et~al.\/}(1995){Hamilton}, {Kim} \&
  {Waleffe}]{hamilton95}
{\sc {Hamilton}, J.~M., {Kim}, J. \& {Waleffe}, F.} 1995 {Regeneration
  mechanisms of near-wall turbulence structures}. {\em J. Fluid Mech.\/} {\bf
  287}, 317.

\bibitem[{Hartmann}(2009)]{hartmann09}
{\sc {Hartmann}, L.} 2009 {\em {Accretion Processes in Star Formation: Second
  Edition}\/}. Cambridge University Press.

\bibitem[{Hawley} \& {Balbus}(1991)]{hawley91}
{\sc {Hawley}, J.~F. \& {Balbus}, S.~A.} 1991 {A powerful local shear
  instability in weakly magnetized disks. II. Nonlinear evolution}. {\em
  Astrophys. J.\/} {\bf 376}, 223.

\bibitem[{Hawley} {\em et~al.\/}(1995){Hawley}, {Gammie} \& {Balbus}]{hawley95}
{\sc {Hawley}, J.~F., {Gammie}, C.~F. \& {Balbus}, S.~A.} 1995 {Local
  three-dimensional magnetohydrodynamic simulations of accretion disks}. {\em
  Astrophys. J.\/} {\bf 440}, 742.

\bibitem[{Hawley} {\em et~al.\/}(1996){Hawley}, {Gammie} \& {Balbus}]{hawley96}
{\sc {Hawley}, J.~F., {Gammie}, C.~F. \& {Balbus}, S.~A.} 1996 {Local
  three-dimensional simulations of an accretion disk hydromagnetic dynamo}.
  {\em Astrophys. J.\/} {\bf 464}, 690.

\bibitem[{H{\'e}non}(1976)]{henon76}
{\sc {H{\'e}non}, M.} 1976 {A two-dimensional mapping with a strange
  attractor}. {\em Commun. Math. Phys.\/} {\bf 50}, 69.

\bibitem[{Herault} {\em et~al.\/}(2011){Herault}, {Rincon}, {Cossu}, {Lesur},
  {Ogilvie} \& {Longaretti}]{Herault2011}
{\sc {Herault}, J., {Rincon}, F., {Cossu}, C., {Lesur}, G., {Ogilvie}, G.~I. \&
  {Longaretti}, P.-Y.} 2011 {Periodic magnetorotational dynamo action as a
  prototype of nonlinear magnetic-field generation in shear flows}. {\em Phys.
  Rev. E\/} {\bf 84}, 036321.

\bibitem[Hernandez {\em et~al.\/}(2005)Hernandez, Roman \& Vidal]{slepc}
{\sc Hernandez, Vicente, Roman, Jose~E. \& Vidal, Vicente} 2005 {SLEPc}: A
  scalable and flexible toolkit for the solution of eigenvalue problems. {\em
  ACM Transactions on Mathematical Software\/} {\bf 31}, 351.

\bibitem[{Hof} {\em et~al.\/}(2008){Hof}, {de Lozar}, {Kuik} \&
  {Westerweel}]{hof08}
{\sc {Hof}, B., {de Lozar}, A., {Kuik}, D.~J. \& {Westerweel}, J.} 2008
  {Repeller or attractor? Selecting the dynamical model for the onset of
  turbulence in pipe flow}. {\em Phys. Rev. Lett.\/} {\bf 101}, 214501.

\bibitem[{Hof} {\em et~al.\/}(2003){Hof}, {Juel} \& {Mullin}]{hof03}
{\sc {Hof}, B., {Juel}, A. \& {Mullin}, T.} 2003 {Scaling of the turbulence
  transition threshold in a pipe}. {\em Phys. Rev. Lett.\/} {\bf 91}, 244502.

\bibitem[{Hof} {\em et~al.\/}(2004){Hof}, {van Doorne}, {Westerweel},
  {Nieuwstadt}, {Faisst}, {Eckhardt}, {Wedin}, {Kerswell} \& {Waleffe}]{hof04}
{\sc {Hof}, B., {van Doorne}, C.~W.~H., {Westerweel}, J., {Nieuwstadt},
  F.~T.~M., {Faisst}, H., {Eckhardt}, B., {Wedin}, H., {Kerswell}, R.~R. \&
  {Waleffe}, F.} 2004 {Experimental observation of nonlinear traveling waves in
  turbulent pipe flow}. {\em {Science}\/} {\bf 305}, 1594.

\bibitem[{Hof} {\em et~al.\/}(2006){Hof}, {Westerweel}, {Schneider} \&
  {Eckhardt}]{hof06}
{\sc {Hof}, B., {Westerweel}, J., {Schneider}, T.~M. \& {Eckhardt}, B.} 2006
  {Finite lifetime of turbulence in shear flows}. {\em Nature\/} {\bf 443}, 59.

\bibitem[{Hwang} \& {Cossu}(2010)]{hwang2010}
{\sc {Hwang}, Y. \& {Cossu}, C.} 2010 {Self-sustained process at large scales
  in turbulent channel flow}. {\em Phys. Rev. Lett.\/} {\bf 105}, 044505.

\bibitem[{Hwang} \& {Cossu}(2011)]{hwang2011}
{\sc {Hwang}, Y. \& {Cossu}, C.} 2011 {Self-sustained processes in the
  logarithmic layer of turbulent channel flows}. {\em Phys. Fluids\/} {\bf 23},
  061702.

\bibitem[{Itano} \& {Toh}(2001)]{itano2001}
{\sc {Itano}, T. \& {Toh}, S.} 2001 {The dynamics of bursting process in wall
  turbulence}. {\em J. Phys. Soc. Jpn.\/} {\bf 70}, 703.

\bibitem[{K{\"a}pyl{\"a}} \& {Korpi}(2011)]{kapyla11}
{\sc {K{\"a}pyl{\"a}}, P.~J. \& {Korpi}, M.~J.} 2011 {Magnetorotational
  instability driven dynamos at low magnetic Prandtl numbers}. {\em Mon. Not.
  R. Astron. Soc.\/} {\bf 413}, 901.

\bibitem[{Kawahara} \& {Kida}(2001)]{kawahara01}
{\sc {Kawahara}, G. \& {Kida}, S.} 2001 {Periodic motion embedded in plane
  Couette turbulence: regeneration cycle and burst}. {\em J. Fluid Mech.\/}
  {\bf 449}, 291.

\bibitem[{Kawahara} {\em et~al.\/}(2012){Kawahara}, {Uhlmann} \& {van
  Veen}]{kawahara12}
{\sc {Kawahara}, G., {Uhlmann}, M. \& {van Veen}, L.} 2012 {The significance of
  simple invariant solutions in turbulent flows}. {\em Annu. Rev. Fluid
  Mech.\/} {\bf 44}, 203.

\bibitem[{Kerswell}(2005)]{kerswell05}
{\sc {Kerswell}, R.~R.} 2005 {Recent progress in understanding the transition
  to turbulence in a pipe}. {\em Nonlinearity\/} {\bf 18}, 17.

\bibitem[{Kerswell} \& {Tutty}(2007)]{kerswell07}
{\sc {Kerswell}, R.~R. \& {Tutty}, O.~R.} 2007 {Recurrence of travelling waves
  in transitional pipe flow}. {\em J. Fluid Mech.\/} {\bf 584}, 69.

\bibitem[{Kim} \& {Moehlis}(2008)]{kim08}
{\sc {Kim}, L. \& {Moehlis}, J.} 2008 {Characterizing the edge of chaos for a
  shear flow model}. {\em Phys. Rev. E\/} {\bf 78}, 036315.

\bibitem[{Knobloch}(1985)]{knobloch85}
{\sc {Knobloch}, E.} 1985 {The stability of non-separable barotropic and
  baroclinic shear flows}. {\em Astrophys. Space Sci.\/} {\bf 116}, 149.

\bibitem[{Knobloch} \& {Moore}(1986)]{knobloch86}
{\sc {Knobloch}, E. \& {Moore}, D.~R.} 1986 {Transition to chaos in
  two-dimensional double-diffusive convection}. {\em J. Fluid Mech.\/} {\bf
  166}, 409.

\bibitem[{Knobloch} \& {Weiss}(1981)]{knobloch81}
{\sc {Knobloch}, E. \& {Weiss}, N.~O.} 1981 {Bifurcations in a model of
  double-diffusive convection}. {\em Phys. Lett. A\/} {\bf 85}, 127.

\bibitem[{Korycansky}(1992)]{korycansky92}
{\sc {Korycansky}, D.~G.} 1992 {Growth and decay of disturbances in stratified
  shear flow in a rotating frame}. {\em Astrophys. J.\/} {\bf 399}, 176.

\bibitem[{Kreilos} \& {Eckhardt}(2012)]{kreilos12}
{\sc {Kreilos}, T. \& {Eckhardt}, B.} 2012 Periodic orbits near the onset of
  chaos in plane couette flow. {\em Chaos\/} {\bf 22}, 047505.

\bibitem[{Lan} \& {Cvitanovi{\'c}}(2008)]{lan08}
{\sc {Lan}, Y. \& {Cvitanovi{\'c}}, P.} 2008 {Unstable recurrent patterns in
  Kuramoto-Sivashinsky dynamics}. {\em Phys. Rev. E\/} {\bf 78}, 026208.

\bibitem[{Landahl}(1980)]{landahl80}
{\sc {Landahl}, M.~T.} 1980 A note on an algebraic instability of inviscid
  parallel shear flows. {\em J. Fluid Mech.\/} {\bf 98}, 243.

\bibitem[{Lebovitz}(2009)]{lebovitz2009}
{\sc {Lebovitz}, N.~R.} 2009 Shear-flow transition: the basin boundary. {\em
  Nonlinearity\/} {\bf 22}, 2645.

\bibitem[{Lebovitz}(2012)]{lebovitz2012}
{\sc {Lebovitz}, N.~R.} 2012 Boundary collapse in models of shear-flow
  transition. {\em Communications in Nonlinear Science and Numerical
  Simulation\/} {\bf 17}, 2095.

\bibitem[{Lesur} \& {Longaretti}(2005)]{lesur05}
{\sc {Lesur}, G. \& {Longaretti}, P.-Y.} 2005 {On the relevance of subcritical
  hydrodynamic turbulence to accretion disk transport}. {\em Astron.
  Astrophys.\/} {\bf 444}, 25.

\bibitem[{Lesur} \& {Longaretti}(2007)]{lesur07}
{\sc {Lesur}, G. \& {Longaretti}, P.-Y.} 2007 {Impact of dimensionless numbers
  on the efficiency of magnetorotational instability induced turbulent
  transport}. {\em Mon. Not. R. Astron. Soc.\/} {\bf 378}, 1471.

\bibitem[{Lesur} \& {Ogilvie}(2008{\natexlab{{\em a\/}}})]{lesur08b}
{\sc {Lesur}, G. \& {Ogilvie}, G.~I.} 2008{\natexlab{{\em a\/}}} {Localized
  magnetorotational instability and its role in the accretion disc dynamo}.
  {\em Mon. Not. R. Astron. Soc.\/} {\bf 391}, 1437.

\bibitem[{Lesur} \& {Ogilvie}(2008{\natexlab{{\em b\/}}})]{lesur08}
{\sc {Lesur}, G. \& {Ogilvie}, G.~I.} 2008{\natexlab{{\em b\/}}} {On
  self-sustained dynamo cycles in accretion discs}. {\em Astron. Astrophys.\/}
  {\bf 488}, 451.

\bibitem[{Lin} \& {Papaloizou}(1996)]{lin96}
{\sc {Lin}, D.~N.~C. \& {Papaloizou}, J.~C.~B.} 1996 {Theory of accretion disks
  II: application to observed systems}. {\em Annu. Rev. Astron. Astrophys.\/}
  {\bf 34}, 703.

\bibitem[{Longaretti} \& {Lesur}(2010)]{longaretti10}
{\sc {Longaretti}, P.-Y. \& {Lesur}, G.} 2010 {MRI-driven turbulent transport:
  the role of dissipation, channel modes and their parasites}. {\em Astron.
  Astrophys.\/} {\bf 516}, 51.

\bibitem[{Lord~Kelvin}(1887)]{kelvin1887}
{\sc {Lord~Kelvin}} 1887 Stability of fluid motion - rectilineal motion of
  viscous fluid between two parallel planes. {\em {Phil. Mag.}\/} {\bf 24},
  188.

\bibitem[{Lynden-Bell} \& {Pringle}(1974)]{lyndenbell74}
{\sc {Lynden-Bell}, D. \& {Pringle}, J.~E.} 1974 {The evolution of viscous
  discs and the origin of the nebular variables.} {\em Mon. Not. R. Astron.
  Soc.\/} {\bf 168}, 603.

\bibitem[{Manneville}(2009)]{manneville09}
{\sc {Manneville}, P.} 2009 {Spatiotemporal perspective on the decay of
  turbulence in wall-bounded flows}. {\em Phys. Rev. E\/} {\bf 79}, 025301.

\bibitem[{Mellibovsky} \& {Eckhardt}(2011)]{mellibovsky11}
{\sc {Mellibovsky}, F. \& {Eckhardt}, B.} 2011 {Takens-Bogdanov bifurcation of
  travelling-wave solutions in pipe flow}. {\em J. Fluid Mech.\/} {\bf 670},
  96.

\bibitem[{Mellibovsky} \& {Eckhardt}(2012)]{mellibovsky12}
{\sc {Mellibovsky}, F. \& {Eckhardt}, B.} 2012 {From travelling waves to mild
  chaos: a supercritical bifurcation cascade in pipe flow}. {\em J. Fluid
  Mech.\/} {\bf 709}, 149.

\bibitem[{Mellibovsky} {\em et~al.\/}(2009){Mellibovsky}, {Meseguer},
  {Schneider} \& {Eckhardt}]{mellibovsky09}
{\sc {Mellibovsky}, F., {Meseguer}, A., {Schneider}, T.~M. \& {Eckhardt}, B.}
  2009 {Transition in localized pipe flow turbulence}. {\em Phys. Rev. Lett.\/}
  {\bf 103}, 054502.

\bibitem[{Miesch} {\em et~al.\/}(2007){Miesch}, {Gilman} \&
  {Dikpati}]{miesch07}
{\sc {Miesch}, M.~S., {Gilman}, P.~A. \& {Dikpati}, M.} 2007 {Nonlinear
  evolution of global magnetoshear instabilities in a three-dimensional
  thin-shell model of the solar tachocline}. {\em ApJ. Supp. Ser.\/} {\bf 168},
  337.

\bibitem[{Moehlis} {\em et~al.\/}(2004{\natexlab{{\em a\/}}}){Moehlis},
  {Eckhardt} \& {Faisst}]{Moehlis2004a}
{\sc {Moehlis}, J., {Eckhardt}, B. \& {Faisst}, H.} 2004{\natexlab{{\em a\/}}}
  {Fractal lifetimes in the transition to turbulence}. {\em Chaos\/} {\bf 14},
  11.

\bibitem[{Moehlis} {\em et~al.\/}(2004{\natexlab{{\em b\/}}}){Moehlis},
  {Faisst} \& {Eckhardt}]{Moehlis2004b}
{\sc {Moehlis}, J., {Faisst}, H. \& {Eckhardt}, B.} 2004{\natexlab{{\em b\/}}}
  {A low-dimensional model for turbulent shear flows}. {\em New Journal of
  Physics\/} {\bf 6}, 56.

\bibitem[{Moehlis} {\em et~al.\/}(2005){Moehlis}, {Faisst} \&
  {Eckhardt}]{Moehlis2005}
{\sc {Moehlis}, J., {Faisst}, H. \& {Eckhardt}, B.} 2005 {Periodic orbits and
  chaotic sets in a low-dimensional model for shear flows}. {\em SIAM J. Appl.
  Dyn. Syst.\/} {\bf 4}, 352.

\bibitem[{Moehlis} {\em et~al.\/}(2002){Moehlis}, {Smith}, {Holmes} \&
  {Faisst}]{moehlis02}
{\sc {Moehlis}, J., {Smith}, T.~R., {Holmes}, P. \& {Faisst}, H.} 2002 {Models
  for turbulent plane Couette flow using the proper orthogonal decomposition}.
  {\em Phys. Fluids\/} {\bf 14}, 2493.

\bibitem[{Moffatt}(1977)]{moffatt77}
{\sc {Moffatt}, H.~K.} 1977 {\em {Magnetic field generation in electrically
  conducting fluids}\/}. Cambridge University Press.

\bibitem[{Monchaux} {\em et~al.\/}(2009){Monchaux}, {Berhanu}, {Aumaitre},
  {Chiffaudel}, {Daviaud}, {Dubrulle}, {Ravelet}, {Fauve}, {Mordant},
  {Petrelis}, {Bourgoin}, {Odier}, {Pinton}, {Plihon} \& {Volk}]{monchaux09}
{\sc {Monchaux}, R., {Berhanu}, M., {Aumaitre}, S., {Chiffaudel}, A.,
  {Daviaud}, F., {Dubrulle}, B., {Ravelet}, F., {Fauve}, S., {Mordant}, N.,
  {Petrelis}, F., {Bourgoin}, M., {Odier}, P., {Pinton}, J.-F., {Plihon}, N. \&
  {Volk}, R.} 2009 The {V}on {K}\'arm\'an {S}odium experiment: turbulent
  dynamical dynamos. {\em Phys. Fluids\/} {\bf 21}, 035108.

\bibitem[{Moore} {\em et~al.\/}(1983){Moore}, {Toomre}, {Knobloch} \&
  {Weiss}]{moore83}
{\sc {Moore}, D.~R., {Toomre}, J., {Knobloch}, E. \& {Weiss}, N.~O.} 1983
  {Period doubling and chaos in partial differential equations for
  thermosolutal convection}. {\em Nature\/} {\bf 303}, 663.

\bibitem[Moxey \& Barkley(2010)]{moxey2010}
{\sc Moxey, {D.} \& Barkley, {D.}} 2010 {Distinct large-scale turbulent-laminar
  states in transitional pipe flow}. {\em Proc. Nat. Acad. Sci.\/} {\bf
  107}~(18), 8091.

\bibitem[{Nagata}(1986)]{nagata86}
{\sc {Nagata}, M.} 1986 Bifurcations in {C}ouette flow between almost
  corotating cylinders. {\em J. Fluid Mech.\/} {\bf 169}, 229.

\bibitem[{Nagata}(1990)]{nagata90}
{\sc {Nagata}, M.} 1990 Three-dimensional finite-amplitude solutions in plane
  {C}ouette flow: bifurcation from infinity. {\em J. Fluid Mech.\/} {\bf 217},
  519.

\bibitem[{Newhouse}(1979)]{newhouse79}
{\sc {Newhouse}, S.~E.} 1979 The abundance of wild hyperbolic sets and
  nonsmooth stable sets for diffeomorphisms. {\em Publ. Math. I.H.E.S.\/} {\bf
  50}, 101.

\bibitem[{Ogilvie} \& {Pringle}(1996)]{ogilvie96}
{\sc {Ogilvie}, G.~I. \& {Pringle}, J.~E.} 1996 {The non-axisymmetric
  instability of a cylindrical shear flow containing an azimuthal magnetic
  field}. {\em Mon. Not. R. Astron. Soc.\/} {\bf 279}, 152.

\bibitem[{Oishi} \& {Mac Low}(2011)]{oishi11}
{\sc {Oishi}, J.~S. \& {Mac Low}, M.-M.} 2011 {Magnetorotational turbulence
  transports angular momentum in stratified disks with low magnetic Prandtl
  number but magnetic Reynolds number above a critical value}. {\em Astrophys.
  J.\/} {\bf 740}, 18.

\bibitem[{Orr}(1907)]{orr1907}
{\sc {Orr}, W.~M.} 1907 The stability or instability of the steady motions of a
  perfect liquid and of a viscous liquid. part {I}: a perfect liquid. {\em
  {Proc. R. Irish Acad. A.}\/} {\bf 27}, 9.

\bibitem[{Ott}(2002)]{ott02}
{\sc {Ott}, E.} 2002 {\em {Chaos in dynamical systems}\/}. Cambridge University
  Press.

\bibitem[{Palis} \& {Takens}(1993)]{palis93}
{\sc {Palis}, J. \& {Takens}, F.} 1993 {\em Hyperbolicity and sensitive chaotic
  dynamics at homoclinic bifurcations\/}. Cambridge University Press.

\bibitem[{Papaloizou} \& {Lin}(1995)]{papaloizou95}
{\sc {Papaloizou}, J.~C.~B. \& {Lin}, D.~N.~C.} 1995 {Theory of accretion disks
  I: angular momentum transport processes}. {\em Annu. Rev. Astron.
  Astrophys.\/} {\bf 33}, 505.

\bibitem[{Peixinho} \& {Mullin}(2006)]{peixinho06}
{\sc {Peixinho}, J. \& {Mullin}, T.} 2006 {Decay of turbulence in pipe flow}.
  {\em Phys. Rev. Lett.\/} {\bf 96}, 094501.

\bibitem[{Pomeau}(1986)]{pomeau86}
{\sc {Pomeau}, Y.} 1986 Front motion, metastability and subcritical
  bifurcations in hydrodynamics. {\em Physica D\/} {\bf 23}, 3.

\bibitem[{Pringle} {\em et~al.\/}(2009){Pringle}, {Duguet} \&
  {Kerswell}]{pringle09}
{\sc {Pringle}, C.~C.~T., {Duguet}, Y. \& {Kerswell}, R.~R.} 2009 {Highly
  symmetric travelling waves in pipe flow}. {\em Phil. Trans. R. Soc. Lond.
  A\/} {\bf 367}, 457.

\bibitem[{Pringle} \& {Kerswell}(2007)]{pringle07}
{\sc {Pringle}, C.~C.~T. \& {Kerswell}, R.~R.} 2007 {Asymmetric, helical, and
  mirror-symmetric traveling waves in pipe flow}. {\em Phys. Rev. Lett.\/} {\bf
  99}, 074502.

\bibitem[{Pringle}(1981)]{pringle81}
{\sc {Pringle}, J.~E.} 1981 {Accretion discs in astrophysics}. {\em Annu. Rev.
  Astron. Astrophys.\/} {\bf 19}, 137.

\bibitem[{Pumir}(1996)]{pumir96}
{\sc {Pumir}, A.} 1996 {Turbulence in homogeneous shear flows}. {\em Phys.
  Fluids\/} {\bf 8}, 3112.

\bibitem[{Rempel} {\em et~al.\/}(2010){Rempel}, {Lesur} \& {Proctor}]{rempel10}
{\sc {Rempel}, E.~L., {Lesur}, G. \& {Proctor}, M.~R.~E.} 2010 {Supertransient
  magnetohydrodynamic turbulence in Keplerian shear flows}. {\em Phys. Rev.
  Lett.\/} {\bf 105}, 044501.

\bibitem[{Reynolds}(1883)]{reynolds83}
{\sc {Reynolds}, O.} 1883 An experimental investigation of the circumstances
  which determine whether the motion of water shall be direct of sinuous and of
  the law of resistance in parallel channels. {\em Phil. Trans. R. Soc.\/} {\bf
  174}, 935.

\bibitem[{Rincon} {\em et~al.\/}(2007){Rincon}, {Ogilvie} \&
  {Proctor}]{rincon07b}
{\sc {Rincon}, F., {Ogilvie}, G.~I. \& {Proctor}, M.~R.~E.} 2007
  {Self-sustaining nonlinear dynamo process in Keplerian shear flows}. {\em
  Phys. Rev. Lett.\/} {\bf 98}, 254502.

\bibitem[{Rincon} {\em et~al.\/}(2008){Rincon}, {Ogilvie}, {Proctor} \&
  {Cossu}]{rincon08}
{\sc {Rincon}, F., {Ogilvie}, G.~I., {Proctor}, M.~R.~E. \& {Cossu}, C.} 2008
  {Subcritical dynamos in shear flows}. {\em Astron. Nachr.\/} {\bf 329}, 750.

\bibitem[{Robinson}(1983)]{robinson83}
{\sc {Robinson}, C.} 1983 Bifurcation to infinitely many sinks. {\em Commun.
  Math. Phys.\/} {\bf 90}, 433.

\bibitem[{Schmid} \& {Henningson}(2000)]{schmid00}
{\sc {Schmid}, P.~J. \& {Henningson}, D.~S.} 2000 {\em Stability and transition
  in shear flows\/}. Springer-Verlag, Berlin.

\bibitem[{Schmiegel}(1997)]{Schmiegel1997}
{\sc {Schmiegel}, A.} 1997 {Fractal stability border in plane couette Flow}.
  {\em Phys. Rev. Lett.\/} {\bf 79}, 5250.

\bibitem[{Schneider} {\em et~al.\/}(2006){Schneider}, {Eckhardt} \&
  {Yorke}]{schneider06}
{\sc {Schneider}, T., {Eckhardt}, B. \& {Yorke}, J.~A.} 2006 Edge of chaos in
  pipe flow. {\em Chaos\/} {\bf 16}, 041103.

\bibitem[{Schneider} {\em et~al.\/}(2007{\natexlab{{\em a\/}}}){Schneider},
  {Eckhardt} \& {Yorke}]{schneider07a}
{\sc {Schneider}, T., {Eckhardt}, B. \& {Yorke}, J.~A.} 2007{\natexlab{{\em
  a\/}}} Turbulence transition and the edge of chaos in pipe flow. {\em Phys.
  Rev. Lett.\/} {\bf 99}, 034502.

\bibitem[{Schneider} \& {Eckhardt}(2008)]{schneider08a}
{\sc {Schneider}, T.~M. \& {Eckhardt}, B.} 2008 {Lifetime statistics in
  transitional pipe flow}. {\em Phys. Rev. E\/} {\bf 78}, 046310.

\bibitem[{Schneider} {\em et~al.\/}(2007{\natexlab{{\em b\/}}}){Schneider},
  {Eckhardt} \& {Vollmer}]{schneider07b}
{\sc {Schneider}, T.~M., {Eckhardt}, B. \& {Vollmer}, J.} 2007{\natexlab{{\em
  b\/}}} {Statistical analysis of coherent structures in transitional pipe
  flow}. {\em Phys. Rev. E\/} {\bf 75}, 066313.

\bibitem[{Schneider} {\em et~al.\/}(2008){Schneider}, {Gibson}, {Lagha}, {de
  Lillo} \& {Eckhardt}]{schneider08b}
{\sc {Schneider}, T.~M., {Gibson}, J.~F., {Lagha}, M., {de Lillo}, F. \&
  {Eckhardt}, B.} 2008 {Laminar-turbulent boundary in plane Couette flow}. {\em
  Phys. Rev. E\/} {\bf 78}, 037301.

\bibitem[{Sim\'o}(1989)]{simo89}
{\sc {Sim\'o}, C.} 1989 In {\em Les M\'ethodes Modernes de la M\'ecanique
  C\'eleste (Goutelas '89)\/} (ed. D.~Benest \& C.~Froeschl\'e), p. 285.
  Editions Fronti\`eres, Gif-sur-Yvette.

\bibitem[{Simon} {\em et~al.\/}(2012){Simon}, {Beckwith} \&
  {Armitage}]{simon12}
{\sc {Simon}, J.~B., {Beckwith}, K. \& {Armitage}, P.~J.} 2012 {Emergent
  mesoscale phenomena in magnetized accretion disc turbulence}. {\em Mon. Not.
  R. Astron. Soc.\/} {\bf 422}, 2685.

\bibitem[{Simon} {\em et~al.\/}(2011){Simon}, {Hawley} \& {Beckwith}]{simon11}
{\sc {Simon}, J.~B., {Hawley}, J.~F. \& {Beckwith}, K.} 2011
  {Resistivity-driven state changes in vertically stratified accretion disks}.
  {\em Astrophys. J.\/} {\bf 730}, 94.

\bibitem[{Skufca} {\em et~al.\/}(2006){Skufca}, {Yorke} \&
  {Eckhardt}]{skufca06}
{\sc {Skufca}, J.~D., {Yorke}, J.~A. \& {Eckhardt}, B.} 2006 {Edge of chaos in
  a parallel shear flow}. {\em Phys. Rev. Lett.\/} {\bf 96}, 174101.

\bibitem[{Smale}(1967)]{smale67}
{\sc {Smale}, S.} 1967 {Differentiable dynamical systems}. {\em Bull. Amer.
  Math. Soc.\/} {\bf 73}, 747.

\bibitem[{Sparrow}(1982)]{sparrow82}
{\sc {Sparrow}, C.} 1982 {\em The Lorenz Equations: Bifurcation, Chaos and
  Strange Attractors\/}, {\em Appl. Math. Sci.\/}, vol.~41. Springer-Verlag.

\bibitem[{Spruit}(2002)]{spruit02}
{\sc {Spruit}, H.~C.} 2002 {Dynamo action by differential rotation in a stably
  stratified stellar interior}. {\em Astron. Astrophys.\/} {\bf 381}, 923.

\bibitem[{Steenbeck} {\em et~al.\/}(1966){Steenbeck}, {Krause} \&
  {R{\"a}dler}]{steenbeck66}
{\sc {Steenbeck}, M., {Krause}, F. \& {R{\"a}dler}, K.-H.} 1966 {Berechnung der
  mittleren LORENTZ-Feldst{\"a}rke $\protect\overline{\vec{v}\times\vec{B}}$
  f{\"u}r ein elektrisch leitendes Medium in turbulenter, durch
  CORIOLIS-Kr{\"a}fte beeinflu{\ss}ter Bewegung}. {\em Z. Naturforschung Teil
  A\/} {\bf 21}, 369.

\bibitem[Sterling {\em et~al.\/}(1999)Sterling, Dullin \& Meiss]{sterling99}
{\sc Sterling, D., Dullin, H.~R. \& Meiss, J.~D.} 1999 Homoclinic bifurcations
  for the {H}\'enon map. {\em Physica D\/} {\bf 134}, 153.

\bibitem[{Stone} {\em et~al.\/}(1996){Stone}, {Hawley}, {Gammie} \&
  {Balbus}]{stone96}
{\sc {Stone}, J.~M., {Hawley}, J.~F., {Gammie}, C.~F. \& {Balbus}, S.~A.} 1996
  {Three-dimensional magnetohydrodynamical simulations of vertically stratified
  accretion disks}. {\em Astrophys. J.\/} {\bf 463}, 656.

\bibitem[{Swift} \& {Wiesenfeld}(1984)]{swift84}
{\sc {Swift}, J.~W. \& {Wiesenfeld}, K.} 1984 {Suppression of period doubling
  in symmetric systems}. {\em Phys. Rev. Lett.\/} {\bf 52}, 705.

\bibitem[{Terquem} \& {Papaloizou}(1996)]{terquem96}
{\sc {Terquem}, C. \& {Papaloizou}, J.~C.~B.} 1996 {On the stability of an
  accretion disc containing a toroidal magnetic field}. {\em Mon. Not. R.
  Astron. Soc.\/} {\bf 279}, 767.

\bibitem[{Tobias} {\em et~al.\/}(2011){Tobias}, {Cattaneo} \&
  {Brummell}]{tobias11}
{\sc {Tobias}, S.~M., {Cattaneo}, F. \& {Brummell}, N.~H.} 2011 {On the
  generation of organized magnetic fields}. {\em Astrophys. J.\/} {\bf 728},
  153.

\bibitem[{Umurhan} \& {Regev}(2004)]{umurhan04}
{\sc {Umurhan}, O.~M. \& {Regev}, O.} 2004 {Hydrodynamic stability of
  rotationally supported flows: Linear and nonlinear 2D shearing box results}.
  {\em Astron. Astrophys.\/} {\bf 427}, 855.

\bibitem[{van Veen} \& {Kawahara}(2011)]{vanveen11}
{\sc {van Veen}, L. \& {Kawahara}, G.} 2011 {Homoclinic tangle on the edge of
  shear turbulence}. {\em Phys. Rev. Lett.\/} {\bf 107}, 114501.

\bibitem[{Velikhov}(1959)]{velikhov59}
{\sc {Velikhov}, E.~P.} 1959 Stability of an ideally conducting liquid flowing
  between cylinders rotating in a magnetic field. {\em Sov. Phys. JETP\/} {\bf
  36}, 1398.

\bibitem[{Viswanath}(2007)]{viswanath07}
{\sc {Viswanath}, D.} 2007 {Recurrent motions within plane Couette turbulence}.
  {\em J. Fluid Mech.\/} {\bf 580}, 339.

\bibitem[{Vollmer} {\em et~al.\/}(2009){Vollmer}, {Schneider} \&
  {Eckhardt}]{Vollmer2009}
{\sc {Vollmer}, J., {Schneider}, T. \& {Eckhardt}, B.} 2009 {Basin boundary,
  edge of chaos and edge state in a two dimensional model}. {\em New Journal of
  Physics\/} {\bf 11}, 013040.

\bibitem[{Waleffe}(1995{\natexlab{{\em a\/}}})]{waleffe95a}
{\sc {Waleffe}, F.} 1995{\natexlab{{\em a\/}}} {Hydrodynamic stability and
  turbulence: beyond transients to a self-sustaining process}. {\em Studies in
  Applied Math.\/} {\bf 95}, 319.

\bibitem[{Waleffe}(1995{\natexlab{{\em b\/}}})]{waleffe95b}
{\sc {Waleffe}, F.} 1995{\natexlab{{\em b\/}}} {Transition in shear flows.
  Nonlinear normality versus non-normal linearity}. {\em Phys. Fluids\/} {\bf
  7}, 3060.

\bibitem[{Waleffe}(1997)]{waleffe97}
{\sc {Waleffe}, F.} 1997 {On a self-sustaining process in shear flows}. {\em
  Phys. Fluids\/} {\bf 9}, 883.

\bibitem[{Waleffe}(1998)]{waleffe98}
{\sc {Waleffe}, F.} 1998 {Three-dimensional coherent states in plane shear
  flows}. {\em Phys. Rev. Lett.\/} {\bf 81}, 4140.

\bibitem[{Waleffe}(2001)]{waleffe01}
{\sc {Waleffe}, F.} 2001 {Exact coherent structures in channel flow}. {\em J.
  Fluid Mech.\/} {\bf 435}, 93.

\bibitem[{Waleffe}(2003)]{waleffe03}
{\sc {Waleffe}, F.} 2003 {Homotopy of exact coherent structures in plane shear
  flows}. {\em Phys. Fluids\/} {\bf 15}, 1517.

\bibitem[{Wang} {\em et~al.\/}(2007){Wang}, {Gibson} \& {Waleffe}]{wang07}
{\sc {Wang}, J., {Gibson}, J. \& {Waleffe}, F.} 2007 {Lower branch coherent
  states in shear flows: transition and control}. {\em Phys. Rev. Lett.\/} {\bf
  98}, 204501.

\bibitem[{Wedin} \& {Kerswell}(2004)]{wedin04}
{\sc {Wedin}, H. \& {Kerswell}, R.~R.} 2004 {Exact coherent structures in pipe
  flow: travelling wave solutions}. {\em J. Fluid Mech.\/} {\bf 508}, 333.

\bibitem[{Willis} {\em et~al.\/}(2013){Willis}, {Cvitanovi{\'c}} \&
  {Avila}]{willis13}
{\sc {Willis}, A.~P., {Cvitanovi{\'c}}, P. \& {Avila}, M.} 2013 Revealing the
  state space of turbulent pipe flow by symmetry reduction. {\em J. Fluid
  Mech.\/} {\bf 721}, 514.

\bibitem[{Willis} \& {Kerswell}(2007)]{willis07}
{\sc {Willis}, A.~P. \& {Kerswell}, R.~R.} 2007 {Critical behavior in the
  relaminarization of localized turbulence in pipe flow}. {\em Phys. Rev.
  Lett.\/} {\bf 98}, 014501.

\bibitem[{Willis} \& {Kerswell}(2009)]{willis2009}
{\sc {Willis}, A.~P. \& {Kerswell}, R.~R.} 2009 {Turbulent dynamics of pipe
  flow captured in a reduced model: puff relaminarization and localized `edge'
  states}. {\em J. Fluid Mech.\/} {\bf 619}, 213.

\bibitem[{Yorke} \& {Alligood}(1983)]{yorke83}
{\sc {Yorke}, J.~A. \& {Alligood}, T.} 1983 {Period doubling cascades of
  attractors: a prerequisite for horseshoes}. {\em Bull. Amer. Math. Soc.\/}
  {\bf 9}, 319.

\bibitem[{Zel'dovich} {\em et~al.\/}(1984){Zel'dovich}, {Ruzmaikin},
  {Molchanov} \& {Sokoloff}]{zeldo84}
{\sc {Zel'dovich}, Y.~B., {Ruzmaikin}, A.~A., {Molchanov}, S.~A. \& {Sokoloff},
  D.~D.} 1984 {Kinematic dynamo problem in a linear velocity field}. {\em J.
  Fluid Mech.\/} {\bf 144}, 1.

\end{thebibliography}

\end{document}